\documentclass[aps,pre,twocolumn,superscriptaddress,nofootinbib,longbibliography,floatfix]{revtex4-2}

\usepackage{amsmath,amssymb,bm}
\usepackage{appendix}
\usepackage{graphicx}
\usepackage{placeins}
\usepackage{mathtools}
\usepackage{xcolor}
\usepackage[normalem]{ulem}
\usepackage{cancel}

\newif\ifshowchanges
\showchangesfalse 

\ifshowchanges
  
  \newcommand{\del}[1]{{\color{black}\sout{#1}}}
  
  \newcommand{\delmath}[1]{\textcolor{black}{\cancel{#1}}}
\else
  
  \newcommand{\del}[1]{}
  
  \newcommand{\delmath}[1]{}
\fi

\newcommand{\dd}{\mathrm{d}}
\newcommand{\ee}{\mathrm{e}}
\newcommand{\ii}{\mathrm{i}}
\newcommand{\Dq}{D_{\!q}}
\newcommand{\phid}{\phi^\dagger}
\newcommand{\Lop}{\mathcal L}

\newcommand{\mean}[1]{\langle #1 \rangle}
\newcommand{\ktwo}{\kappa_2}

\begin{document}

\title{ Branching stochastic mechanics: Clustering and connected correlations within a branching-process representation of the Schrödinger equation}

\author{Eric Dumonteil}
\email{eric.dumonteil@cea.fr}
\affiliation{Université Paris-Saclay, CEA\\ Institut de Recherche sur les Lois Fondamentales de l'Univers, Gif-sur-Yvette, France}

\author{Benoît Bischoff}
\affiliation{Université Paris-Saclay, CEA\\ Institut de Recherche sur les Lois Fondamentales de l'Univers, Gif-sur-Yvette, France}
\affiliation{Universit\'e Paris-Saclay, Ecole Normale Supérieure Paris-Saclay,
Gif-sur-Yvette, France}

\author{Alain Letourneau}
\affiliation{Université Paris-Saclay, CEA\\ Institut de Recherche sur les Lois Fondamentales de l'Univers, Gif-sur-Yvette, France}

\author{Loïc Thulliez}
\affiliation{Université Paris-Saclay, CEA\\ Institut de Recherche sur les Lois Fondamentales de l'Univers, Gif-sur-Yvette, France}

\author{Corentin Doutre}
\affiliation{Université Paris-Saclay, CEA\\ Institut de Recherche sur les Lois Fondamentales de l'Univers, Gif-sur-Yvette, France}

\begin{abstract}

Can finite-range correlations hide in the statistics of an extended quantum state? The Schr\"odinger-Nagasawa transform represents the wave function by positive forward and backward diffusion fields whose product is the Born density. We promote them to branching superprocesses, \(\Phi_F\) and \(\Phi_B\): diffusion samples stochastic paths, whereas Bohm/Fisher-controlled branching generates genealogies of alternative continuations. With rates evaluated on the prescribed Born density, their means reproduce Schr\"odinger dynamics exactly. The connected sector exhibits supercritical, critical, and subcritical clustering in confinement and has critical dimension \(d_c=2\) in free space. Stationary eigenmodes remain extended; subcritical clusters acquire the reduced de~Broglie scale. We then let the branching rate respond to the fluctuating product \(\Phi_F\Phi_B\). In the reciprocal basis, the fields equal a smooth reference \(R\) plus centered fluctuations \(\psi_F,\psi_B\), defining the signed kernel \(C_{\rm FB}(x,y)=\mathbb E_\omega[\psi_F(x)\psi_B(y)]\). On the anticorrelated branch, \(\rho_{\rm BSM}(x)=-C_{\rm FB}(x,x)\) is the  positive organized connected weight.  Stationary matching of this weight to the Born profile is imposed as a closure condition, while off-diagonal decay defines the correlation range. The pair equation splits into collective and relative sectors: spectral cancellation selects the Born collective mode, while suppression of the leading density fluctuation selects the anticorrelated source channel. With relative diffusivity \(D_{\rm eff}\) and positive relaxation rate \(\mu_{\rm FB}\), correlations have screening length \(\xi_{\rm FB}=\sqrt{D_{\rm eff}/\mu_{\rm FB}}\).  Within this reciprocal closure, an extended Born-shaped connected weight and a finite correlation range can coexist in one stochastic kernel.
\end{abstract}


\maketitle

\section{Introduction}
\label{sec:introduction}

Can finite-range correlations hide in the statistics of a quantum
state whose wave function remains extended? Within standard quantum
mechanics the question cannot be formulated: the pure state of a
single particle is fully specified by its wave function, and an
extended stationary mode leaves no room for an additional, shorter
scale of spatial organization. The question becomes meaningful once
the wave function is embedded in a larger statistical object, as the
low-order moment of an ensemble whose higher moments are not fixed by
the mean. The purpose of the present paper is to construct such an
embedding and to characterize the information carried by the higher
moments.

The starting point of the present work is the observation that the
Schr\"odinger equation can be written exactly in a form that already
has the structure of a first-moment evolution. Reformulations of
nonrelativistic quantum mechanics in terms of real variables have a
long history
\cite{Madelung1927,Bohm1952,Schrodinger1931,Furth1933,Nelson1966,
fenyes_wahrscheinlichkeitstheoretische_1952}. In the polar
representation,
$\psi(x,t)=R(x,t)\exp[\ii S(x,t)/\hbar]$,
where $x$ denotes position, $t$ time, $\psi$ the
wave function, $R\geq0$ its amplitude, $S$ the real action field,
$\ii^2=-1$, and $\hbar$ the reduced Planck constant, the
Schr\"odinger equation is equivalent to a continuity equation for the
Born density $\rho=R^2$ and a Hamilton-Jacobi equation corrected by
the Bohm potential,
\begin{equation}
Q[\rho]
=
-\frac{\hbar^2}{2m}
\frac{\nabla^2\sqrt{\rho}}
{\sqrt{\rho}} .
\label{eq:bohm-intro}
\end{equation}
Here $m$ is the particle mass and $\nabla^2$ is the
spatial Laplacian.

A less familiar construction, due to Nagasawa
\cite{Nagasawa1993,Nagasawa2000}, goes one step further. Defining
\[
\phi=R\,e^{-S/\hbar},
\qquad
\phi^\dagger=R\,e^{S/\hbar},
\qquad
\rho=\phi^\dagger\phi ,
\]
one obtains, under the usual regularity, phase, and boundary conditions,
the reciprocal diffusion-reaction pair
\begin{equation}
\partial_t\phi=\mathcal F[\rho]\,\phi,
\qquad
\partial_t\phi^\dagger=-\mathcal F[\rho]\,\phi^\dagger,
\label{eq:nagasawa-intro}
\end{equation}
with
\begin{equation}
\mathcal F[\rho]
=
\frac{\hbar}{2m}\nabla^2
+
\frac{V+2Q[\rho]}{\hbar},
\label{eq:F-intro}
\end{equation}
where $V(x,t)$ is the external scalar potential and
$\mathcal F[\rho]$ denotes the displayed density-dependent
diffusion-reaction operator. Here and below, $\rho$ denotes the
Born density of the prescribed Schr\"odinger state.

The Schr\"odinger-Nagasawa (SN) equations
\eqref{eq:nagasawa-intro}, together with
$\rho=\phi^\dagger\phi=R^2$ and the usual phase and boundary
conditions, are exactly equivalent to the Schr\"odinger dynamics.
At this stage nothing stochastic has been added: $\phi$ and
$\phi^\dagger$ are smooth deterministic fields carrying the same
information as the wave function. Their form, however, is highly
suggestive. Each equation has precisely the structure of the
first-moment equation of a diffusion-reaction process, and hence of
a suitably chosen local branching-killing process, with one-sector
net rate controlled by
$(V+2Q[\rho])/\hbar$.

This observation is the pivot of the present construction. A
first-moment equation does not determine the stochastic process that
produces it: infinitely many stochastic laws may share the same mean
while differing in their quadratic variation, genealogy, connected
correlations, and higher cumulants. Once the SN fields are recognized
as first moments, it is therefore natural to ask what is obtained by
retaining the branching process itself rather than only its mean.
The wave function then fixes the one-point sector, while the higher
statistical orders remain available to carry information not contained
in that sector.

A canonical measure-valued completion associated with local branching
is obtained by promoting each SN field to a Dawson-Watanabe
superprocess
\cite{ikeda_nagasawa_watanabe_1968_I,Dawson1993,Etheridge2000,
legall_1999}, whose square-root branching noise is generated by the
same local reproduction events that produce the mean evolution. We
refer to this promotion as the \emph{stochastic lift}, and to the
resulting pair of reciprocal superprocesses
$(\Phi_F,\Phi_B)$, admitting the SN fields as first moments, as
\emph{branching stochastic mechanics} (BSM). In earlier work with
Ikeda and Watanabe, Nagasawa had also contributed to the
probabilistic foundations of branching Markov processes
\cite{ikeda_nagasawa_watanabe_1968_I,
ikeda_nagasawa_watanabe_1968_II,
ikeda_nagasawa_watanabe_1969_III,nagasawa_1968_sign}.
The probabilistic theory of branching processes and Nagasawa's later
real-diffusion formulation of quantum mechanics therefore constitute
two closely related strands of his work; the present paper brings
these structures together.

The physical significance of retaining the full hierarchy is familiar
from branching systems. Such stochastic processes couple the transport of individuals, particles, or walkers to birth-death mechanisms and arise across physics and biology, from epidemic outbreaks and their spatial spread \cite{dumonteil_spatial_2013}, population ecology and species dispersal, to reaction-diffusion systems and neutron transport in multiplying media. At the level of the
mean, the expected density of branching random walkers obeys a closed,
linear diffusion-reaction equation. The stochastic ensemble summarized
by this mean, however, also carries a genealogy: descendants sharing a
common ancestor are statistically correlated, and these connected
correlations can organize the two-point sector in ways that are
completely invisible to the one-point density. In prototypical
birth-death-diffusion processes, this genealogical organization
produces spatial patterns and strongly inhomogeneous realizations even when
the ensemble-averaged density remains smooth
\cite{Zhang1990,Young2001,Houchmandzadeh2008,Houchmandzadeh2009,
ferte_clusters_2023}.

Neutron transport in multiplying media provides a particularly clear
physical example of this separation between statistical orders. There,
the mean neutron density describes only the first level of the branching
process, whereas the second and higher moments retain common fission
ancestry, spatial correlations, fluctuations, and departures from
ergodic sampling that leave no trace in the mean
\cite{BellGlasstone1970,Pazsit2008}. Rigorous large-generation
asymptotics have recently been obtained for all population moments of
bounded neutron-transport branching processes, including the linear
growth of the second moment at criticality and the associated survival
and Yaglom limits \cite{dumonteil_moment_2025}. Neutron clustering,
which is also of direct interest for nuclear safety, has been identified
in Monte Carlo criticality simulations
\cite{Dumonteil2014}, characterized in confined geometries both for
freely evolving populations and for branching systems with an imposed
fixed-population constraint \cite{Zoia2014,DeMulatier2015}, and observed
experimentally through neutron correlations in a zero-power reactor
\cite{Dumonteil2021}. In this setting, clustering is not a numerical
artifact: it is the spatial organization of correlated descendants
generated by the branching genealogy. The one-point density may remain
smooth and extended while the connected two-point measure develops a
much shorter relative scale. This provides the central intuition for
the role assigned below to the connected sector of the lifted quantum
representation.

In the BSM representation, an elementary branch represents a
\emph{possible continuation}: a locally admissible stochastic
continuation of the dynamics compatible with the prescribed drift,
diffusion, boundary data, and action functional. The term ``possible''
is used here in this restricted operational sense. These continuations
are bookkeeping degrees of freedom of the stochastic representation
and are not interpreted as hidden material particles. Within this
representation, diffusion and branching play complementary exploratory
roles: diffusion explores the possible paths of each continuation,
whereas branching explores the possible continuations themselves -- the
creation, reweighting, and termination of locally admissible
alternatives.

Branching is essential here because these alternatives are not sampled
independently: possible continuations generate further possible
continuations. The stochastic lift therefore constructs more than a
distribution of independent paths. It defines a measure-valued ensemble
that retains genealogy, so that descendants issued from a common
continuation remain statistically related even after their paths have
separated. The connected moments are precisely the statistical record
of this inherited dependence. In this sense, the superprocess provides
a measure on the possible continuations while preserving the
correlations generated by the fact that possibilities themselves beget
possibilities.

The two reciprocal sectors have complementary roles. The forward field
propagates possible continuations from the initial data, whereas the
backward field weights their compatibility with the conjugate
conditioning and boundary constraints. After the stochastic lift, we
denote by $\Phi_F(x,t)$ and $\Phi_B(x,t)$ the realized positive forward
and backward branching fields, and by $\mathbb E_\omega$ the average over
their stochastic realizations. Their common reciprocal representatives
are decomposed as
\begin{equation}
\ee^{S/\hbar}\Phi_F=R+\psi_F,
\qquad
\ee^{-S/\hbar}\Phi_B=R+\psi_B,
\label{eq:intro-centered-fields}
\end{equation}
with
$\mathbb E_\omega[\psi_F]=\mathbb E_\omega[\psi_B]=0$.
The signed connected reciprocal kernel is
\begin{equation}
C_{\rm FB}(x,y,t)
=
\mathbb E_\omega
\!\left[\psi_F(x,t)\psi_B(y,t)\right],
\label{eq:CFB-intro}
\end{equation}
and the  organized connected weight is defined on its anticorrelated branch by
\begin{equation}
\rho_{\rm BSM}(x,t)
=
-C_{\rm FB}(x,x,t).
\label{eq:rhoBSM-intro}
\end{equation}
This sign convention assigns positive weight to the organized sector
selected by $\psi_B\simeq-\psi_F$.  Within the centered reference, the mean reciprocal product satisfies
\begin{equation}
\begin{aligned}
\underbrace{\mathbb E_\omega[\Phi_F\Phi_B]}_
 {\substack{\text{mean reciprocal}\\\text{product}}}
&=\underbrace{R^2}_{\text{Born reference}}
+\underbrace{C_{\rm FB}(x,x,t)}_
 {\substack{\text{signed connected}\\\text{correction}}}
\\
&=\rho-\rho_{\rm BSM}.
\end{aligned}
\label{eq:intro-product-budget}
\end{equation}
 While the normalized mean product is a natural candidate probability density
when its integral is finite and strictly positive, this interpretation
is not pursued here. The quantity $\rho_{\rm BSM}$ instead denotes an
organized connected weight, not a probability density: on the
anticorrelated branch, it measures the deficit of the mean product
relative to the Born reference. Initially the independent reciprocal fields
give $C_{\rm FB}=0$. Stationary organization is matched to the prescribed
Born profile through $\rho_{\rm BSM}^{\star}=\rho$.

The SN correspondence fixes the first-moment generator of this
representation, but it does not uniquely determine its higher
stochastic structure. The branching quadratic variation -- the local
rate at which the branching martingale accumulates variance -- the
pair-source normalization, and the joint forward-backward covariance
constitute additional second-order data. When the local reaction rate
is evaluated on a prescribed Born density, the two marginal mean fields
reproduce the SN reference dynamics exactly, while the connected
moments describe genealogy and clustering within the chosen branching
ensemble.

The introductory question can now be formulated precisely. Can the
signed connected kernel in Eq.~\eqref{eq:CFB-intro} acquire a diagonal
$-C_{\rm FB}(x,x,t)$ equal to the extended Born profile while developing a
finite correlation range away from it? In this formulation, the
 collective weight and the relative correlations are different
projections of one stochastic two-field observable. The neutron
analogy suggests how such a separation can arise: the one-point sector
need not reveal the spatial organization carried by correlated
descendants in the two-point sector.

This dictionary between branching processes and quantum mechanics must
be distinguished from its well-known numerical counterpart. Diffusion
Monte Carlo (DMC) methods sample the Wick-rotated Schr\"odinger
equation, $t\rightarrow\ii\tau$, by means of an ensemble of branching
random walkers
\cite{kalos_monte_1962,anderson_randomwalk_1975,
ceperley_introduction_2004,schimansky-geier_quantum_1997}. These
methods themselves descend historically from neutron-transport Monte
Carlo \cite{metropolis_monte_1949,spanier_monte_1969}. The lineage is
shared, but the status assigned to branching fluctuations is different.
In DMC, branching is primarily a numerical device: correlations between
walkers and finite-population fluctuations appear as population-control
biases to be reduced
\cite{hetherington_observations_1984,umrigar_diffusion_1993,
assaraf_diffusion_2000,nemec_diffusion_2010}, and walker genealogies
are mostly analyzed as a source of statistical error
\cite{del_moral_feynman-kac_2004,kosztin_introduction_1996}. Here, by
contrast, the Nagasawa pair is an exact real-time rewriting of
Schr\"odinger dynamics rather than an imaginary-time projection scheme,
and the connected fluctuations of the branching ensemble are promoted
from numerical nuisance to candidate carriers of physical structure,
as they are in neutron transport. The analogy pursued here is therefore
structural rather than algorithmic.

A caveat is in order before proceeding. Real-variable and
trajectory-based formulations of quantum mechanics face well-known
difficulties involving nodes, phase quantization, conditioning, and
nonlocal correlations. In particular, the Wallstrom objection shows
that hydrodynamic or stochastic variables alone do not recover the full
Schr\"odinger theory unless the appropriate single-valuedness or
circulation conditions are also imposed \cite{Wallstrom1994}. The
present construction does not attempt to assign a unique underlying
stochastic trajectory to a particle, nor does it claim to resolve this
objection. The Schr\"odinger-Nagasawa equivalence, together with its
phase and boundary conditions, is imposed first, and the branching
hierarchy is built on top of it.

The language of possible continuations also has a precise continuum
meaning. Assigning a vanishing bookkeeping mass to each elementary
branch while increasing the branching frequency at fixed quadratic
variation yields the standard Dawson-Watanabe measure-valued limit
\cite{Dawson1993,Perkins2002,Etheridge2000}: the limiting object is a
random measure over possible continuations, not a finite population of
hidden particles. Two possible continuations sampled independently
from the same mean field carry no connected correlation; such
correlations require common branching ancestry, a common stochastic
source, or subsequent forward-backward compatibility feedback. Two
second-order objects accordingly organize the analysis: one-sector
connected functions, which record genealogy within a single random
measure, and the signed connected kernel $C_{\rm FB}$, which records
reciprocal compatibility between the two reduced fields.

The paper develops the construction in two stages. In the first stage,
the branching rate is evaluated on the prescribed Born density. The two
marginal mean fields then reproduce the Schr\"odinger-Nagasawa dynamics
exactly, while the connected hierarchy becomes exactly solvable. A
dimensionless control parameter comparing reaction and diffusion
separates an extended, ergodic sector from a clustered connected sector.
In confinement, the hierarchy recovers the supercritical, critical, and
subcritical regimes of branching random walks; in free space its
marginal behavior has the critical dimension $d_c=2$. Stationary
Schr\"odinger eigenmodes remain in the extended sector, as required by
ordinary quantum mechanics. On the subcritical branch of the frozen
hierarchy, by contrast, connected pairs acquire a finite relative size
set by the reduced de~Broglie wavelength of the corresponding kinetic
scale, and the condition for clustering takes the form of the standard
semiclassical crossover. The frozen hierarchy therefore identifies the
natural quantum length associated with a localized connected sector,
while showing at the same time that an ordinary stationary
Schr\"odinger state is not driven onto that branch by the one-sector
dynamics alone.

In the second stage, the Bohm/Fisher contribution is allowed to respond
to the fluctuating reciprocal product of the two branching fields. The
forward and backward sectors must then be treated jointly, and the signed
connected kernel $C_{\rm FB}$ becomes the central dynamical object. We
derive its evolution equation and separate the collective diagonal
$\rho_{\rm BSM}=-C_{\rm FB}(x,x)$ from its relative profile.  Matching the connected weight to the prescribed Born profile at
reciprocal saturation is imposed through the diagonal
condition $\rho_{\rm BSM}^{\star}=\rho$, while positive
semidefiniteness of the joint
source covariance, together with suppression of the leading
realization-level density fluctuation, favors an anticorrelated
forward-backward channel.

The crucial point is that the diagonal and relative sectors need not
share the same relaxation scale. If the dressed diagonal-preserving
relative response acquires a positive infrared relaxation rate
$\mu_{\rm FB}$, the off-diagonal correlations are screened over the
length
\begin{equation}
\xi_{\rm FB}^{2}
=
\frac{D_{\rm eff}}{\mu_{\rm FB}} ,
\label{eq:xiFB-intro}
\end{equation}
where $D_{\rm eff}$ is the dressed relative diffusivity of the same
 mode. The coefficient $\mu_{\rm FB}$ is an effective parameter of
the infrared closure; its dynamical selection is not derived here.
A response-field treatment of the dressed reciprocal coefficient is
developed separately~\cite{bischoff_branching_2026}.
The result established below is conditional on stationary diagonal
matching and a positive relative relaxation rate: an extended
Born-shaped connected weight and a finite-range relative profile
can coexist within the specified reciprocal closure. The frozen hierarchy supplies
the natural de~Broglie scale associated with such a localized connected
sector,  while the reciprocal pair equation identifies the channel in which
a finite relative scale can arise under the stated closure assumptions.

The mathematical status of these steps should be distinguished. The
marginal frozen processes are Dawson-Watanabe superprocesses, rigorously
defined by their martingale problems; the local SPDE notation is used
only when the corresponding random measures admit densities. The
connected-moment hierarchy is exact for a local Markov branching process
with prescribed rates, while its use with $Q[\rho]$ frozen on a
prescribed Born density constitutes the frozen quantum-statistical
benchmark developed here. The nonlinear Bohm/Fisher feedback is a
separate regulated extension, because the Bohm functional cannot be
applied directly to rough measure-valued realizations and must instead
be evaluated on a smoothed reciprocal product. Thus the exact
superprocess hierarchy, the frozen quantum-statistical construction,
and the interacting reciprocal closure have distinct logical status
throughout the paper.

Once the feedback is introduced, the relevant second-order object is the
signed connected reciprocal kernel. Recovery of the prescribed quantum
state is the stationary diagonal condition
\begin{equation}
-C_{\rm FB}^{\star}(x,x)=\rho(x),
\end{equation}
equivalently $\rho_{\rm BSM}^{\star}=\rho$. The remaining dynamical problem
is to determine whether the full joint law can simultaneously maintain
this diagonal matching and generate a positive finite-range relative
response.

The paper is organized as follows. Section~\ref{sec:nagasawa} develops
the Schr\"odinger-Nagasawa diffusion pair and the branching dictionary.
Section~\ref{sec:bbm} constructs the superprocess limit, derives the
associated moment hierarchy, and introduces the two genealogical
forward-backward superprocesses and their gauge-covariant covariance.
Sections~\ref{sec:confined} and~\ref{sec:dimension} analyze the
prescribed-rate clustering regimes in confinement and in free space.
Section~\ref{sec:reciprocal-overlap} specifies the joint
forward-backward second-moment law, identifies the anticorrelated sign
favored by suppression of the linear realization-level density noise,
derives the exact diagonal/relative decomposition, states the screened
solution of the positive-gap infrared regime, and describes the
transport of correlated branches. Finally, Sec.~\ref{sec:discussion}
distinguishes the exact, effective, and conditional steps, examines
stationary matching of the pair-kernel diagonal, and identifies the
remaining open problems requiring a self-consistent dynamical closure.

\section{Schr\"odinger-Nagasawa diffusion pair}
\label{sec:nagasawa}

We consider a spinless particle of mass \(m\) in $d$ spatial dimensions, with Hamiltonian
\begin{equation}
\widehat H
=
-\frac{\hbar^2}{2m}\nabla^2+V(x,t)
\end{equation}
-- where \(\widehat H\) is the Hamiltonian operator and
\(V(x,t)\) is the external scalar potential --, and wave function
\begin{equation}
\psi(x,t)=R(x,t)\exp\!\left[\frac{\ii S(x,t)}{\hbar}\right].
\end{equation}
We introduce the two real fields
\begin{equation}
\phi
=
R\,\ee^{-S/\hbar},
\qquad
\phid
=
R\,\ee^{S/\hbar},
\qquad
\rho
=
\phid\phi
=
R^2 .
\label{eq:phi-def}
\end{equation}
Under the usual regularity, phase, and boundary assumptions, the
Schr\"odinger equation and its complex conjugate are equivalent to the
pair
\begin{align}
\partial_t\phi
&=
\Dq\nabla^2\phi
+
\left(
\frac{V}{\hbar}
+
\frac{2Q[\rho]}{\hbar}
\right)\phi,
\label{eq:forward}\\
\partial_t\phid
&=
-\Dq\nabla^2\phid
-
\left(
\frac{V}{\hbar}
+
\frac{2Q[\rho]}{\hbar}
\right)\phid,
\label{eq:backward}
\end{align}
where
\begin{equation}
\Dq=\frac{\hbar}{2m}.
\label{eq:Dq}
\end{equation}
The derivation is recalled in Appendix~\ref{app:nagasawa}.

The opposite signs in Eqs.~\eqref{eq:forward} and
\eqref{eq:backward} are the real-time remnant of complex conjugation.
The resulting system is a pair of adjoint diffusion-reaction equations
and is closely related to reciprocal-process and Doob-transformed
diffusion constructions \cite{Nagasawa1993,Nagasawa2000}. On a nodal
domain where the weighting fields have a fixed sign, the backward field
may be viewed as weighting the forward diffusion. A global identification
with a positive Doob $h$-transform is more delicate in the presence of
nodes and must be supplemented by the phase and circulation conditions
required for full equivalence with quantum mechanics.

In compact form, the pair recovers Eqs.~\eqref{eq:nagasawa-intro} and \eqref{eq:F-intro}, with the non-linear dependence on
\(\rho\) being a consequence of replacing the complex-valued evolution by a pair of coupled real diffusion equations.

The Bohm potential can also be written as the variational derivative of
the Fisher information
\cite{Frieden1998,hall_schrodinger_2002},
\begin{equation}
I[\rho]
=
\int\dd^dx\,
\frac{(\nabla\rho)^2}
{\rho},
\qquad
\frac{2Q[\rho]}{\hbar}
=
\frac{\hbar}{4m}
\frac{\delta I}{\delta\rho}.
\label{eq:fisher}
\end{equation}
The Bohm contribution to the diffusion-reaction rate is therefore fixed
by the local Fisher geometry of the Born density.

\subsection{First-moment branching representation}

For a prescribed density
\(\rho\), Eq.~\eqref{eq:forward} has the form of a
linear diffusion-reaction equation. It has the same first-moment
structure as a branching Brownian motion or as the equation for the mean
neutron density $n$ in a multiplying medium
\cite{BellGlasstone1970,Pazsit2008},
\begin{equation}
\partial_t n
=
\Dq\nabla^2 n+\beta n,
\qquad
\beta
=
\lambda(\nu_1-1),
\label{eq:bbm-mean}
\end{equation}
where $\lambda$ is the event branching rate and
\begin{equation}
\nu_1
=
\mathbb E[K]
=
\sum_k k\,p_k
\end{equation}
is the first offspring moment, with $p_k=\Pr(K=k)$.
In this neutron transport setting, $K$ is the number of descendant neutrons emitted in a neutron-induced fission event. 

In the present quantum branching representation, $K$ counts locally distinct
outgoing continuations generated by an effective branching event. These
continuations are bookkeeping degrees of freedom of the stochastic
representation; they are not actual physical particles. Unlike neutron
transport, which has a single population sector, the quantum representation
carries two conjugate sectors, forward and backward. We use ``one-sector''
throughout for statements concerning either real field taken separately, as
opposed to paired quantities built from
\(\rho=\phid\phi\). Reading
Eq.~\eqref{eq:forward} in the form \eqref{eq:bbm-mean} identifies the
one-sector net rate as
\begin{equation}
\beta_{\rm q}(x,t)
=
\frac{V(x,t)+2Q[\rho](x,t)}{\hbar} .
\label{eq:betaq}
\end{equation}
The branching terminology must be understood as an effective
branching-killing representation of possible continuations. The conserved physical quantity is not the population
represented by either real field separately, but their bilinear product.
At this stage,
\(\rho\) is a prescribed smooth density, so that
\(Q[\rho]\) is well defined. Section~\ref{sec:bbm} promotes each marginal mean field to a
positive superprocess. Only the later feedback of \(Q\) on their
regulated stochastic product, which couples the two rough measures, is
treated as an effective interacting extension.

The correspondence between neutron transport and the quantum branching representation is
summarized in Table~\ref{tab:dictionary}. 

\begin{table}[t]
\caption{
\label{tab:dictionary}
Structural dictionary between neutron transport in fissile media and the
branching representation of quantum mechanics -- referred to as branching stochastic mechanics (BSM). The Born density is
\(\rho=\phid\phi\). After the stochastic lift, the centered reciprocal
fields define
\(C_{\rm FB}=\mathbb E_\omega[\psi_F\psi_B]\) and the  organized connected weight \(\rho_{\rm BSM}=-C_{\rm FB}(x,x)\). Stationary reciprocal
matching requires \(\rho_{\rm BSM}^{\star}=\rho\). The last two rows state the structural
correspondence obtained in the prescribed-rate frozen hierarchy.}
\footnotesize
\renewcommand{\arraystretch}{1.25}
\begin{ruledtabular}
\begin{tabular}{|p{0.38\columnwidth}|p{0.57\columnwidth}|}
\hline
\textbf{Neutron transport} &
\textbf{BSM} \\
\hline

neutrons outgoing from a fission event &
locally admissible possible continuations (degrees of freedom) \\
\hline

mean neutron density $n(x,t)$ &
one-sector mean compatibility weight $\phi$ or $\phid$ \\
\hline

diffusion coefficient $D$ &
$\Dq=\hbar/(2m)$ \\
\hline

net rate $\beta=\lambda(\nu_1{-}1)$ &
one-sector rate $\beta_{\rm q}=(V+2Q)/\hbar$ in the fixed energy gauge \\
\hline

offspring-multiplicity fluctuations, which set the
finite connected-pair source &
local branching multiplicity and the corresponding
connected-pair source; its coefficient is derived with the
measure-valued limit in Sec.~\ref{sec:bbm} \\
\hline

fixed-population constraint (when imposed) &
absent from the analytic superprocess limit; independently, the Born
density satisfies \(\int\rho=1\), while stationary reciprocal matching
requires \(\int\rho_{\rm BSM}=1\) \\
\hline

branching criticality &
zero principal exponent of $\Dq\nabla^2+\beta_{\rm q}$ \\
\hline

genealogical clustering associated with subcriticality in unconstrained
branching &
localized connected regime of the frozen quantum branching representation (particle-like organization) \\
\hline

extended genealogical exploration associated with supercriticality in
unconstrained branching &
extended ergodic regime of the frozen quantum branching representation
(field-like organization) \\
\hline
\end{tabular}
\end{ruledtabular}
\end{table}

\subsection{Energy gauge and one-sector growth rate}

A potential energy is defined only up to an additive constant. Under
\begin{equation}
V(x,t)\longrightarrow V(x,t)+c ,
\end{equation}
the Schr\"odinger equation is unchanged provided the wave function acquires
an overall time-dependent phase,
\begin{equation}
\psi
\longrightarrow
\ee^{-\ii ct/\hbar}\psi ,
\end{equation}
which describes the same physical state. In the polar representation
\eqref{eq:phi-def}, this phase is carried entirely by the action,
\begin{equation}
R\longrightarrow R,
\qquad
S\longrightarrow S-ct .
\end{equation}
The shift of $S$ in turn rescales the two real fields by conjugate
exponential factors,
\begin{equation}
\phi
\longrightarrow
\ee^{+ct/\hbar}\phi,
\qquad
\phid
\longrightarrow
\ee^{-ct/\hbar}\phid ,
\label{eq:gauge-fields}
\end{equation}
so that the amplitude, and hence the Born density, is left invariant:
\begin{equation}
\phid\phi
\longrightarrow
\phid\phi
=
\rho .
\end{equation}
The one-sector rate, however, is not invariant. It transforms as
\begin{equation}
\beta_{\rm q}
\longrightarrow
\beta_{\rm q}+\frac{c}{\hbar} .
\label{eq:gauge-beta}
\end{equation}

The absolute growth or decay rate of either real field separately is
therefore gauge dependent. Only the paired dynamics and observables built
from \(\rho=\phid\phi\)
, and, after the lift, from the gauge-invariant product
\(\Phi_F\Phi_B\), are invariant. Statements concerning the sign of
$\beta_{\rm q}$, and the corresponding one-sector branching
classification, require a fixed energy convention. Throughout this paper, the energy reference is fixed as follows: in the
infinite-well benchmark, $V=0$ inside the well, the confinement being
imposed by the boundary conditions (Dirichlet walls) rather than by a
finite potential term; in free space, $V(x)\to0$ as $|x|\to\infty$. All
subsequent references to the sign of $\beta_{\rm q}$ are understood in
this fixed gauge.

More generally, when the one-sector rate is spatially dependent or the
geometry allows leakage, branching criticality is not determined by the
pointwise condition $\beta_{\rm q}=0$. It corresponds to the vanishing
of the principal growth exponent of the operator
\begin{equation}
\mathcal L_{\rm q}
=
\Dq\nabla^2+\beta_{\rm q}(x).
\label{eq:quantum-growth-operator}
\end{equation}
The simpler criterion $\beta_{\rm q}=0$ applies to a homogeneous rate in
a non-leaking geometry whose diffusion operator has a zero fundamental
eigenvalue.

\subsection{Conservation of the Born density}

Multiplying Eq.~\eqref{eq:forward} by $\phid$, multiplying
Eq.~\eqref{eq:backward} by $\phi$, and adding the two equations gives
\begin{align}
\partial_t\rho
&=
\Dq
\left(
\phid\nabla^2\phi
-
\phi\nabla^2\phid
\right)
\nonumber\\
&=
\Dq\nabla\cdot
\left(
\phid\nabla\phi
-
\phi\nabla\phid
\right).
\label{eq:rho-continuity-fields}
\end{align}
Introducing the current
\begin{equation}
\mathbf j
=
\Dq
\left(
\phi\nabla\phid
-
\phid\nabla\phi
\right),
\label{eq:current-fields}
\end{equation}
one recovers
\begin{equation}
\partial_t\rho
+\nabla\cdot\mathbf j=0.
\label{eq:rho-continuity}
\end{equation}
Using Eq.~\eqref{eq:phi-def}, the current reduces to the usual quantum
current,
\begin{equation}
\mathbf j
=
\rho\,
\frac{\nabla S}{m}.
\label{eq:current-bohm}
\end{equation}
For vanishing normal current, Dirichlet boundary conditions, or
sufficiently rapid decay at infinity,
\begin{equation}
\frac{\dd}{\dd t}
\int\dd^dx\,\rho(x,t)
=
0.
\end{equation}
With the usual normalization,
\begin{equation}
\int\dd^dx\,\rho(x,t)
=
\int\dd^dx\,\phid(x,t)\phi(x,t)
=
1.
\label{eq:mass}
\end{equation}
Neither $\int\phi$ nor $\int\phid$ is separately conserved; the conserved physical measure is the paired Born density $\rho$. 

\subsection{Stationary states and reciprocal spectral weighting}

Consider a stationary eigenstate
\begin{equation}
\psi_n(x,t)
=
u_n(x)\,
\ee^{-\ii E_nt/\hbar},
\qquad
\widehat H u_n=E_nu_n,
\label{eq:stationary-state}
\end{equation}
where $u_n$ may be chosen real. We denote by \(Q_n\) the
Bohm potential associated with this eigenstate on any one of its nodal
domains. Away from its nodes, the
Hamilton-Jacobi equation gives
\begin{equation}
V(x)+Q_n(x)=E_n.
\label{eq:stationary-HJ}
\end{equation}

Because a real excited eigenfunction changes sign, the polar
representation must be defined separately on its nodal domains. On each
domain, $u_n$ has a fixed sign and one may take
\begin{equation}
R_n=|u_n|.
\end{equation}
After absorbing the constant phase of the nodal domain into the reciprocal
normalization
\begin{equation}
\phi\longrightarrow C\phi,
\qquad
\phid\longrightarrow C^{-1}\phid,
\end{equation}
where \(C>0\) is a constant reciprocal normalization,
the real fields can be written locally as
\begin{equation}
\phi_n
=
R_n\,\ee^{E_nt/\hbar},
\qquad
\phid_n
=
R_n\,\ee^{-E_nt/\hbar},
\qquad
\rho_n
=
R_n^2
=
u_n^2.
\label{eq:stationary-overlap}
\end{equation}
The two fields carry opposite spectral exponents, which cancel exactly in
their product. These opposite exponents are the spectral counterpart of
the gauge transformation Eq.~\eqref{eq:gauge-fields}: the eigenvalue fixes
the repartition of the growth between the two conjugate sectors, exactly
as an energy offset would. The eigenvalue $E_n$ therefore fixes the
reciprocal spectral weighting between the two conjugate sectors, while
the Born density \(\rho_n=R_n^2\) remains stationary.

The nodal-domain qualification is essential.  At a node, the polar
representation is singular and the local diffusion representation must
be completed by the phase-matching, circulation, or regularization
conditions required to reconstruct the global Schr\"odinger state. A signed extension of branching Markov process theory, closer in spirit to the sign structure encountered here across nodal domains, was already considered by Nagasawa \cite{nagasawa_1968_sign}.

In the infinite well, $V=0$ in the bulk and
\begin{equation}
Q_n=E_n
\end{equation}
on each nodal domain. Each member of the Nagasawa pair $(\phi,\phid)$ then has the same
diffusion-reaction operator structure as an imaginary-time
Schr\"odinger equation with a constant spectral shift. This is an
operator-level analogy, not a Wick rotation of the physical time: the two
conjugate equations occur simultaneously, and their opposite spectral
weights reconstruct an exact real-time Schr\"odinger state.

\section{Branching ensemble and moment hierarchy}
\label{sec:bbm}

\subsection{Measure-valued scaling}

We consider a family of local branching systems indexed by an integer
$N$. The $N$-th system is initialized with $Z_N(t=0)$ elementary
continuations, with $Z_N(0)$ of order $N$, and each continuation carries
the bookkeeping mass $1/N$. Branching and killing events make the number
$Z_N(t)$ of continuations alive at time $t$ fluctuate around this scale. If
$X_i^{(N)}(t)$ denotes its position, the empirical measure is
\begin{equation}
\varrho_N(t)
=
\frac{1}{N}
\sum_{i=1}^{Z_N(t)}
\delta_{X_i^{(N)}(t)} .
\label{eq:empirical}
\end{equation}
Here \(\delta_X\) denotes the Dirac point mass at \(X\).
The elementary continuations remain distinct at the microscopic level,
while their individual bookkeeping weight vanishes as $N\to\infty$. We assume that the
initial empirical measures converge in distribution to a finite measure
$\varrho_0$. When the limiting mean measure is absolutely continuous, we
write
\begin{equation}
n(x,t)
=
\frac{\mathbb E_\omega[\varrho_t(\dd x)]}{\dd x},
\label{eq:mean-measure-density}
\end{equation}
where $\mathbb E_\omega$ denotes an average over realizations of the
branching process.

To avoid confusion between the two notations, we reserve
\(\rho=\phid\phi\) for the smooth Born density
associated with the
prescribed Schr\"odinger-Nagasawa solution, whereas $\varrho_N$ (and
$\varrho$ in the limit) denotes the random one-sector empirical branching
measure. In the frozen-rate construction,
\(\rho\) determines the prescribed
coefficient
\(Q[\rho]\), while stochastic averages and connected moments are
taken over $\varrho$; in particular, its mean density $n$ is a one-sector
quantity and should not be identified with either the Born density
\(\rho\).

Since $\varrho_N$ is a positive measure, its direct identification with a
one-sector Schr\"odinger-Nagasawa field is understood locally on a nodal
domain where that field has a fixed sign, after the reciprocal
normalization discussed in Sec.~\ref{sec:nagasawa}. The reconstruction of
a global excited Schr\"odinger state additionally requires the appropriate
matching of phases and boundary conditions across nodal surfaces.

Let $\lambda_N(x,t)$ be the elementary event rate in the $N$-th system,
and let $K_N$ be the number of descendants replacing a continuation at an
event. We denote
\begin{equation}
\nu_{1,N}
=
\mathbb E[K_N],
\qquad
\nu_{2,f,N}
=
\mathbb E[K_N(K_N-1)]
\label{eq:offspring-moments-N}
\end{equation}
the first and second factorial offspring moments. The measure-valued limit is obtained by taking $N\to\infty$ under the
high-frequency scaling
\begin{equation}
\lambda_N\to\infty,
\qquad
\frac{\lambda_N}{N}\to\lambda_0,
\label{eq:superprocess-scaling}
\end{equation}
with $\lambda_0$ finite.

We further assume the standard near-critical scaling
\begin{equation}
\lambda_N(x,t)
\bigl[\nu_{1,N}(x,t)-1\bigr]
\xrightarrow[N\to\infty]{}
\beta_{\rm q}(x,t).
\label{eq:near_critical_scaling}
\end{equation}
Since $\lambda_N\to\infty$ while the macroscopic rate
$\beta_{\rm q}$ remains finite, Eq.~\eqref{eq:near_critical_scaling}
implies $\nu_{1,N}\longrightarrow 1$. This near-criticality concerns only the mean offspring bias: the
offspring variance may remain of order unity, so that individual events
can still correspond to genuine deaths or duplications. Elementary
events therefore become increasingly frequent and asymptotically
unbiased at the level of their first moment, while the product of their
small mean bias and their large event rate remains finite and reproduces
the prescribed Schr\"odinger-Nagasawa rate $\beta_{\rm q}$.

The Schr\"odinger-Nagasawa correspondence fixes this macroscopic
one-sector drift,
\begin{equation}
\beta_{\rm q}(x,t)
=
\frac{V(x,t)+2Q[\rho](x,t)}{\hbar},
\label{eq:betaq-superprocess}
\end{equation}
in the energy gauge specified in Sec.~\ref{sec:nagasawa}. It does not
determine separately the microscopic event rate $\lambda_N$ and the
offspring law. In particular, $\beta_{\rm q}$ is a net growth or killing
rate; it should not be identified with the rate of elementary branching
events.

The finite quadratic variation is controlled by a second, independent
scaling combination. Defining
\begin{equation}
\sigma_N^2
=
\mathbb E\!\left[(K_N-1)^2\right],
\end{equation}
we impose
\begin{equation}
\frac{\lambda_N}{N}\,
\sigma_N^2
\xrightarrow[N\to\infty]{}
\kappa_2 .
\label{eq:qv-scaling}
\end{equation}
Using the exact identity
\begin{equation}
\mathbb E\!\left[(K_N-1)^2\right]
=
\nu_{2,f,N}-\nu_{1,N}+1,
\label{eq:jump-factorial-relation}
\end{equation}
together with $\nu_{1,N}\to1$ and
$\lambda_N/N\to\lambda_0$, the contribution proportional to
$1-\nu_{1,N}$ vanishes in the near-critical limit. Hence
\begin{equation}
\frac{\lambda_N}{N}\,
\nu_{2,f,N}
\xrightarrow[N\to\infty]{}
\kappa_2 .
\label{eq:factorial-scaling}
\end{equation}
Equations~\eqref{eq:near_critical_scaling} and \eqref{eq:qv-scaling} display an
important separation. The wave function fixes the first-moment drift
$\beta_{\rm q}$, whereas the quadratic-variation coefficient $\kappa_2$ is
not determined by the Schr\"odinger equation alone. It depends on the
microscopic branching convention, or equivalently on the normalization
chosen for the second-moment sector. This freedom sets the overall amplitude of the correlations,
but does not affect their normalized spatial structure or the spectral
clustering criterion obtained from the frozen evolution operator.

The integer $N$ therefore indexes only the microscopic discretization of
the measure-valued representation and fixes the elementary weight $1/N$.
It should not be confused with the finite particle number maintained in
a fixed-population branching model. In the limit
\eqref{eq:superprocess-scaling}, the microscopic number of elementary
continuations is sent to infinity, while the finite quadratic-variation
coefficient $\kappa_2$ survives. Thus the connected sector remains
nontrivial even though no finite microscopic population count survives
the limit. The limiting initial measure may nevertheless have a finite
total mass $\int\varrho_0(\dd x)$, which can enter correlation amplitudes
but is not a controlled particle number.

\subsection{Martingale problem and first moment}

We first prescribe the Bohm/Fisher rate by evaluating it on a given
Schr\"odinger mean field. In this frozen-rate setting,
$\beta_{\rm q}(x,t)$ is an external coefficient of the branching process,
rather than a functional of the fluctuating random measure itself.
The resulting one-sector drift generator is denoted by
\begin{equation}
\Lop\equiv\Dq\nabla^2+\beta_{\rm q}.
\label{eq:superprocess-generator-definition}
\end{equation}
For any
smooth test function $f$, the limiting martingale problem is written in the standard Dawson-Watanabe form \cite{Dawson1993,Perkins2002,Etheridge2000}
\begin{align}
M_t^f
&=
\langle\varrho_t,f\rangle
-
\langle\varrho_0,f\rangle
\nonumber\\
&\quad
-
\int_0^t\dd s\,
\left\langle
\varrho_s,
\Dq\nabla^2 f+\beta_{\rm q}f
\right\rangle ,
\label{eq:superprocess-mart}
\end{align}
where
\begin{equation}
\langle\mu,f\rangle
=
\int f(x)\,\mu(\dd x).
\end{equation}
Its predictable quadratic variation is
\begin{equation}
\left\langle M^f\right\rangle_t^{\rm qv}
=
\int_0^t\dd s\,
\left\langle
\varrho_s,
\kappa_2 f^2
\right\rangle ,
\label{eq:superprocess-qv}
\end{equation}
and polarization gives, for another test function \(g\),
the corresponding covariation
$\langle M^f,M^g\rangle_t^{\rm qv}$ with $f^2$ replaced by $fg$.
Whenever the limiting measure admits a density, the same
superprocess can be read in the physicists' SPDE notation
\begin{equation}
\partial_t\varrho
=
\Lop\varrho
+
\sqrt{\ktwo\varrho}\,\xi,
\label{eq:superprocess-spde}
\end{equation}
where \(\xi\) is a centered space-time white noise with
\(\langle\xi(x,t)\xi(y,t')\rangle=\delta^{(d)}(x-y)\delta(t-t')\).
Equivalently, the branching source
\(\eta_{\rm br}=\sqrt{\ktwo\varrho}\,\xi\) has the conditional
covariance
\begin{equation}
\left\langle
\eta_{\rm br}(x,t)\eta_{\rm br}(y,t')\mid\varrho
\right\rangle
=
\ktwo\varrho(x,t)\delta^{(d)}(x-y)\delta(t-t').
\label{eq:superprocess-conditional-noise}
\end{equation}
The random field itself, rather than its mean, therefore fixes the
instantaneous noise strength.  Equation~\eqref{eq:superprocess-spde} is
the density form of the martingale problem, not a truncation at second
order: its positive solutions retain the complete branching genealogy
and all higher cumulants.
The superscript ``qv'' distinguishes predictable quadratic variation from
the expectation brackets used elsewhere in the paper. The independent Brownian
motions of the elementary continuations generate the diffusion operator
in the drift. Their direct contribution to the empirical-measure noise is
suppressed by $1/N$, whereas the accelerated branching fluctuations retain
the finite quadratic variation \eqref{eq:superprocess-qv}.

Taking the expectation of Eq.~\eqref{eq:superprocess-mart} gives
\begin{equation}
\partial_t n
=
\Lop n .
\label{eq:superprocess-first-moment}
\end{equation}
This is the level at which the branching construction reproduces one
member of the Schr\"odinger-Nagasawa diffusion pair. Together with the
conjugate backward mean, it reconstructs the Born density
\(\rho=\phid\phi\). The independent frozen reciprocal fields have
\(C_{\rm FB}=0\). The  organized connected weight
\(\rho_{\rm BSM}=-C_{\rm FB}(x,x)\) is generated only by the interacting
joint dynamics and is matched to \(\rho\) at stationary reciprocal
saturation.

The frozen-rate qualification is essential. Up to this point, the
Schr\"odinger-Nagasawa rate
\(\beta_{\rm q}=(V+2Q[\rho])/\hbar\)
is evaluated on a prescribed smooth
Born density \(\rho\) and therefore acts as an external coefficient of the
linear superprocess. 
Formally, making the Bohm/Fisher rate state dependent amounts
to replacing the prescribed Born density by a fluctuating density.
A one-sector notation \(Q[\varrho]\) makes this distinction visible, but
is not the reciprocal BSM prescription: the random measure \(\varrho\)
belongs to only one genealogical sector and is generally too rough for
the Bohm functional.  The paired construction instead follows the
sequence
\[
Q[\rho]
\longrightarrow Q[\rho_\omega]
\longrightarrow Q[\rho_{\omega,\mathrm{reg}}],
\qquad
\rho_\omega=\Phi_B\Phi_F,
\]
where the regulated reciprocal product is defined in
Sec.~\ref{sec:effective-system}.  This feedback couples the two marginal
superprocesses into an interacting measure-valued extension, whose moment hierarchy no longer closes order by order.
The leading
consequences of this Bohm/Fisher feedback are considered separately in
Sec.~\ref{sec:reciprocal-overlap}.

The measure-valued limit gives a precise mathematical meaning to the
language of possible continuations. It propagates a random measure over
locally admissible continuations, whose mean obeys the one-sector
Schr\"odinger-Nagasawa equation and whose quadratic variation supplies
additional information beyond that mean.

This structure can be read against the path-integral representation.
At the level of the mean, Eq.~\eqref{eq:superprocess-first-moment}
admits the standard Feynman-Kac form: the average over branching
realizations reduces to a Wiener integral over diffusive paths
weighted by the exponentiated rate $\exp[\int_0^t\beta_{\rm q}\,\dd s]$
-- a sum over possible paths, with the essential difference that the
sum here runs over the real diffusive paths of the two reciprocal
fields, the phase being carried by the forward-backward pairing
rather than by an oscillatory weight. The branching representation
departs from this picture only at the next statistical order. Instead
of weighting independent paths, it unravels the exponential weight
into actual birth-death events, whose genealogy correlates
continuations descending from a common ancestor: possible
continuations emerge from other possible continuations. This
genealogical structure, encoded in the connected source of
Eq.~\eqref{eq:G-bbm}, is invisible to any Feynman-Kac weighting of
independent paths, and is precisely what the second-moment hierarchy
adds to the path-integral picture.

\subsection{Statistical averages}

Several notions of average occur in what follows and must be kept
distinct. For a quantum observable \(O\) represented by
the operator \(\widehat O\), we write
\begin{equation}
\langle O\rangle_\psi
=
\langle\psi|\widehat O|\psi\rangle
\end{equation}
for the standard quantum expectation value, evaluated with the Born
density when $O$ is a configuration-space observable. When the context
is unambiguous, the subscript $\psi$ is omitted. We use \(\mathbb E_\omega\) for the expectation over realizations \(\omega\) of the full spatial branching process, including the stochastic branching trajectories, event times, and offspring outcomes. At finite \(N\), both Brownian motion and branching contribute to the randomness of the empirical measure. In the superprocess scaling used here, the direct empirical-measure noise generated by independent Brownian motions vanishes, while diffusion remains in the deterministic generator
\(\Dq\nabla^2\); the finite martingale quadratic variation is supplied by the accelerated branching fluctuations. We reserve \(\mathbb E[\cdot]\) without the index \(\omega\) for averages over the offspring variable \(K\) at a single branching event.

A one-sector branching average supplies the forward or backward
compatibility field.  The BSM observable belongs to the centered joint
sector: the signed covariance
\(C_{\rm FB}=\mathbb E_\omega[\psi_F\psi_B]\) gives
\(\rho_{\rm BSM}=-C_{\rm FB}(x,x)\).  Stationary equality
\(\rho_{\rm BSM}^{\star}=\rho\) expresses the Born matching of this
organized sector.  One-sector, forward-backward, single-realization, and
realization-averaged observables are kept distinct below.

\subsection{Connected second moment and genealogical source}

The quadratic variation retained in the superprocess limit generates fluctuations beyond the one-sector mean. Writing
\begin{equation}
\delta\varrho=\varrho-n
\end{equation}
for the fluctuation of the limiting random measure around its mean, we
define, whenever the covariance measure admits a density, the connected
two-point function
\begin{equation}
G_{\rm br}(x,y,t)
=
\mathbb E_\omega
\left[
\delta\varrho(x,t)\,
\delta\varrho(y,t)
\right].
\label{eq:G-def}
\end{equation}
The subscript ``br'' emphasizes that this is, at this stage, the generic
genealogical covariance of a single local branching measure. For a prescribed local rate and a vanishing limiting initial covariance, Itô’s product formula together with the martingale covariation gives
(Appendix~\ref{app:bbm})
\begin{equation}
\begin{aligned}
\partial_tG_{\rm br}(x,y,t)
&=
\left[
\Lop_x+\Lop_y
\right]
G_{\rm br}(x,y,t)
\\
&\quad
+
\kappa_2\,
\delta^{(d)}(x-y)\,
n(x,t).
\end{aligned}
\label{eq:G-bbm}
\end{equation}
The subscripts on \(\Lop_x\) and \(\Lop_y\) indicate that
the generator \(\Lop\) acts on the \(x\) and \(y\) argument,
respectively.

The same local genealogical source was obtained for branching
neutron transport by comparing the master equation with its stochastic
diffusion form \cite{dechenaux_percolation_2022}.  Here it follows
directly from the superprocess SPDE~\eqref{eq:superprocess-spde}; no
auxiliary stochastic field is required.

Equation~\eqref{eq:G-bbm} follows directly from the quadratic variation
of the positive one-sector branching measure.  In particular, the local
pair source is linear in the one-sector mean density \(n\),
\begin{equation}
{\cal S}_2(x,t)
=
\kappa_2 n(x,t).
\end{equation}
At the level of the underlying measure-valued process, this linear
quadratic variation is the characteristic signature of multiplicative
branching noise: the amount of fluctuation generated locally is
proportional to the random mass present there.

We retain instead the multiplicative SPDE itself.  Its
conditional source is proportional to the realized mass
\(\varrho\); the mean \(n=\mathbb E_\omega\varrho\) appears only
after the It\^o product has been averaged, which is why the source of
Eq.~\eqref{eq:G-bbm} is \(\ktwo n\).

\subsection{Forward-backward genealogical processes and gauge covariance}
\label{sec:reciprocal-noise-embedding}

Equation~\eqref{eq:G-bbm} is a two-points one-sector statement.  We now
apply the same genealogical construction to both members of the
Schr\"odinger-Nagasawa pair.  On a nodal domain, let \(\Phi_F\) and
\(\Phi_B\) be positive stochastic branching densities whose
frozen-background means are
\begin{equation}
\mathbb E_\omega[\Phi_F]=\phi,
\qquad
\mathbb E_\omega[\Phi_B]=\phi^\dagger .
\label{eq:FB-superprocess-means}
\end{equation}
The subscripts \(F\) and \(B\) denote forward and backward time
orientation.  On a finite interval \([0,T]\), the backward process is
read in the increasing variable \(s=T-t\), through
\(\check\Phi_B(x,s)=\Phi_B(x,T-s)\).

We write \(\nu_2\equiv\ktwo/2\) for the branching normalization in a
common reciprocal mass unit and define its original-basis coefficients
by
\begin{equation}
\nu_F=\nu_2\ee^{-S/\hbar},
\qquad
\nu_B=\nu_2\ee^{S/\hbar}.
\label{eq:gauge-covariant-branching-rates}
\end{equation}
The coefficients \(\nu_F\) and \(\nu_B\) are the branching
normalizations in the original forward and backward mass units.  They
are representation coefficients, not observable numbers of possible
continuations.

In physicists' notation, the two marginal martingale problems may be
written
\begin{equation}
\begin{aligned}
\partial_t\Phi_F
&=
\mathcal F[\rho]\,\Phi_F
+
\sqrt{2\nu_F\Phi_F}\,\xi_F,
\\
\partial_s\check\Phi_B
&=
\mathcal F[\rho](x,T-s)\,\check\Phi_B
+
\sqrt{2\nu_B\check\Phi_B}\,\check\xi_B .
\end{aligned}
\label{eq:frozen-FB-superprocess-SPDE}
\end{equation}
Here \(\xi_F\) and \(\check\xi_B\) are independent space-time white
noises in the bare theory.  The adjective \emph{bare}, and the
superscript \((0)\) used below, mean that the two marginal genealogies
have not yet been linked by a forward-backward cross covariance or
dressed by the fluctuating Bohm/Fisher feedback.  The square roots
contain the realized fields, so
Eq.~\eqref{eq:frozen-FB-superprocess-SPDE} describes demographic
branching rather than an externally imposed additive noise.

The separate normalization of \(\Phi_F\) and \(\Phi_B\) is not
physical: the energy-gauge transformation~\eqref{eq:gauge-fields}
rescales them oppositely while leaving their product invariant.  It is
therefore useful to collect the same genealogical masses in the
gauge-invariant reciprocal basis
\begin{equation}
\Phi_S
\equiv
\begin{pmatrix}
\Phi_{S,F}\\
\Phi_{S,B}
\end{pmatrix}
=
\begin{pmatrix}
\ee^{S/\hbar}\Phi_F\\
\ee^{-S/\hbar}\Phi_B
\end{pmatrix},
\qquad
\mathbb E_\omega[\Phi_S]
=
R
\begin{pmatrix}
1\\
1
\end{pmatrix}.
\label{eq:gauge-invariant-populations}
\end{equation}
This introduces no additional stochastic process: \(\Phi_S\) is the
same pair expressed in the common branching-mass unit used in
Eq.~\eqref{eq:gauge-covariant-branching-rates}.  If a Feller mass is
rescaled as \(X'=aX\), then
\(dX=\sqrt{2\nu X}\,dW\) becomes
\(dX'=\sqrt{2(a\nu)X'}\,dW\).  The physical drift bias associated with
\(\nabla S/m\) remains in the forward and backward generators;
Eq.~\eqref{eq:gauge-covariant-branching-rates} should not be read as a
velocity-dependent branching law.

With
\(\eta_S=(\eta_{S,F},\eta_{S,B})^T\), the exact conditional bare
covariance in the common reciprocal basis is
\begin{equation}
\left\langle
\eta_S(x,t)\eta_S^T(y,t')
\mid\Phi_S
\right\rangle_0
=
2\Gamma_S^{(0)}[\Phi_S]\,
\delta^{(d)}(x-y)\delta(t-t'),
\label{eq:conditional-bare-covariance}
\end{equation}
where
\begin{equation}
\Gamma_S^{(0)}[\Phi_S]
=
\nu_2
\begin{pmatrix}
\Phi_{S,F}&0\\
0&\Phi_{S,B}
\end{pmatrix}.
\label{eq:bare-genealogical-covariance}
\end{equation}
Thus each marginal source is proportional to its own realized
population and the bare cross channel vanishes.  In the original
forward-backward variables,
\begin{equation}
\Gamma_{\eta,FB}^{(0)}[\Phi]
=
\begin{pmatrix}
\nu_F\Phi_F&0\\
0&\nu_B\Phi_B
\end{pmatrix}.
\label{eq:bare-covariance-original-basis}
\end{equation}
Under a reciprocal rescaling
\(\Phi_F\mapsto\ee^\chi\Phi_F\),
\(\Phi_B\mapsto\ee^{-\chi}\Phi_B\), the coefficients transform as
\(\nu_F\mapsto\ee^\chi\nu_F\) and
\(\nu_B\mapsto\ee^{-\chi}\nu_B\).  Equation
\eqref{eq:bare-covariance-original-basis} then transforms exactly as
the covariance of the two rescaled noises, whereas \(\Phi_S\) and
Eq.~\eqref{eq:bare-genealogical-covariance} remain invariant.
For the time-dependent energy-gauge shift
\(\chi(t)=ct/\hbar\), differentiating the reciprocal rescaling adds
the drift terms \(\pm\dot\chi\,\Phi_{F,B}\).  These are precisely the
shifts already contained in
\(\beta_{\rm q}\mapsto\beta_{\rm q}+c/\hbar\), so covariance and drift
transform consistently.

For the quadratic construction, introduce the reference-subtracted
fields
\begin{equation}
\begin{aligned}
\psi_F
&=
\Phi_{S,F}-R
=
\ee^{S/\hbar}(\Phi_F-\phi),
\\
\psi_B
&=
\Phi_{S,B}-R
=
\ee^{-S/\hbar}(\Phi_B-\phi^\dagger).
\end{aligned}
\label{eq:regular-FB-fields-noise}
\end{equation}
At short times the two marginal populations remain close to the smooth
reference, \(|\psi_{F,B}|\ll R\), and the expansion about
\(\Phi_S=(R,R)^T\) is controlled.  This statement fixes the regime in
which the bare quadratic kernel is derived.  It does not assume that
\(\psi_F\) and \(\psi_B\) remain small or independent during the
subsequent evolution.

After the interacting dynamics is switched on, the same reduced fields
resolve the reciprocal correlations generated around the fixed one-point
background.  We define their signed connected kernel by
\begin{equation}
C_{\rm FB}(x,y,T)
=
\mathbb E_\omega\left[
\psi_F(x,T)\psi_B(y,T)
\right].
\label{eq:connected-FB-kernel}
\end{equation}
The organized branch is anticorrelated, so its diagonal covariance is
negative.  The  organized connected weight is therefore defined with the sign
convention
\begin{equation}
\rho_{\rm BSM}(x,T)
=
-C_{\rm FB}(x,x,T).
\label{eq:correlated-pair-observable}
\end{equation}
This definition gives positive weight to the correlated sector selected
by reciprocal locking.

The realization-level reciprocal product obeys the exact identity
\begin{equation}
\rho_\omega
\equiv
\Phi_B\Phi_F
=
(R+\psi_B)(R+\psi_F)
=
\rho
+
R(\psi_F+\psi_B)
+
\psi_F\psi_B.
\label{eq:random-product-expansion}
\end{equation}
Centering removes the linear contribution upon ensemble averaging and
gives
\begin{equation}
\mathbb E_\omega[\rho_\omega]
=
R^2
+
C_{\rm FB}(x,x,T)
=
\rho-\rho_{\rm BSM}.
\label{eq:averaged-product-budget}
\end{equation}
The smooth contribution \(R^2=\rho\) carries the initial weight of the
uncorrelated reference sector, whereas \(\rho_{\rm BSM}\) measures the
 deficit of the mean product relative to that reference.  Accordingly,
\(\mathbb E_\omega[\rho_\omega]\)  quantifies the mean reciprocal co-occupation.   The organized connected weight is instead encoded by
the signed diagonal
\(\rho_{\rm BSM}(x,T)=-C_{\rm FB}(x,x,T)\).

The same connected kernel can reproduce the Born profile on its signed
diagonal while retaining a nontrivial dependence on the relative
coordinate \(x-y\).  These complementary projections make
\(C_{\rm FB}\) the central second-order observable of the construction.
If
\begin{equation}
-C_{\rm FB}(x,x,T)
\longrightarrow
R^2(x)
\qquad
(T\longrightarrow\infty),
\end{equation}
or equivalently \(\rho_{\rm BSM}\to\rho\), then
\(\mathbb E_\omega[\rho_\omega]\to0\). In this stationary regime, the
entire Born weight is represented by the correlated sector: its diagonal
follows the Born density, while its off-diagonal dependence may retain
a nontrivial relative structure.

At finite observation time \(T\), the reciprocal co-occupation
\(\rho_\omega(x,T)=\Phi_F(x,T)\Phi_B(x,T)\) is nonnegative and has
expectation \(\rho(x,T)-\rho_{\rm BSM}(x,T)\) by
Eq.~\eqref{eq:averaged-product-budget}. As exact stationary matching
is approached,  this co-occupation may become concentrated in increasingly
rare surviving configurations, in analogy with the critical catastrophe
 of branching processes \cite{DeMulatier2015}; this possibility does not
follow from the vanishing mean alone. In this interpretation,
and anticipating the discussion of Sec.~\ref{sec:discussion}, the
terminal backward field weights the forward configurations and thereby
selects the surviving sector in detector-normalized observables; within
this sector, the nonzero product \(\rho_\omega(x,T)\) represents a local
density of possible forward continuations weighted by their contribution
to the final measurement. At exact stationary matching, the nonnegative
 unconditioned product, if defined at the matched limit, has zero
expectation and therefore
 vanishes almost surely at each fixed \(x\); its normalized version
requires a separate limiting prescription.
At the bare level the two genealogical drivers are independent, so
\begin{equation}
C_{\rm FB}^{(0)}(x,y;t,t')
=
0.
\label{eq:bare-CFB-zero}
\end{equation}
The initial condition therefore contains no correlated-pair
observable.  A nonzero \(C_{\rm FB}\) is generated only after the
Bohm/Fisher interaction has dressed the joint dynamics.  The fields
\(\psi_F,\psi_B\) are the centered dynamical coordinates whose joint
connected sector is constructed by the evolution; the smooth one-point
profile remains carried by $R$.

Using
\(\Phi_{S,F}=R+\psi_F\) and
\(\Phi_{S,B}=R+\psi_B\), the exact conditional covariance decomposes as
\begin{equation}
\Gamma_S^{(0)}[\Phi_S]
=
\underbrace{
\nu_2R\,\mathbb I
}_{\displaystyle\widehat\Gamma_S^{(0)}}
+
\underbrace{
\nu_2
\begin{pmatrix}
\psi_F&0\\
0&\psi_B
\end{pmatrix}
}_{\mathclap{\substack{\text{multiplicative}\\
\text{branching vertex}}}}.
\label{eq:background-plus-branching-vertex}
\end{equation}
Here \(\mathbb I\) is the \(2\times2\) identity in the ordered
\((F,B)\) reciprocal basis.  The hatted matrix
\begin{equation}
\widehat\Gamma_S^{(0)}
=
\nu_2R\,\mathbb I
\label{eq:bare-quadratic-reference-kernel}
\end{equation}
is the covariance evaluated on the short-time smooth reference.  It is
 the bare quadratic stochastic kernel of the reference expansion.  The second term preserves the exact
field dependence of the underlying square-root genealogy away from
the reference.

The decomposition in
Eq.~\eqref{eq:background-plus-branching-vertex} is algebraically exact,
but the use of its first term as a closed quadratic kernel is a
reference expansion.  It is controlled initially when
\(|\psi_{F,B}|/R\ll1\). At later times the full covariance is the dressed
 quantity associated with the interacting law. The full multiplicative
remainder belongs to the underlying branching process; its resummation
is not undertaken here. At the effective covariance level, departures
from the reference kernel can be parameterized by a correlation
correction,
\begin{equation}
\widehat\Gamma_S^{\rm eff}(T)
=
\widehat\Gamma_S^{(0)}
+
\Sigma_C(T).
\label{eq:dressed-correlation-kernel-link}
\end{equation}
Thus \(\widehat\Gamma_S^{(0)}\) seeds the initially independent
fluctuations, while \(C_{\rm FB}\) is an output of the interacting
evolution rather than a bare input.

\subsection{Positivity, nodal domains, and regulated feedback}
\label{sec:noise-positivity-nodes}

The
stochastic fields of this construction are the positive superprocess
densities themselves.  Their square-root noises vanish whenever the
corresponding realized population vanishes, not merely when its mean
does.  The construction is local to nodal domains, where the
Schr\"odinger-Nagasawa weights have fixed sign; the usual phase and
boundary matching is still required across nodes.  The Bohm potential
also remains singular on a raw nodal or distribution-valued realization,
so positivity alone does not define the fully interacting problem.

No extra spatial smoothing is used in the frozen marginal
superprocesses.  When the Bohm/Fisher functional is allowed to respond
to \(\rho_\omega\), however, it is evaluated with the fixed support and
smoothing prescription of the projected closure.  This regularization
belongs to the effective interacting extension and is not a new
genealogical scale.

\subsection{Regulated interacting BSM extension}
\label{sec:effective-system}

To define the interacting extension while retaining the
marginal square-root noises, let \(K_{\ell_{\rm reg}}\) be a fixed,
positive, normalized mollifier of width \(\ell_{\rm reg}\) on a
reference nodal domain \(\Omega_R\), and set
\begin{equation}
\rho_{\omega,\mathrm{reg}}
=
\mathbf 1_{\Omega_R}
\bigl(K_{\ell_{\rm reg}}*\Phi_B\bigr)
\bigl(K_{\ell_{\rm reg}}*\Phi_F\bigr).
\label{eq:regulated-reciprocal-product}
\end{equation}
Here \(*\) denotes spatial convolution; the mollifier and
support are fixed closure data, not stochastic variables. Let
\(\xi_B^{\leftarrow}(x,t)\equiv-\check\xi_B(x,T-t)\) denote the
reverse-time backward white noise written with the physical-time label.
The interacting extension used to organize the closure is then written
compactly as
\begin{equation}
\begin{aligned}
\partial_t\Phi_F-\mathcal F[\rho_{\omega,\mathrm{reg}}]\Phi_F
&=\sqrt{2\nu_F\Phi_F}\,\xi_F,\\
\partial_t\Phi_B+\mathcal F[\rho_{\omega,\mathrm{reg}}]\Phi_B
&=\sqrt{2\nu_B\Phi_B}\,\xi_B^{\leftarrow},\\
\rho_\omega&=\Phi_B\Phi_F .
\end{aligned}
\label{eq:effective-FB-system}
\end{equation}
The sign of \(\xi_B^{\leftarrow}\) makes the second
line equivalent to the reverse-time SPDE
\eqref{eq:frozen-FB-superprocess-SPDE} and does not affect its
covariance.
Equation~\eqref{eq:effective-FB-system} is an effective regulated
extension because the nonlinear functional
\(Q[\rho_{\omega,\mathrm{reg}}]\) couples rough branching measures; no
global existence claim is made for the unregularized coupled SPDE.
Its two frozen marginals, by contrast, are the superprocesses defined
by the martingale problem.

At the bare level the two white drivers are independent, equivalently
\begin{equation}
\left\langle
\eta_{S,F}(x,t)\eta_{S,B}(y,t')
\mid\Phi_S
\right\rangle_0=0 .
\label{eq:bare-FB-cross-zero}
\end{equation}
A nonzero forward-backward cross covariance is an admissible joint
extension of the organized sector and is introduced only in
Sec.~\ref{sec:overlap-nucleation}.  In response-field
notation the same reference covariance is the bare half-kernel,
\begin{equation}
\widehat\Gamma_S^{(0)}
=
\nu_2R\,\mathbb I,
\end{equation}
and its dressed counterpart is written
\begin{equation}
\widehat\Gamma_S^{\rm eff}
=
\widehat\Gamma_S^{(0)}+\Sigma_C.
\label{eq:dressed-noise-paperI}
\end{equation}
Here \(\Sigma_C\) denotes the correlation self-energy generated by the
interacting Bohm/Fisher dynamics.  Thus the bare quadratic source has no
forward-backward cross entry, while such an entry may emerge in the
dressed kernel.  

\subsection{Numerical diagnostics: the BSM-MC implementation}
\label{sec:bsm-mc-conventions}

The analysis below is accompanied by numerical results of two distinct
kinds. In the frozen-rate hierarchy of
Secs.~\ref{sec:confined} and~\ref{sec:dimension}, Monte Carlo estimates
serve as benchmarks of exact moment formulas. Once the Bohm/Fisher
feedback is activated (Sec.~\ref{sec:reciprocal-overlap}), the simulations instead serve as
diagnostics of the specified effective closures: they test and
visualize the projected reciprocal sector, but they are not used as an
independent derivation of the joint forward-backward law. Throughout
the paper, BSM-MC denotes the projected branching Monte Carlo
implementation of the BSM hierarchy defined in
Appendix~\ref{app:mc-protocol}. Despite the shared branching-walker
lineage, it should not be confused with conventional imaginary-time
Diffusion Monte Carlo: BSM-MC samples the real-time
forward-backward pair and its connected diagnostics, rather than
projecting onto a ground state. The same conventions apply to all
simulation figures.

\section{Clustering transition in a confined geometry}
\label{sec:confined}

The frozen construction of Sec.~\ref{sec:bbm} provides the starting
point for the clustering analysis developed here. Once the Born profile
is prescribed, $Q[\rho]$ becomes an external coefficient and the bare
joint law factorizes into independent forward and backward marginal
superprocesses. For a stationary profile, reversing the backward time
places both marginals under the same time-independent one-sector
generator; in the homogeneous benchmark considered below, its
branching rate is constant. The connected hierarchy of either marginal,
and in particular Eq.~\eqref{eq:G-bbm}, is then precisely the
one-sector branching-diffusion hierarchy used in neutron transport
theory. We may therefore apply the corresponding confined-geometry
results \cite{Pazsit2008,Zoia2014,DeMulatier2015} to either reciprocal
sector separately. The forward-backward coupling is restored only in
Sec.~\ref{sec:reciprocal-overlap}, when the Bohm/Fisher rate is allowed
to respond to the fluctuating reciprocal product.

\subsection{Control parameter and the three regimes}

Building on results in the field of stochastic neutron transport theory \cite{Zoia2014,DeMulatier2015}, we use a one-dimensional interval $[-L,L]$ with reflecting boundaries as
an analytically solvable benchmark for the frozen branching hierarchy. The later comparison with the Dirichlet infinite well is
spectral only; no equivalence between the two boundary conditions is
assumed. The diffusion Green function from an initial
position \(x_0\) is
\begin{equation}
G_D(x,t|x_0)=\frac{1}{2L}
+\frac{1}{L}\sum_{n\ge 1}c_n(x)\,c_n(x_0)\,
\ee^{-n^2t/\tau_D},
\label{eq:green-reflect}
\end{equation}
with $c_n(z)=\cos[\tfrac{n\pi}{2}(1+z/L)]$ and
\begin{equation}
\tau_D=\frac{4L^2}{\pi^2\Dq}=\frac{8mL^2}{\pi^2\hbar} ,
\label{eq:tauD}
\end{equation}
the diffusive time of the interval (the inverse of the fundamental quantum
frequency\(E_1/\hbar\) of the Dirichlet infinite-well benchmark
on the same interval, with $E_1=\hbar^2\pi^2/(8mL^2)$). We also introduce
a second characteristic time associated with the branching process. In the
infinite-well benchmark, the Bohm potential is constant in the bulk. For
$V=0$, the Schr\"odinger-Nagasawa rate is therefore
\begin{equation}
    \frac{1}{\tau_c}
    =
    \frac{2Q}{\hbar}.
\end{equation}
This definition corresponds to the frozen mean-field regime. In the full
branching problem, however, the Bohm potential is evaluated on the
fluctuating density field. The effective rate controlling the correlated
sector may therefore become time dependent and signed once the feedback
of the correlations is included; note that \(\tau_c\)itself carries the
sign of the effective rate and is negative in the subcritical sector. In
the frozen hierarchy used below, we keep this rate constant and define
\begin{equation}
a
=
-\frac{\tau_D}{\tau_c}
\label{eq:a}
\end{equation}
Since \(\tau_D>0\) and \(a=-\tau_D/\tau_c\), one has \(a>0\iff\tau_c<0\); consequently, \(\sqrt{-4\Dq\tau_c}\) is real in the subcritical clustered regime. The branching classification associated with the sign
of $a$ is established below and summarized in
Table~\ref{tab:regimes}. A time-dependent
effective control parameter belongs to
the feedback problem discussed in Sec.~\ref{sec:reciprocal-overlap}.

A distinction with finite-population branching models is important
here. The stationary pair separation obtained below has the same
functional structure as that found in branching systems with an
imposed fixed-population constraint \cite{DeMulatier2015}. In those
systems, the constraint introduces an explicit dependence on the
finite population size, so that genealogical clustering weakens as
that population is increased. In the present measure-valued
hierarchy, by contrast, that microscopic population parameter has
already been removed by the superprocess limit: $N,\lambda_N\to\infty$
while the quadratic variation remains finite. The microscopic
branching dependence of the connected pair source is then absorbed
into $\ktwo$, which controls the amplitude of the connected
correlations, whereas their normalized spatial structure is controlled
by the macroscopic parameter $a=-\beta_{\rm q}\tau_D$.

The relevant observable is the mean-square separation of correlated
branches, with relative distance \(r=|x-y|\),
\begin{equation}
\mean{r^2}_{x,y}
=
\frac{
\int\!\!\int
(x-y)^2G_{\rm br}(x,y,t)\,\dd x\,\dd y
}{
\int\!\!\int
G_{\rm br}(x,y,t)\,\dd x\,\dd y
}.
\label{eq:r2def}
\end{equation}
For the nonnegative local source and vanishing initial covariance
considered here, the Green-function representation of
Appendix~\ref{app:bbm} gives \(G_{\rm br}(x,y,t)\geq0\), so that it
can be used directly as the pair weight in Eq.~\eqref{eq:r2def}.
Inserting the mode expansion \eqref{eq:green-reflect} into the
Green-function solution of Eq.~\eqref{eq:G-bbm} for a uniform initial
mean density $n(x,0)=c_0$ (Appendix~\ref{app:bbm}), the long-time
limit of Eq.~\eqref{eq:r2def} splits into three regimes:
\begin{align}
&a>0:&
\mean{r^2}_{x,y}^{t\to\infty}&=-4\Dq\tau_c
\left[1-\frac{2}{\pi}\sqrt{\frac{2}{a}}
\tanh\!\Big(\frac{\pi}{2}\sqrt{\frac{a}{2}}\Big)\right],
\label{eq:r2-bbm}\\
&a=0:&
\mean{r^2}_{x,y}^{t\to\infty}&=\frac{2L^2}{3},
\label{eq:r2crit}\\
&a<0:&
\mean{r^2}_{x,y}^{t\to\infty}&=\frac{2L^2}{3} .
\label{eq:r2sup}
\end{align}

In the subcritical regime, $a>0$, the effective rate acts as a net
killing term. For $a\gg1$, the bracket in Eq.~\eqref{eq:r2-bbm} tends
to one and the cluster size $\sqrt{-4\Dq\tau_c}\ll L$ decouples from
the system size: a localized correlated sector has formed. In a
finite unconstrained branching population, this localization has a
simple genealogical interpretation: as the population approaches
extinction, independent lineages progressively disappear and the
surviving particles become increasingly dominated by descendants
sharing recent common ancestors. The same mechanism has a different
continuum reading in the present measure-valued representation, in
which $N$ is a microscopic discretization sent to infinity rather
than a population parameter: branching supplies a finite local source
of genealogically correlated pairs through $\ktwo$, diffusion tends
to separate their descendants, and the finite genealogical time scale
$|\beta_{\rm q}|^{-1}$ suppresses old, widely separated pair
histories before diffusion can explore the whole system. The
resulting connected pair measure therefore acquires a finite relative
length.

In the supercritical regime, $a<0$, the exponentially growing factors
cancel between the numerator and the denominator of
Eq.~\eqref{eq:r2def}. The ensemble keeps exploring the whole interval
and remains ergodic.

At criticality, that is for $a=0$, the spatial spread still saturates
at the independent-configuration value \eqref{eq:r2crit},
continuously matching the $a\to0^+$ limit of Eq.~\eqref{eq:r2-bbm}.
However, as shown explicitly in Appendix~\ref{app:bbm}, the zero-mode
contribution to the correlation function itself grows linearly in
time. Denoting this contribution by $G_{\rm br}^{(0)}$,
Eq.~\eqref{eq:Gbr-modes} gives at criticality
\begin{equation}
\frac{G_{\rm br}^{(0)}(t)}{c_0^2}
=
\frac{\ktwo\,t}{2Lc_0}.
\label{eq:critical-zero-mode}
\end{equation}
This is the measure-valued counterpart of the
``critical catastrophe'' of confined branching processes
\cite{Zoia2014,DeMulatier2015}. Within the generic frozen hierarchy,
this growth corresponds to a slow accumulation of recurrent
common-ancestry pairs.

\begin{figure}[!htbp]
\includegraphics[width=\columnwidth]{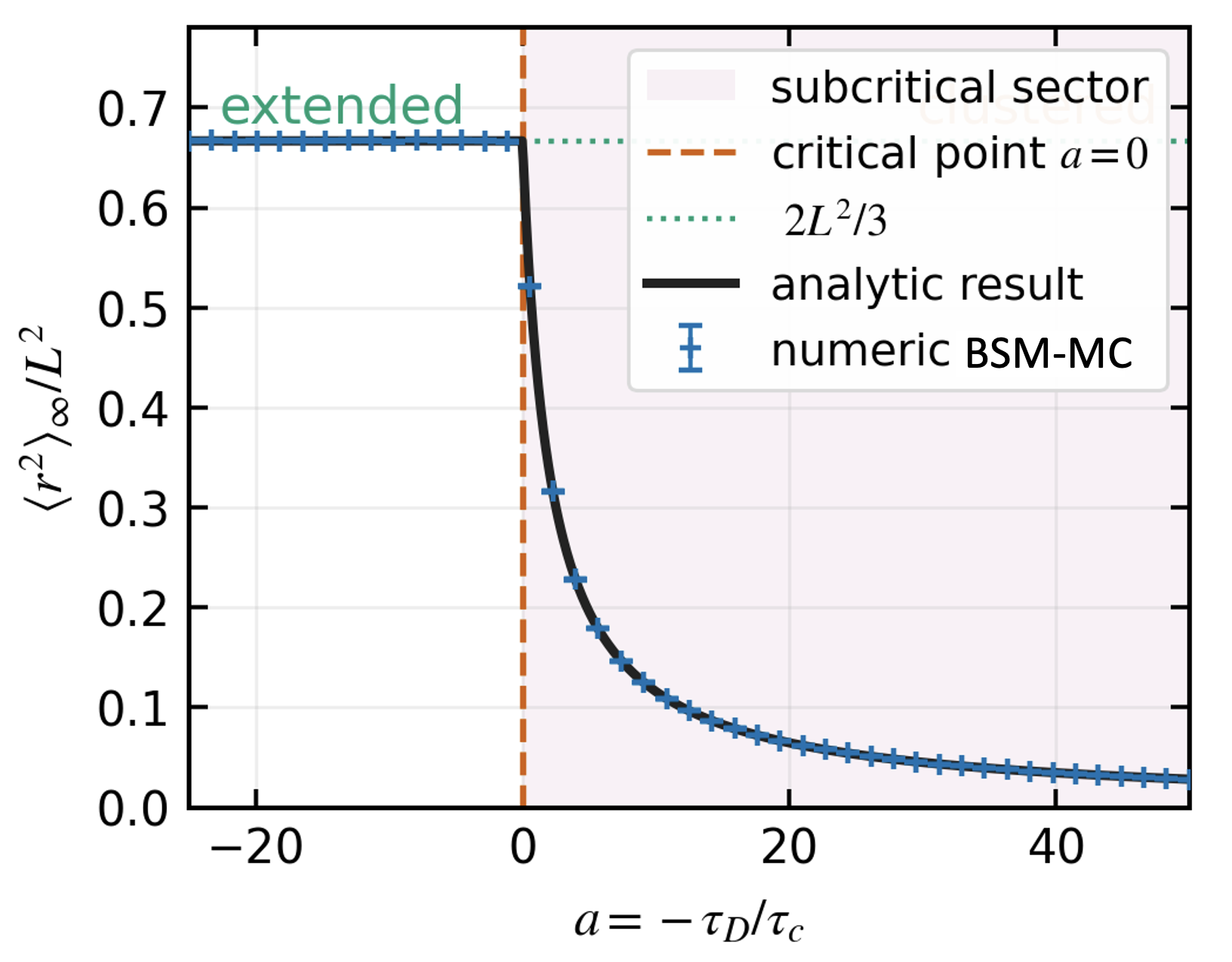}
\caption{
Benchmark of the frozen-rate hierarchy in the reflecting interval:
long-time pair separation $\mean{r^2}_\infty/L^2$ as a function of the
reduced control parameter $a=-\tau_D/\tau_c$. Solid line: analytic
result, Eqs.~\eqref{eq:r2-bbm}--\eqref{eq:r2sup}; symbols: BSM-MC
estimates (error bars: ensemble statistics). For $a\le0$ (extended
sector) the separation saturates at the independent-configuration
value $2L^2/3$ (dotted line); for $a>0$ the correlated pairs form
clusters whose size decouples from the system size as $a$ increases.
}
\label{fig:a-transition}
\end{figure}

The three regimes are summarized in Table~\ref{tab:regimes} and
benchmarked against BSM-MC in Fig.~\ref{fig:a-transition}.

\begin{table}[t]
\caption{\label{tab:regimes}The three regimes of the confined branching
ensemble.}
\begin{ruledtabular}
\begin{tabular}{llll}
regime & $a$ & $\mean{r^2}_\infty$ & sector \\
\hline
supercritical & $a<0$ & $2L^2/3$ & extended, ergodic (field-like)\\
critical & $a=0$ & $2L^2/3$ & marginal, $G_{\rm br}^{(0)}\!\propto\! t$ \\
subcritical & $a>0$ & Eq.~\eqref{eq:r2-bbm}
& localized, clustered (particle-like) \\
\end{tabular}
\end{ruledtabular}
\end{table}

\subsection{Quantum interpretation of the control parameter}
\label{sec:abohm}

The quantum expression of the control parameter is obtained by combining
its definition $a=-\beta_{\rm q}\tau_D$ with $\tau_D=\hbar/E_1$ and the
prescribed Schr\"odinger-Nagasawa reaction rate of
Eq.~\eqref{eq:betaq}. In the frozen linear hierarchy,
\(Q[\rho]\), and
hence $\beta_{\rm q}$, is prescribed by the Born density \(\rho\) rather
than recomputed from the fluctuating measure $\varrho$.
For $V=0$, the constant-rate projection maps the spatially dependent rate onto its
constant component, obtained by averaging with the Born-density weight,\(Q[\rho](x)\mapsto\mean{Q}\), and gives
\begin{equation}
\overline{\beta}_{\rm q}
=
\frac{2\mean{Q}}{\hbar},
\qquad
a
=
-\overline{\beta}_{\rm q}\tau_D
=
-\frac{2\mean{Q}}{E_1}
=
-\frac{16mL^2}{\pi^2\hbar^2}\,\mean{Q},
\label{eq:aQ}
\end{equation}
where $\mean{Q}$ is the density-weighted mean of the Bohm potential,
\(\mean{Q}\equiv\int\dd^dx\,\rho(x)Q[\rho](x)\) for the
normalization \(\int\dd^dx\,\rho=1\).

The physical content of this relation is exposed by the
velocity-variance decomposition derived in
Appendix~\ref{app:velocity}. We denote by
\(\widehat v=-\ii\hbar\nabla/m\) the velocity operator,
\(\widehat p=m\widehat v\) the momentum operator, and by
\(\sigma_v^2=\langle\widehat v^2\rangle_\psi-
\langle\widehat v\rangle_\psi^2\) its quantum variance. For a state
$\psi=R\,\ee^{\ii S/\hbar}$,
\begin{equation}
\sigma_v^2
=
\mean{v^2}-\mean{v}^2
=
\frac{2}{m}\mean{Q}
+
\sigma^2_{v_B},
\label{eq:sigmav}
\end{equation}
where
\begin{equation}
v_B=\frac{\nabla S}{m}
\end{equation}
is the Bohmian velocity, equivalently the probability-current velocity,
and
\(\sigma_{v_B}^2\equiv\langle v_B^2\rangle-\langle v_B\rangle^2\)
is its variance.  Moreover,
\begin{equation}
\frac{2}{m}\mean{Q}
=
\mean{u^2},
\qquad
u=\frac{\hbar}{m}\frac{\nabla R}{R},
\end{equation}
with $u$ the osmotic velocity. The terminology ``current velocity''
reflects the identity \(j=\rho v_B\), which distinguishes $v_B$ from the
osmotic contribution $u$.

Equivalently, decomposing the local energy per unit Born weight,
\begin{equation}
E_{\rm tot}(x)
=
\frac{(\nabla S)^2}{2m}
+
\frac{\hbar^2}{2m}\frac{(\nabla R)^2}{R^2}
+
V(x),
\label{eq:energy}
\end{equation}
the second term is the Fisher, or osmotic, share of the kinetic energy.
It is the purely quantum contribution, and Eq.~\eqref{eq:aQ} shows that
$a$ measures minus twice its density-weighted mean in units of $E_1$.
Three consequences follow.

(i) \emph{Quantum states are non-subcritical in the frozen hierarchy.}
For any sufficiently regular normalized confined profile, with boundary
terms that vanish,
\begin{equation}
\mean{Q}
=
\frac{\hbar^2}{2m}
\int(\nabla R)^2\,\dd x
\geq 0.
\end{equation}
Consequently, $a\leq0$. Nontrivial stationary infinite-well modes have
$\mean{Q}>0$ and therefore belong to the extended supercritical sector,
consistently with unitarity. For the $n$-th mode of the well,
$a_n=-2n^2$.

(ii) \emph{Localized overlap requires an additional reciprocal sector.}
A unitary Schr\"odinger state in a confining box has $\mean{Q}\geq0$ and
therefore remains on the non-subcritical side of the frozen hierarchy.
A localized reciprocal-overlap sector cannot therefore be inferred from
the sign of the frozen one-sector rate alone. It requires the additional
forward-backward second-moment law of Sec.~\ref{sec:reciprocal-overlap}, in which an
anti-correlated covariance can nucleate the overlap and a positive dressed relative gap screens its shape. The self-consistent mechanism selecting the covariance strength and the gap lies with the full interacting dynamics beyond the present second-moment closure.

(iii) \emph{The subcritical scale provides the natural localization
scale of the connected sector.}
On the subcritical side, let the effective kinetic scale $E_c>0$ be
defined by
\begin{equation}
\overline{\beta}_{\rm q}
=
-\frac{2E_c}{\hbar},
\qquad
\tau_c
=
-\frac{\hbar}{2E_c}.
\end{equation}
The condition $a\gg1$ then reads, up to geometry-dependent numerical
factors,
\begin{equation}
L\sqrt{2mE_c}\gg\hbar,
\end{equation}
which is the usual semiclassical condition that the characteristic
action be large compared with $\hbar$.
In the same regime, Eq.~\eqref{eq:r2-bbm} gives
\begin{equation}
\mean{r^2}^{t\to\infty}_{x,y}
\longrightarrow
\frac{\hbar^2}{mE_c}
=
2\lambda_{\rm dB}^{2},
\qquad
\lambda_{\rm dB}
=
\frac{\hbar}{\sqrt{2mE_c}},
\end{equation}
so that $a\gg1$ is equivalent to $\lambda_{\rm dB}\ll L$: the branch on
which the connected sector becomes localized selects, without any
additional length scale, the reduced de~Broglie wavelength associated
with the effective kinetic scale $E_c$. The parameter controlling
the strength of clustering on the subcritical side therefore takes the
form of the standard semiclassical crossover parameter.

This correspondence should be understood as a consistency benchmark,
not as a statement that an ordinary stationary Schr\"odinger state is
already localized at second order. By (i), a genuine Schr\"odinger
state remains on the extended side of the frozen hierarchy, and a
nonzero pair-source intensity $\ktwo$ only populates the connected
sector without modifying the prescribed Bohm/Fisher rate: the first
moment retains the ordinary Schr\"odinger dynamics, while the branching
fluctuations open an additional statistical sector of possible
continuations. The limitation is structural. In the frozen hierarchy
the connected sector is propagated with the same rate
$\overline{\beta}_{\rm q}$ as the mean [Eq.~\eqref{eq:G-bbm}], so an
extended one-point sector cannot coexist with a localized connected
sector: by (i) and (ii), such coexistence requires a reciprocal channel
carrying its own relaxation rate, decoupled from the one-body spectral
scale.

The benchmark instead identifies the relative scale that
would follow if the interacting reciprocal dynamics generated a
finite-range response: if its relaxation inherited the kinetic scale of
the reference state, the range would be of de~Broglie order.  The
interacting construction of Sec.~\ref{sec:reciprocal-overlap} replaces
the frozen coefficient by feedback from
\(\rho_\omega=\Phi_B\Phi_F\), derives the centered reciprocal pair
equation, and separates its collective diagonal from its relative profile.
The coefficient \(\ktwo\) fixes the genealogical source strength; the
finite-amplitude saturation law and the selection of a finite relative
range are then central open problems for the full self-consistent joint
dynamics.

\subsection{Minimal quantum consistency benchmark: infinite-well eigenmodes}
\label{sec:spectral-benchmark}
The infinite well provides a parameter-free consistency check for the
present branching representation: stationary Schr\"odinger eigenstates
must remain extended rather than being turned into localized connected
clusters. For $x\in[-L,L]$,
\begin{equation}
k_n
=
\frac{n\pi}{2L},
\qquad
E_n
=
\frac{\hbar^2k_n^2}{2m}
=
\frac{\hbar^2\pi^2n^2}{8mL^2}.
\label{eq:well-spectrum}
\end{equation}
For a real stationary eigenmode,
\begin{equation}
\psi_n(x,t)
=
u_n(x)\ee^{-\ii E_nt/\hbar},
\end{equation}
the polar representation is understood separately on each nodal domain.
Writing $R_n=|u_n|$ on such a domain, the
Schr\"odinger-Nagasawa fields are
\begin{equation}
\phi_n
=
R_n\ee^{E_nt/\hbar},
\qquad
\phid_n
=
R_n\ee^{-E_nt/\hbar},
\qquad
\phid_n\phi_n
=
R_n^2
=
u_n^2.
\label{eq:well-nagasawa-fields}
\end{equation}
In the bulk, $V=0$ and $Q_n=E_n$ on each nodal domain. In the energy
gauge fixed in Sec.~\ref{sec:nagasawa}, the prescribed one-sector rate
is therefore
\begin{equation}
\beta_{{\rm q},n}
=
\frac{2E_n}{\hbar}.
\end{equation}
Since $\tau_D=\hbar/E_1$, the corresponding control parameter is
\begin{equation}
a_n
=
-\beta_{{\rm q},n}\tau_D
=
-2\frac{E_n}{E_1}
=
-2n^2.
\label{eq:quantum-a-n}
\end{equation}
Every stationary eigenmode of the well consequently lies on the
supercritical, extended side of the frozen-rate hierarchy, and
increasingly so at high energy: the Fisher share of the kinetic energy,
which drives the branching exploration, grows with $n$. The conclusion
is also insensitive to the reference scale. If the control parameter is
instead referred to a single nodal domain of size $2L/n$, whose
diffusive time is $\tau_D/n^2$, one finds $a^{\rm dom}=-2$ for every
mode, so that each nodal cell sits at the same finite distance on the
supercritical side of the hierarchy.

Inserting this spectrally transferred rate into the reflecting-interval
benchmark of Sec.~\ref{sec:confined}, with uniform initial mean
density, the normalized connected pair separation saturates at the
extended value
\begin{equation}
\mean{r^2}_{\infty,n}
=
\frac{2L^2}{3},
\end{equation}
independently of $n$. Consistently with the spectral-only status of
the comparison stated in Sec.~\ref{sec:confined}, this number
characterizes the benchmark hierarchy driven at the eigenmode rate,
not a pair dispersion evaluated in the Dirichlet eigenstate itself;
the parameter-free content of the check is the sign $a_n<0$, not the
saturation value.

This benchmark establishes a limited but important consistency result:
when the rate prescribed by a stationary Schr\"odinger eigenstate is
inserted into the frozen hierarchy, the connected pair sector remains
extended, consistently with ordinary quantum mechanics. The stationary dispersion of a self-consistently selected reciprocal sector involves, in addition, the joint forward-backward response of Sec.~\ref{sec:reciprocal-overlap} and its diagonal-preserving gap.

\subsection{Pair localization and standard Heisenberg uncertainty}
\label{sec:heisenberg}

The stationary pair-separation formula derived in
Eq.~\eqref{eq:r2-bbm} is an exact result of the linear
diffusion-branching moment theory with constant coefficients. In the present context, its
interpretation requires distinguishing a normalized empirical average
within one branching realization, an average over realizations, and a
quantum expectation value evaluated with the Born density.

For a realization $\omega$ containing $N_\omega(t)$ possible
continuations at positions \(X_i^\omega(t)\), the normalized empirical configuration average is
\begin{equation}
\mean{O}_{\omega}
=
\frac{1}{N_\omega(t)}
\sum_{i=1}^{N_\omega(t)}
O\!\left(X_i^\omega(t)\right).
\end{equation}
Its average over independent realizations of the branching process is
simply
\begin{equation}
\mathbb E_\omega
\left[
\mean{O}_{\omega}
\right].
\end{equation}
Neither quantity should generically be identified with a quantum
expectation value, since a one-sector branching measure is not by itself
the Born density. The physical one-body expectation is instead
\begin{equation}
\mean{O}_{\psi}
=
\int\dd x\,
\rho(x,t)\,O(x),
\qquad
\rho=\phid\phi.
\end{equation}
The independent frozen reciprocal fields reconstruct the smooth product
\(\rho=\phid\phi\) through their first moments and have vanishing connected
cross covariance.  The interacting joint theory generates
\(C_{\rm FB}=\mathbb E_\omega[\psi_F\psi_B]\), from which the  organized connected weight is obtained as \(\rho_{\rm BSM}=-C_{\rm FB}(x,x)\).  Its equality
with the Born profile is the stationary diagonal matching condition tested
in Sec.~\ref{sec:reciprocal-overlap}.

The second moment of the branching theory probes a different object:
the relative separation of connected pairs of possible continuations.
The pair observable
$\langle r^2\rangle_{\rm pair}\equiv\mean{r^2}_{x,y}$ is defined by
Eq.~\eqref{eq:r2def}, using the nonnegative genealogical kernel
$G_{\rm br}$ as its pair weight. Equivalently, it is obtained by
selecting a connected pair from the full ensemble of branching
realizations, each realization being weighted by the number of
connected pairs it contributes. It is not, in general, the uniform
average over realizations of a separately normalized
within-realization dispersion.

The two localization statements are therefore built on distinct
observables. In the clustered regime $a\gg1$,
\begin{equation}
\mean{r^2}_{\rm pair}\ll L^2,
\end{equation}
while the one-point mean density of the branching ensemble remains
extended over the interval, with statistical dispersion of order
$L^2$. This expresses restricted ergodic exploration in the connected
sector without implying localization of the one-point sector -- and,
a fortiori, without implying localization of a Born density, since by
Sec.~\ref{sec:abohm} no unitary state reaches the subcritical regime
of the frozen hierarchy. The frozen hierarchy alone thus generates no
crossover of the one-point sector; the fluctuating-rate extension of
Sec.~\ref{sec:reciprocal-overlap}
instead identifies the reciprocal channel through which correlation feedback can screen the connected sector.

The standard Heisenberg inequality follows independently from the
Fisher information \(I[\rho]\) of the Born density, introduced in
Eq.~\eqref{eq:fisher}. For a sufficiently regular one-dimensional
density, or componentwise in higher dimensions,
\begin{equation}
\mean{Q}_\psi
=
\frac{\hbar^2}{8m}
I[\rho],
\end{equation}
 
where $\mean{Q}_\psi$ is the density-weighted mean of the Bohm potential,
written $\mean{Q}$ in Sec.~\ref{sec:abohm}.  The identity concerns the Bohm
functional evaluated on the smooth Born density.  The corresponding
realization-wise nonlinear functional obeys, in general,
 \begin{equation}
 \mathbb E_\omega\!\left[Q[\rho_\omega]\right]
 \neq
 Q\!\left[\mathbb E_\omega[\rho_\omega]\right].
 \label{eq:bohm-nonlinear-average-distinction}
 \end{equation}
 In the interacting BSM sector, the Bohm/Fisher feedback is therefore treated
 through the pair dynamics, whereas the Fisher-Cram\'er-Rao identity used
 here remains the standard one-body statement for the matched Born density.
The
velocity-variance decomposition \eqref{eq:sigmav} gives
\begin{equation}
\sigma_v^2
=
\frac{2}{m}\mean{Q}_\psi
+
\sigma_{v_B}^2
\geq
\frac{2}{m}\mean{Q}_\psi,
\end{equation}
and hence the momentum variance
\(\sigma_{p,\psi}^2\equiv
\langle\widehat p^2\rangle_\psi-\langle\widehat p\rangle_\psi^2\),
\begin{equation}
\sigma_{p,\psi}^2
=
m^2\sigma_v^2
\geq
\frac{\hbar^2}{4}
I[\rho].
\end{equation}
Using the Cram\'er-Rao inequality
\begin{equation}
\sigma_{x,\psi}^2
I[\rho]
\geq
1,
\end{equation}
one recovers for the position and momentum standard
deviations \(\Delta x\equiv\sigma_{x,\psi}\) and
\(\Delta p\equiv\sigma_{p,\psi}\)
\begin{equation}
\Delta x\,\Delta p
\geq
\frac{\hbar}{2}.
\label{eq:heisenberg}
\end{equation}

The two statements must therefore be kept distinct. The
Cram\'er-Rao-Fisher argument gives the standard quantum uncertainty
bound for the one-body Born density, whereas Eq.~\eqref{eq:r2-bbm}
describes the relative localization of connected possible
continuations within the branching ensemble. Their common dependence
on $\hbar$ originates from the same Schr\"odinger diffusion
coefficient and Fisher geometry, but the two quantities belong to
different statistical orders. However, the clustered pair sector also carries
its own $\hbar$-scale product: by Sec.~\ref{sec:abohm}, its
stationary size and the subcritical momentum scale
$p_c=\sqrt{2mE_c}$ combine into
\(r_{\rm cl}\,p_c\sim\hbar\), where
\(r_{\rm cl}\equiv\sqrt{\langle r^2\rangle_{\rm pair}}\). This is a
dimensional echo of the same Fisher geometry at second statistical
order, not an operator uncertainty bound, and not a derivation of the
Heisenberg inequality which involves one-point observables.

\section{Dimensional effects: free space}
\label{sec:dimension}

We now remove the confinement and consider the ensemble in
$\mathbb{R}^d$, in the gauge $V\equiv0$ specified in
Sec.~\ref{sec:nagasawa}. For the spatially uniform frozen rate considered
here, the free propagator is
\begin{equation}
G_D(x,t|x_0)
=
(4\pi\Dq t)^{-d/2}
\exp\left[-\frac{(x-x_0)^2}{4\Dq t}\right].
\end{equation}

For a uniform initial concentration $c_0$ -- the free-space counterpart
of the uniform confined density used in Sec.~\ref{sec:confined} -- the
normalized connected correlation function, obtained from the frozen pair kernel $G_{\rm br}$
by dividing by the product of the two uniform mean densities
(Appendix~\ref{app:free}), reads
\begin{equation}
g_{\rm br}(r,t)
=
\frac{\ktwo}{c_0}
\int_0^t\dd t'\,
\ee^{\beta_{\rm q}(t'-t)}
\frac{
\exp\left[-r^2/(8\Dq t')\right]
}{
(8\pi\Dq t')^{d/2}
},
\label{eq:gfree}
\end{equation}
where $r=|x-y|$. This expression is the free-space analogue of the
confined pair-correlation kernel: a pair is created locally and the two
descendants then diffuse independently during the genealogical time
$t'$.

The long-time behavior depends on the sign of $\beta_{\rm q}$ and on the
spatial dimension. In the subcritical regime, $\beta_{\rm q}<0$, the
amplitude of the correlation normalized by the squared mean density grows
as the mean population decays. Its spatial profile, however, approaches a
stationary clustered shape after normalization by its integrated pair
weight. In the supercritical regime, $\beta_{\rm q}>0$, the normalized
correlation is progressively washed out by the extended exploration
field.

\begin{figure*}[!htbp]
\includegraphics[width=\textwidth]
{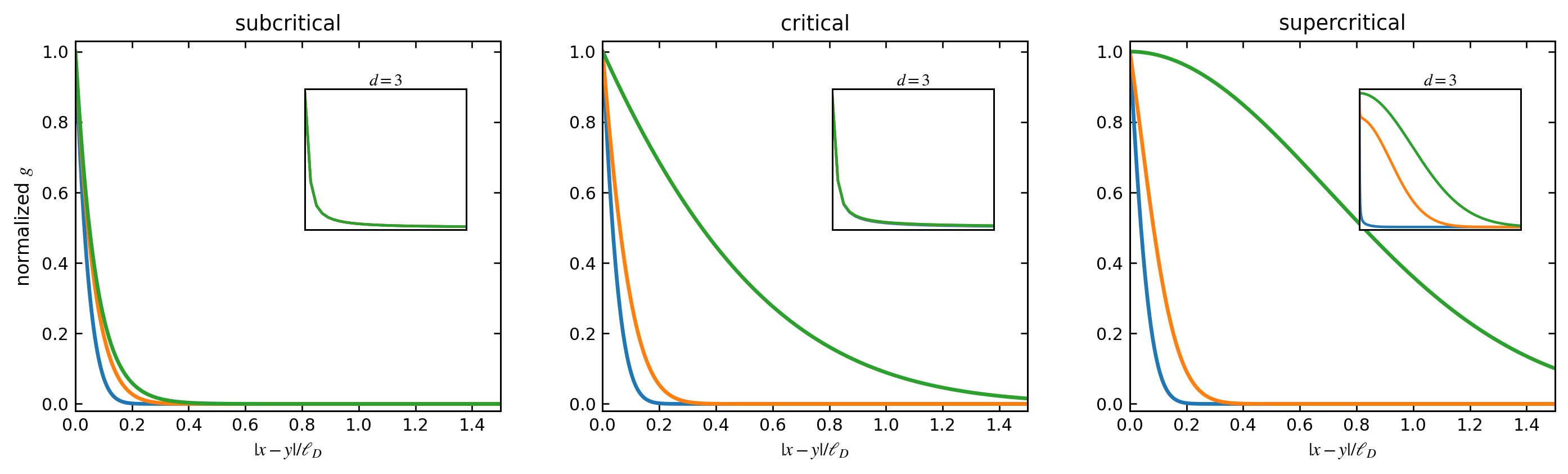}
\caption{
Normalized connected correlation profiles
\(g_{\rm br}(r,t)/g_{\rm br}(0,t)\) in the three frozen-rate regimes, generated
from Eq.~\eqref{eq:gfree}; the relative distance is measured in units
of the diffusive length $\ell_D=\sqrt{4\Dq t_{\rm max}}$,
where \(t_{\rm max}\) is the largest displayed time. In each
panel, the three curves correspond to increasing times (blue to
green). Main panels: $d=1$; insets: corresponding $d=3$ profiles. In
the subcritical regime the normalized profile converges to a
stationary clustered shape; at criticality the correlation range
grows without bound for $d\le2$ but saturates for $d>2$, illustrating
the critical dimension $d_c=2$; in the supercritical regime the
normalized correlation is progressively washed out. The contact
behavior at $r=0$ is controlled by the microscopic regularization of
the local pair source.
}
\label{fig:freespace}
\end{figure*}

At criticality, $\beta_{\rm q}=0$, the infrared behavior is dimension
dependent. For a fixed nonzero separation, the integral diverges at large
genealogical time for $d\leq2$, whereas its long-time contribution remains
finite for $d>2$. Thus the infrared critical dimension of the free
branching cloud is~\cite{Zhang1990,Houchmandzadeh2008}
\begin{equation}
d_c=2.
\end{equation}
The contact singularity at $r=0$ is a distinct short-time effect of the
local pair source and requires the usual microscopic or coarse-graining
regularization. The distinction between subcritical, critical, and
supercritical frozen rates is therefore dimension independent, whereas
the marginal case inherits the standard dimensional sensitivity of
branching random walks. The three regimes and the dimensional sensitivity
at criticality are shown in Fig.~\ref{fig:freespace}.

In the subcritical phase, the relative spread -- defined, by
translation invariance, as the $g_{\rm br}$-weighted mean-square
relative separation,
$\mean{r^2}(t)=\int\dd^dr\,r^2g_{\rm br}(r,t)\big/\!\int\dd^dr\,g_{\rm br}(r,t)$
-- is obtained in closed form,
\begin{align}
\mean{r^2}(t)
&=
4d\Dq
\frac{
\bar\tau_c-\ee^{-t/\bar\tau_c}(t+\bar\tau_c)
}{
1-\ee^{-t/\bar\tau_c}
}
\nonumber\\
&\xrightarrow[t\to\infty]{}
4d\Dq\bar\tau_c ,
\label{eq:r2free}
\end{align}
with
\begin{equation}
\bar\tau_c=-\frac{1}{\beta_{\rm q}}>0
\end{equation}
the positive decay time of the subcritical sector. It coincides with the
absolute value of the signed time $\tau_c$ used in
Sec.~\ref{sec:confined},
\begin{equation}
\bar\tau_c=|\tau_c|=-\tau_c
\qquad
\text{for}
\qquad
\beta_{\rm q}<0.
\end{equation}
This is the free-space counterpart of the confined clustered regime. It
shows explicitly that the stationary relative size is fixed by the
balance between diffusion and the effective subcritical rate. The overall
amplitude of the connected correlation may continue to evolve, but this
amplitude cancels in the normalized pair-separation observable
\eqref{eq:r2free}.
Two consistency remarks follow. First, for $d=1$ the stationary value
$4\Dq\bar\tau_c$ coincides with the deep-subcritical limit
$-4\Dq\tau_c$ of the confined benchmark, Eq.~\eqref{eq:r2-bbm} at
$a\gg1$: once the cluster is smaller than the system, the boundaries
become irrelevant and the confined and free clusters are the same
object. Second, in the kinetic parametrization of
Sec.~\ref{sec:abohm}, $\beta_{\rm q}=-2E_c/\hbar$, the stationary
spread reads
\begin{equation}
\mean{r^2}^{t\to\infty}
=
d\,\frac{\hbar^2}{mE_c}
=
2d\,\lambda_{\rm dB}^{\,2},
\end{equation}
so that the de~Broglie localization of the clustered sector found in
confinement extends to free space in any dimension, with the factor
$d$ counting the independent relative directions.

Together with Sec.~\ref{sec:confined}, this completes the frozen
benchmark. A genuine Schr\"odinger state remains on the extended side of
the frozen hierarchy; nevertheless, whenever an effective subcritical
rate is supplied to the connected pair sector, its stationary relative
scale is automatically of de~Broglie order. The significance of this
result is therefore structural: the frozen hierarchy identifies the
natural quantum length associated with a localized connected sector,
without providing the dynamical mechanism that drives the physical
system onto that branch.

The next section asks whether such a scale can instead emerge from the
fluctuating forward-backward dynamics, through a reciprocal relaxation
of the connected sector, without requiring the one-sector
Schr\"odinger-Nagasawa rate itself to become subcritical.

\section{Formation and structure of the reciprocal overlap sector}
\label{sec:reciprocal-overlap}

The stochastic extension of the Schr\"odinger-Nagasawa pair introduces
a joint statistical dynamics for the centered forward and backward
fields.  The central object of this section is the equal-time signed
connected reciprocal kernel
\begin{equation}
C_{\rm FB}(x,y,t)
=
\mathbb E_\omega
\left[
\psi_F(x,t)\psi_B(y,t)
\right]
\label{eq:full-FB-pair-kernel}
\end{equation}
where the product is formed within each realization before the ensemble
average is taken.  Centering makes this expression a connected
covariance.

The anticorrelated branch has $C_{\rm FB}(x,x,t)<0$.  Its positive organized connected weight is
\begin{equation}
\rho_{\rm BSM}(x,t)
=
-C_{\rm FB}(x,x,t),
\label{eq:physical-density-from-pair-kernel}
\end{equation}
while its dependence on the separation \(x-y\) resolves the spatial
range of the reciprocal overlap.

The collective and relative structures therefore belong to the same
signed connected object.  The spectrally matched component controls the
 organized connected weight, while the off-diagonal dependence carries the
relative organization generated by the coupled forward-backward
dynamics.

We consider the stationary Schr\"odinger-Nagasawa spectral family
\begin{equation}
\phi_n(x)
=
R_n(x)\ee^{-S_n(x)/\hbar}
\qquad
\phi_n^\dagger(x)
=
R_n(x)\ee^{S_n(x)/\hbar}
\label{eq:reference-FB-family}
\end{equation}
with
\begin{equation}
\phi_n^\dagger(x)\phi_n(x)
=
R_n^2(x)
=
\rho_n(x).
\label{eq:reference-born-density}
\end{equation}

Throughout this section,
\(\Phi_F\) and \(\Phi_B\) denote the complete stochastic Nagasawa
fields.  Their joint evolution progressively constructs
\(C_{\rm FB}\).  Fluctuation variables around the selected spectral
state are introduced only to resolve the onset and sign of this
reciprocal organization.

\subsection{Reciprocal pair equation}
\label{sec:exact-pair-equation}

In the common increasing-time representation, the complete stochastic
fields obey the coupled forward-backward equations with
\begin{equation}
\rho_\omega(x,t)
=
\Phi_F(x,t)\Phi_B(x,t)
\end{equation}
in the Bohm/Fisher feedback.  The deterministic generator is first
separated into its stationary reference and density-dependent
feedback,
\begin{equation}
\mathcal L_a[\rho_\omega]
=
\mathcal L_a^{(0)}
+
\delta\mathcal L_a[\rho_\omega]
\qquad
a\in\{F,B\},
\label{eq:generator-decomposition}
\end{equation}
where \(\mathcal L_a^{(0)}\) is evaluated on the selected stationary
Schr\"odinger-Nagasawa state.  In the common reciprocal basis the
selected reference is time independent, so it satisfies
\(\mathcal L_a^{(0)}R=0\).  Subtracting this reference equation from
the complete dynamics, after absorbing the reciprocal phase weights,
and writing \(\psi_a=\Phi_{S,a}-R\) gives the centered equations
\begin{align}
d\psi_F(x,t)
&=
\mathcal L_F^{(0)}\psi_F(x,t)\dd t
+
\delta\mathcal L_F[\rho_\omega]\Phi_{S,F}(x,t)\dd t
\nonumber\\
&\qquad\qquad\qquad\qquad\qquad\qquad\qquad+
dM_F(x,t),
\\
d\psi_B(y,t)
&=
\mathcal L_B^{(0)}\psi_B(y,t)\dd t
+
\delta\mathcal L_B[\rho_\omega]\Phi_{S,B}(y,t)\dd t
\nonumber\\
&\qquad\qquad\qquad\qquad\qquad\qquad\qquad
+
dM_B(y,t).
\label{eq:full-stochastic-FB-system}
\end{align}
The one-point centering condition is imposed throughout this connected
closure, so the reference \(R\) continues to carry the prescribed
Schr\"odinger-Nagasawa first moment: the feedback term is understood
with its ensemble mean removed,
\begin{equation}
\delta\mathcal L_a[\rho_\omega]\Phi_{S,a}
\;\longmapsto\;
\delta\mathcal L_a[\rho_\omega]\Phi_{S,a}
-
\mathbb E_\omega\!\left[
\delta\mathcal L_a[\rho_\omega]\Phi_{S,a}
\right],
\label{eq:centering-projection}
\end{equation}
which enforces \(\mathbb E_\omega[\psi_a]=0\) at all times.  This
projection leaves the connected pair dynamics unchanged, because the
subtracted mean multiplies a centered field in
Eq.~\eqref{eq:pair-bohm-feedback} below and therefore averages to
zero.
The deterministic forward and backward generators satisfy the
reciprocal adjoint relation
\begin{equation}
\mathcal L_B
=
-\mathcal L_F^\dagger.
\label{eq:reciprocal-adjoint-relation}
\end{equation}
Their joint quadratic variation is defined by
\begin{equation}
d
\left\langle
M_a(x),M_b(y)
\right\rangle_t
=
\mathcal N_{ab}(x,y,t)\dd t
\qquad
a,b\in\{F,B\},
\label{eq:joint-effective-qv}
\end{equation}
where \(\mathcal N_{\rm FB}\) denotes the instantaneous stochastic
forward-backward cross-covariance kernel.
Applying It\^o's product rule to \(\psi_F(x,t)\psi_B(y,t)\), and then
averaging, gives
\begin{equation}
\partial_tC_{\rm FB}
=
\left(
\mathcal L_F^{(0,x)}
+
\mathcal L_B^{(0,y)}
\right)
C_{\rm FB}
+
\mathcal J_{\rm FB},
\label{eq:pair-equation-compact}
\end{equation}
where
\begin{equation}
\mathcal J_{\rm FB}
=
\mathcal N_{\rm FB}
+
\mathcal B_{\rm FB}
\label{eq:pair-total-source}
\end{equation}
and
\begin{align}
\mathcal B_{\rm FB}(x,y,t)
&=
\mathbb E_\omega
\left[
\left(
\delta\mathcal L_F^{(x)}\Phi_{S,F}(x)
\right)
\psi_B(y)
\right]
\nonumber\\
&\quad+
\mathbb E_\omega
\left[
\psi_F(x)
\left(
\delta\mathcal L_B^{(y)}\Phi_{S,B}(y)
\right)
\right].
\label{eq:pair-bohm-feedback}
\end{align}

The kernel \(\mathcal J_{\rm FB}\) collects the interaction terms
responsible for the formation and reorganization of the reciprocal
overlap.  A negative cross quadratic variation directly injects the
negative covariance selected by the anticorrelated branch.  The term
\(\mathcal B_{\rm FB}\) feeds the accumulated correlations back through
the density dependence of the Bohm/Fisher drift.

The reference operator
\(\mathcal L_F^{(0)}+\mathcal L_B^{(0)}\) therefore carries the
reciprocal spectral propagation, while
\(\mathcal J_{\rm FB}\) carries the correlation dynamics.
\subsection{Nucleation and reciprocal locking}
\label{sec:overlap-nucleation}

The onset of reciprocal organization can be resolved locally around the
selected spectral state by writing
\begin{equation}
\Phi_F
=
\phi_n+\delta\Phi_F,
\qquad
\Phi_B
=
\phi_n^\dagger+\delta\Phi_B,
\end{equation}
and introducing the common reciprocal normalization
\begin{equation}
\psi_F
=
\ee^{S_n/\hbar}\delta\Phi_F,
\qquad
\psi_B
=
\ee^{-S_n/\hbar}\delta\Phi_B.
\label{eq:regular-fields}
\end{equation}
The fields \(\psi_F\) and \(\psi_B\) will be used below to resolve the
dynamically constructed reciprocal sector.  They are reduced fields relative
to the selected one-point background, not a second demographic population
added to that background.  At short times they are weak, and evaluating the
square-root genealogical covariance on the smooth reference is therefore the
appropriate bare expansion.

The corresponding realization-level pair-density variation is
\begin{equation}
\delta\rho_\omega
=
R_n(\psi_F+\psi_B)
+
\psi_F\psi_B,
\label{eq:density-regular-general}
\end{equation}
where \(\delta\rho_\omega\equiv\rho_\omega-\rho_n\).  Its ensemble
average is
\begin{align}
\mathbb E_\omega[\delta\rho_\omega]
&=
R_n\mathbb E_\omega[\psi_F+\psi_B]
+
\mathbb E_\omega[\psi_F\psi_B].
\label{eq:averaged-density-regular-general}
\end{align}
The reduced fields remain centered on the selected one-point background,
\begin{equation}
\mathbb E_\omega[\psi_F]
=
\mathbb E_\omega[\psi_B]
=0.
\label{eq:regular-fields-centering}
\end{equation}
The connected reciprocal kernel on the diagonal is then
\begin{equation}
C_{\rm FB}(x,x,T)
\equiv
\mathbb E_\omega
\!\left[\psi_F(x,T)\psi_B(x,T)\right],
\label{eq:CFB-correlated-observable}
\end{equation}
Thus \(C_{\rm FB}\) is a signed connected quantity.  On the
anticorrelated branch it is negative on the diagonal and  the organized connected weight is \(\rho_{\rm BSM}=-C_{\rm FB}(x,x,T)\).  The ensemble-averaged
realization-level product follows the exact budget
\(\mathbb E_\omega[\rho_\omega]=\rho_n-\rho_{\rm BSM}\).

Accordingly,
\begin{equation}
U
=
\psi_F+\psi_B
\label{eq:linear-density-channel}
\end{equation}
is the linear density channel, while \(\psi_F\psi_B\) is the first
nonlinear reciprocal contribution.  Cancellation of the realization-level
linear channel leaves the bilinear connected contribution in
Eq.~\eqref{eq:CFB-correlated-observable}.

At the minimal level used here, the associated Bohm/Fisher feedback is
resolved through the first variation
\begin{equation}
\delta Q
=
-{\hbar^2\over4m}
\left[
{1\over R_n}
\nabla^2
\left(
{\delta\rho_\omega\over R_n}
\right)
-
{\nabla^2R_n\over R_n^3}
\delta\rho_\omega
\right].
\label{eq:deltaQ}
\end{equation}
Higher functional variations contribute to the finite-amplitude nonlinear
feedback and lie beyond the local onset expansion used here.

We decompose the backward reduced field into the component carried by the
forward field and an orthogonal residual,
\begin{equation}
\psi_B(x,t)
=
\alpha(t)\psi_F(x,t)
+
w(x,t),
\label{eq:alpha-r-decomposition}
\end{equation}
with
\begin{equation}
\int\dd x\,
\mathbb E_\omega
\left[
\psi_F(x,t)w(x,t)
\right]
=0.
\label{eq:r-orthogonality}
\end{equation}

We define the integrated overlap and the corresponding sector weights,
\begin{align}
P(t)
&=
\int\dd x\,
\mathbb E_\omega
\left[
\psi_F\psi_B
\right],
\\
S_F(t)
&=
\int\dd x\,
\mathbb E_\omega
\left[
\psi_F^2
\right],
\\
S_B(t)
&=
\int\dd x\,
\mathbb E_\omega
\left[
\psi_B^2
\right],
\\
S_w(t)
&=
\int\dd x\,
\mathbb E_\omega
\left[
w^2
\right].
\end{align}
The orthogonality condition gives
\begin{equation}
P
=
\alpha S_F,
\qquad
\alpha
=
{P\over S_F},
\label{eq:P-alpha-SF}
\end{equation}
and the exact decomposition
\begin{equation}
S_B
=
\alpha^2S_F
+
S_w.
\label{eq:exact-projected-budget}
\end{equation}
The normalized projected overlap fraction is therefore
\begin{equation}
X_{\rm FB}(t)
=
{P^2(t)\over S_F(t)S_B(t)},
\label{eq:projected-overlap-fraction}
\end{equation}
with
\begin{equation}
1-X_{\rm FB}
=
{S_w\over S_B}.
\label{eq:residual-overlap-fraction}
\end{equation}
The Cauchy-Schwarz inequality gives
\begin{equation}
0
\leq
X_{\rm FB}
\leq
1.
\label{eq:overlap-fraction-bound}
\end{equation}
The limit \(X_{\rm FB}\to1\) therefore corresponds to the disappearance
of the orthogonal component \(w\) and the formation of a fully projected
reciprocal sector.  It does not, by itself, imply the disappearance of the
one-point reference \(R_n\).

We collect the centered fields of the common reciprocal basis into
the doublet \(\psi=(\psi_F,\psi_B)^T\), referred to as the reduced
reciprocal basis, and denote by \(\mathbb N_\psi^{\rm eff}\) the
coincident-point covariance of its dressed martingale sources.  This
kernel is not derived here; it is parameterized by the most general
local form compatible with the structure established in
Sec.~\ref{sec:reciprocal-noise-embedding}.  The reciprocal
representation weights both marginals by the same reference \(R\), so
the two diagonal entries are equal, and positive semidefiniteness
bounds the cross channel by the marginals.  The cross channel is
written with a nonpositive sign, anticipating the dynamical selection
mechanism established below.  Hence
\begin{equation}
\mathbb N_\psi^{\rm eff}
=
q
\begin{pmatrix}
1&-\epsilon\\
-\epsilon&1
\end{pmatrix},
\qquad
q\geq0,
\qquad
0\leq\epsilon\leq1,
\label{eq:local-reciprocal-covariance}
\end{equation}
where \(q\) is the dressed marginal noise power and \(\epsilon\) is
the instantaneous forward-backward source anticorrelation.  In the
bare limit the covariance reduces to the genealogical kernel of
Eq.~\eqref{eq:bare-genealogical-covariance} evaluated on the
reference, \(q=\nu_2R\) and \(\epsilon=0\); the dressing consists in
renormalizing \(q\) and opening the anticorrelated cross channel
\(\epsilon>0\).

Equation~\eqref{eq:local-reciprocal-covariance} is the
reference-field projection of a field-dependent covariance that makes
the connection with a response-field description explicit.  The
minimal local completion that preserves the two genealogical marginal
sources of Eq.~\eqref{eq:bare-genealogical-covariance} while opening
an FB channel is
\begin{equation}
\Gamma_S[\Phi_S;\epsilon]
=
\nu_2
\begin{pmatrix}
\Phi_{S,F}&-\epsilon\sqrt{\Phi_{S,F}\Phi_{S,B}}\\
-\epsilon\sqrt{\Phi_{S,F}\Phi_{S,B}}&\Phi_{S,B}
\end{pmatrix}.
\label{eq:pedagogical-dressed-Gamma}
\end{equation}
Equivalently, its FB dressing is isolated by
\begin{equation}
\Gamma_S
=
\Gamma_S^{(0)}
-
\epsilon\nu_2\sqrt{\Phi_{S,F}\Phi_{S,B}}
\left(\begin{smallmatrix}0&1\\1&0\end{smallmatrix}\right).
\label{eq:pedagogical-Gamma-antidiagonal}
\end{equation}
The first term on the right-hand side is the bare genealogical
covariance of Eq.~\eqref{eq:bare-genealogical-covariance}: genealogy
populates its diagonal because each sector branches in proportion to
its own realized mass.  The second term has support only on the
antidiagonal and is nonpositive for \(\epsilon>0\); it provides an
admissible local representation of the reciprocal dressing of the
joint law, correlating the forward and backward martingales
negatively without replacing either marginal branching process.
Evaluated on the reference \(\Phi_S=(R,R)^T\),
Eq.~\eqref{eq:pedagogical-dressed-Gamma} reduces to
Eq.~\eqref{eq:local-reciprocal-covariance} with \(q=\nu_2R\).

At short times the reduced fields are close to the smooth reference.  The
initial half-kernel is therefore obtained by evaluating the square-root
genealogical covariance on that reference,
\begin{equation}
\widehat\Gamma_S^{(0)}
=
\nu_2R\,\mathbb I,
\label{eq:short-time-reference-covariance}
\end{equation}
so that
\begin{equation}
\mathbb N_\psi^{(0)}
=
2\widehat\Gamma_S^{(0)}
=
2\nu_2R\,\mathbb I.
\label{eq:bare-full-noise-kernel}
\end{equation}
This corresponds to \(q_0=2\nu_2R\) and \(\epsilon_0=0\).  The factor
\(R\) is the reference evaluation of the genealogical amplitude.  The
initial stochastic sources are independent, and the deterministic
dynamics places both reciprocal fields on the common reference $R$
before the noise is introduced.  A finite \(\epsilon\) parametrizes a forward-backward cross entry of the dressed effective kernel generated by projection of the interacting reciprocal dynamics; the bare microscopic martingales retain zero cross quadratic variation.

This distinction becomes transparent in the symmetric and
antisymmetric channels
\begin{equation}
U=\psi_F+\psi_B,
\qquad
W=\psi_B-\psi_F.
\label{eq:UW-fields}
\end{equation}
Their martingale sources follow by linearity, \(dM_U=dM_F+dM_B\) and
\(dM_W=dM_B-dM_F\), and the effective channel covariances are defined,
as in Eq.~\eqref{eq:joint-effective-qv}, through the joint quadratic
variation
\begin{equation}
d\left\langle M_a,M_b\right\rangle_t
=
N_{ab}^{\rm eff}\,\dd t,
\qquad
a,b\in\{U,W\}.
\label{eq:UW-channel-covariances}
\end{equation}
Bilinearity of the quadratic variation, applied to the convention of
Eq.~\eqref{eq:local-reciprocal-covariance}, gives
\begin{align}
N_{UU}^{\rm eff}
&=
2q(1-\epsilon),
\label{eq:NUU-effective}
\\
N_{WW}^{\rm eff}
&=
2q(1+\epsilon),
\label{eq:NWW-effective}
\\
N_{UW}^{\rm eff}
&=
0.
\label{eq:NUW-effective}
\end{align}
The cross covariance vanishes because the two diagonal entries of
Eq.~\eqref{eq:local-reciprocal-covariance} are equal: the reciprocal
F\(\leftrightarrow\)B symmetry makes \((U,W)\) the principal axes of
the local source covariance, with eigenvalues \(2q(1\mp\epsilon)\).

The physical roles of the two channels are fixed by
Eq.~\eqref{eq:density-regular-general}: \(U\) is the linear density
fluctuation, whose product with \(R_n\) is the only first-order
contribution to \(\delta\rho_\omega\), whereas \(W\) is phase-like at
linear order and enters the density only through the bilinear term
\(\psi_F\psi_B=(U^2-W^2)/4\).  A stationary Born profile therefore
requires that stochastic injection into the \(U\) channel be
suppressed.  For a nonzero marginal noise power \(q\), the condition
\(N_{UU}^{\rm eff}\to0\) fixes the stationary source correlation to
\begin{equation}
\epsilon_\star=1.
\label{eq:epsilon-stationary}
\end{equation}
Anticorrelation does not remove the noise, but reroutes it.  At
\(\epsilon_\star=1\) the source covariance has rank one, the density channel receives no further kicks,
\begin{equation}
N_{UU}^{\rm eff}\longrightarrow0,
\label{eq:NUU-locking-condition}
\end{equation}
and the entire source power is carried by the antisymmetric channel, \(N_{WW}^{\rm eff}\to4q\): every local
upward fluctuation of one reciprocal field is compensated by a
downward fluctuation of the other, preserving the collective diagonal
while the relative correlations continue to evolve.

The instantaneous locking of the sources translates into the
accumulated diagnostics of
Eqs.~\eqref{eq:P-alpha-SF}--\eqref{eq:projected-overlap-fraction}:
the evolved fields reach \(\psi_B\simeq-\psi_F\) when the projection
diagnostics simultaneously give \(\alpha\to-1\) and
\(X_{\rm FB}\to1\).  On this branch
\(C_{\rm FB}=\mathbb E_\omega[\psi_F\psi_B]\) is negative on the
diagonal, and the convention \(\rho_{\rm BSM}=-C_{\rm FB}(x,x)\)
assigns positive weight to the organized paired sector.
The cancellation in Eq.~\eqref{eq:NUU-locking-condition} concerns the full
effective covariance of the reciprocal fields.  It should therefore not be
identified with the disappearance of the initial reference kernel taken in
isolation.

The reduced forward-backward cross covariance associated with
Eq.~\eqref{eq:local-reciprocal-covariance} is
\begin{equation}
\mathcal N_{\rm FB}^{(\psi)}
=
-\epsilon q.
\label{eq:reduced-cross-covariance}
\end{equation}
Reducing the stochastic source of the linear realization-level channel
therefore requires a negative forward-backward cross covariance.  The
coefficients \(\epsilon(t)\) and \(\alpha(t)\) characterize two
different levels of the dynamics: the former measures the instantaneous
correlation of the dressed stochastic sources, whereas the latter measures
the accumulated projection of the evolved fields through
\begin{equation}
P(t)
=
\alpha(t)S_F(t).
\end{equation}

To connect the instantaneous source covariance with the accumulated overlap,
we write the reduced dynamics schematically as
\begin{align}
d\psi_F
&=
\left(
L_F\psi_F+\mathcal B_F
\right)\dd t
+
dM_F,
\\
d\psi_B
&=
\left(
L_B\psi_B+\mathcal B_B
\right)\dd t
+
dM_B,
\label{eq:reduced-sector-dynamics}
\end{align}
where \(L_F,L_B\) are the reciprocal linear drifts and
\(\mathcal B_F,\mathcal B_B\) are the corresponding Bohm/Fisher feedback
terms.  It\^o's product rule gives
\begin{equation}
{dP\over dt}
=
N_{\rm FB}(t)
+
B_{\rm FB}(t),
\label{eq:P-noise-bohm}
\end{equation}
where
\begin{equation}
N_{\rm FB}(t)
=
\int\dd x\,
\mathcal N_{\rm FB}^{(\psi)}(x,x,t),
\label{eq:NFB-definition}
\end{equation}
and
\begin{equation}
B_{\rm FB}(t)
=
\int\dd x\,
\mathbb E_\omega
\left[
\mathcal B_F(x,t)\psi_B(x,t)
+
\psi_F(x,t)\mathcal B_B(x,t)
\right].
\label{eq:BFB-definition}
\end{equation}
The reciprocal adjoint relation cancels the linear drift contribution after
integration, under the boundary conditions of the selected spectral problem.

\begin{figure}[!t]
\includegraphics[width=\columnwidth]
{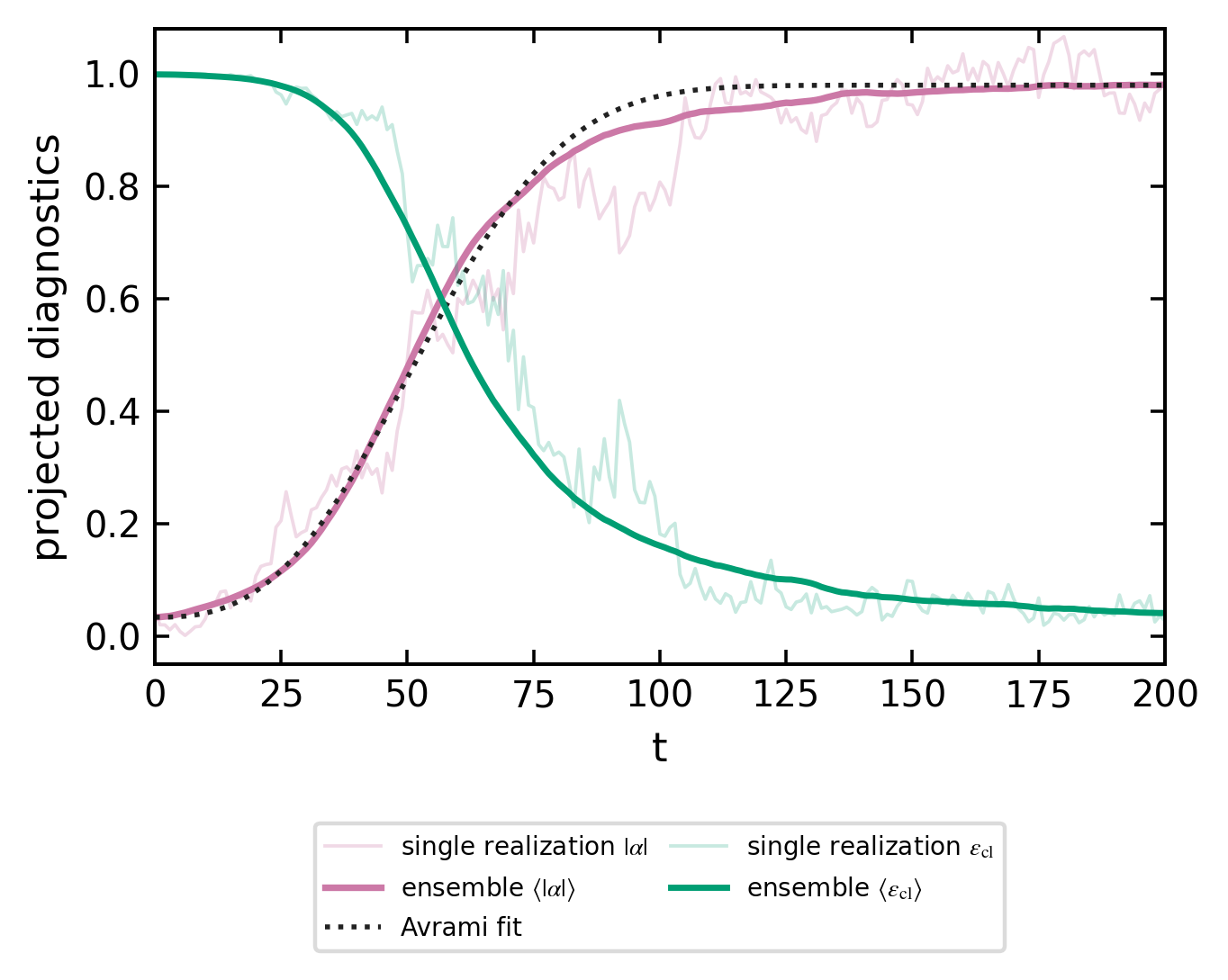}
\caption{
Formation of the projected forward-backward sector in BSM-MC
(Appendix~\ref{app:mc-protocol}).  The magnitude of the signed
realization-wise projection diagnostic
\(|\alpha_{\rm phys}(t)|\) grows toward unity while the corresponding
relative residual weight
\(\varepsilon_{{\rm cl},\omega}(t)\) decreases.  These numerical
quantities are the realization-wise counterparts of the reciprocal locking
and residual depletion described by
Eqs.~\eqref{eq:alpha-r-decomposition}--\eqref{eq:overlap-fraction-bound}.
Transparent curves show one stochastic realization and solid curves show
ensemble averages of the realization-wise diagnostics.  The dotted line is
a Johnson-Mehl-Avrami-Kolmogorov (JMAK) fit of the ensemble-averaged
\(|\alpha_{\rm phys}(t)|\), used as a phenomenological description of the
observed nucleation-and-saturation kinetics.
}
\label{fig:alpha-closure-onset}
\end{figure}

At the onset of reciprocal organization the Bohm/Fisher term is built from
the correlations already accumulated.  If a negative cross covariance is
dynamically generated and the feedback contribution is initially
subleading, it supplies the first direct injection into the overlap,
\begin{equation}
{dP\over dt}
\simeq
N_{\rm FB}(t)
<0,
\end{equation}
selecting
\begin{equation}
P<0,
\qquad
\alpha<0.
\end{equation}
Within the projected numerical diagnostic, the observed approach toward
\begin{equation}
\alpha\to-1,
\qquad
X_{\rm FB}\to1
\end{equation}
describes progressive locking of the two stochastic Nagasawa sectors and
depletion of their orthogonal component.  This is a reorganization of the
connected sector around the selected one-point background.
This conversion is quantified in Fig.~\ref{fig:alpha-closure-onset}.  The
growth of the numerical projection diagnostic toward unit magnitude is
accompanied by the depletion of the realization-wise orthogonal component,
providing the numerical counterpart of the projected-overlap budget derived
above.

\begin{figure*}[!t]
\includegraphics[width=\textwidth]
{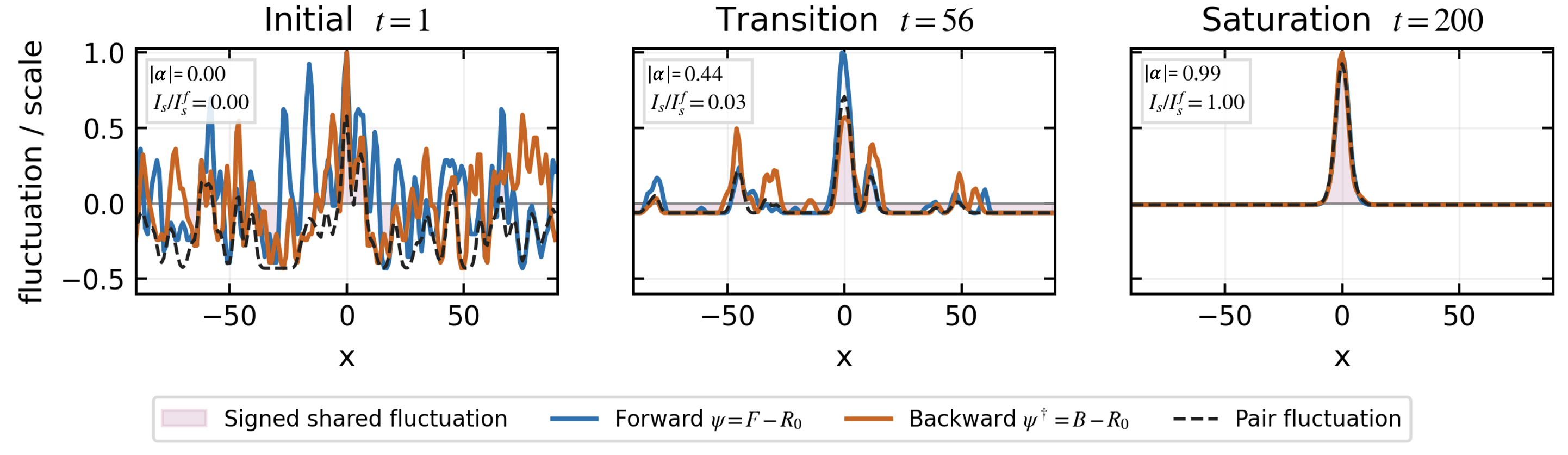}
\caption{
BSM-MC diagnostic of the progressive structuring of the reciprocal
forward-backward overlap sector (Appendix~\ref{app:mc-protocol}).  Each
panel shows, at increasing times ($t=1$, $56$, $200$), the forward and
backward numerical reduced fields \(\psi_F\) and \(\psi_B\), measured
with respect to the reference background as specified in
Appendix~\ref{app:mc-protocol}, their pair contribution (dashed), and the
signed shared component (shaded).  Insets: measured projection coefficient
\(|\alpha_{\rm phys}|\) and shared-field intensity \(I_s\), normalized by
its saturation value \(I_s^{f}\) (Appendix~\ref{app:mc-protocol}).
Initially the two sectors are weakly correlated
(\(|\alpha_{\rm phys}|\simeq0\)); during reciprocal organization they
develop an increasingly dominant shared spatial component, and at the late
time displayed the numerical projection approaches unit magnitude while the
orthogonal component is strongly suppressed.  The figure therefore
illustrates the real-space formation of the projected reciprocal sector
described by Eqs.~\eqref{eq:alpha-r-decomposition}
and~\eqref{eq:projected-overlap-fraction}.
}
\label{fig:projected-alpha-overlap}
\end{figure*}

The same transition can be followed directly in real space.
Figure~\ref{fig:projected-alpha-overlap} shows the progressive spatial
organization of the two reciprocal reduced fields.  The initially weakly
correlated fields develop an increasingly dominant shared component as the
orthogonal component is depleted, providing the spatial counterpart of the
integrated projection diagnostics of Fig.~\ref{fig:alpha-closure-onset}.

\subsection{Collective reciprocal mode and Born density}
\label{sec:collective-born-sector}

The nonlinear pair dynamics contains a collective diagonal dependence
and a relative off-diagonal dependence.  In center-of-mass and
relative coordinates, \(X=(x+y)/2\) and \(r=x-y\), we resolve them by
defining the normalized relative profile
\begin{equation}
F(X,r,t)
=
{C_{\rm FB}(X,r,t)\over C_{\rm FB}(X,0,t)},
\label{eq:relative-profile-definition}
\end{equation}
wherever the diagonal is nonzero, so that, exactly and without loss
of generality,
\begin{equation}
C_{\rm FB}(X,r,t)
=
-\rho_{\rm BSM}(X,t)\,F(X,r,t),
\label{eq:collective-relative-factorization}
\end{equation}
with \(F(X,0,t)=1\) and \(\rho_{\rm BSM}(X,t)=-C_{\rm FB}(X,0,t)\)
the positive collective diagonal weight on the anticorrelated branch.
In the locally homogeneous infrared regime, where the collective
background varies slowly over the relative correlation range, we neglect
the residual dependence of the relative profile on $X$ and use
\begin{equation}
F(X,r,t)\simeq F(r,t).
\label{eq}
\end{equation}
The pair kernel then takes the collective$\times$relative form used in
the remainder of this section.

The collective spatial structure follows directly from the spectrum of
the one-body part of Eq.~\eqref{eq:pair-equation-compact}.  The reciprocal
similarity transformation preserves this spectrum, so it may be written in
the original forward-backward representation as follows.

Let
\begin{equation}
\mathcal L_F^{(0)}\phi_j
=
\lambda_j\phi_j
\label{eq:forward-spectrum}
\end{equation}
and, by reciprocal adjunction,
\begin{equation}
\mathcal L_B^{(0)}\phi_j^\dagger
=
-\lambda_j\phi_j^\dagger.
\label{eq:backward-spectrum}
\end{equation}

The associated pair modes
\begin{equation}
K_{jk}(x,y)
=
\phi_j(x)\phi_k^\dagger(y)
\label{eq:pair-spectral-modes}
\end{equation}
satisfy
\begin{equation}
\left(
\mathcal L_F^{(0,x)}
+
\mathcal L_B^{(0,y)}
\right)
K_{jk}
=
(\lambda_j-\lambda_k)K_{jk}.
\label{eq:pair-spectrum}
\end{equation}

Stationarity requires cancellation of the reciprocal spectral rates.
Every spectrally matched pair \(j=k\) therefore belongs to the
zero-eigenvalue sector of the reciprocal generator,
\begin{equation}
\left(
\mathcal L_F^{(0,x)}
+
\mathcal L_B^{(0,y)}
\right)
K_{jj}
=
0.
\label{eq:matched-reciprocal-zero-mode}
\end{equation}
The equality $j=k$ expresses energy conservation between the forward and
backward members of a reciprocal pair.  The prepared
Schr\"odinger-Nagasawa state fixes the nondegenerate energy sector $n$.
Modes $K_{jj}$ with $j\neq n$ belong to different one-body energies and
require an external energy transfer or a change of boundary data.  Mixed
modes $K_{jk}$ with $j\neq k$ carry the nonzero reciprocal rate
$\lambda_j-\lambda_k$ and cannot form a stationary collective component.
Consequently, within the closed selected sector, the unique stationary
collective shape is
\begin{equation}
K_n(x,y)
=
\phi_n(x)\phi_n^\dagger(y),
\label{eq:selected-matched-kernel}
\end{equation}
which belongs identically to the zero-eigenvalue sector of the
reciprocal generator.

In the common reciprocal basis, the opposite phase weights of
$\phi_n$ and $\phi_n^\dagger$ are removed and both selected one-body
factors reduce to the amplitude $R_n$.  The collective pair mode is
therefore represented by $R_n(x)R_n(y)$.  Choosing another mode $R_j$
would replace the prepared energy $E_n$ by $E_j$ and is excluded by the
closed energy-conserving evolution considered here.

Its diagonal is
\begin{equation}
K_n(x,x)
=
\phi_n(x)\phi_n^\dagger(x)
=
R_n^2(x).
\label{eq:matched-kernel-diagonal}
\end{equation}

A scalar coefficient multiplying \(K_n\) changes the total collective
weight while preserving its spatial form.  For the normalized organized
sector,
\begin{equation}
\int\dd x\,
-C_{\rm FB}(x,x,t)
=
1
\label{eq:pair-normalization}
\end{equation}
and
\begin{equation}
\int\dd x\,
R_n^2(x)
=
1.
\end{equation}
The stationary collective factor therefore has unit weight and spatial
profile
\begin{equation}
G^{\star}(x)
=
K_n(x,x)
=
R_n^2(x),
\label{eq:collective-pair-component}
\end{equation}
while the signed connected component is
\begin{equation}
C_{\rm FB}^{\rm coll}(x,y)
=
-K_n(x,y).
\label{eq:signed-collective-pair-component}
\end{equation}
Its diagonal gives
\begin{equation}
-C_{\rm FB}^{\rm coll}(x,x)
=
R_n^2(x)
=
\rho_n(x).
\label{eq:born-diagonal}
\end{equation}

Thus \(\rho_n\) is the positive diagonal weight of the natural reciprocal
zero mode.  The  organized connected weight
\(\rho_{\rm BSM}=-C_{\rm FB}(x,x)\) equals this Born profile when the
remaining interaction-generated components have zero net diagonal
contribution or compensate within the full signed kernel.  Spectral
cancellation identifies the admissible spatial profile, while the joint
dynamics fixes its stationary weight.

The cancellation occurs before the choice of energy gauge,
\begin{equation}
\lambda_n-\lambda_n
=
0,
\end{equation}
while the energy gauge represents the same selected state with
vanishing individual spectral rates.

During the transient, the interaction kernel \(\mathcal J_{\rm FB}\) can
change the collective weight and populate relative deformations.  Energy
conservation keeps the stationary collective projection in the selected
$n$ sector, and reciprocal spectral cancellation fixes its spatial shape
to $K_n$.  The exact diagonal equation determines its time-dependent
coefficient.

The Bohm functional also satisfies the scale invariance
\begin{equation}
Q[c\rho]
=
Q[\rho]
\qquad
c>0,
\end{equation}
which gives
\begin{equation}
DQ[\rho_n]\rho_n
=
0,
\label{eq:bohm-ward-identity}
\end{equation}
where $DQ[\rho_n]$ denotes the first functional variation of \(Q\) at \(\rho_n\).
The collective density-rescaling direction therefore carries no
intrinsic local restoring rate from the Bohm functional.

Reciprocal spectral cancellation and energy conservation therefore fix
the collective spatial mode.  The identity
Eq.~\eqref{eq:bohm-ward-identity} removes an intrinsic Bohm restoring rate
along its density-rescaling direction.  After the exact diagonal
contribution has been separated, the cross-noise and Bohm/Fisher feedback
act through the diagonal-preserving relative equation derived below.

\subsection{Diagonal and relative overlap dynamics}
\label{sec:relative-overlap-sector}
We now derive the coupled dynamics of the two factors of the
collective-relative decomposition,
Eqs.~\eqref{eq:relative-profile-definition}
and~\eqref{eq:collective-relative-factorization}.  The derivation
proceeds in three steps.  First, the exact pair equation is separated
into a diagonal equation for \(\rho_{\rm BSM}\) and a
normalization-preserving equation for \(F\).  Second, outside the
microscopic overlap core the relative feedback is closed by a local
infrared expansion, which defines the coefficients \(\mu_{\rm FB}\)
and \(D_{\rm eff}\).  Third, the stationary relative profile is
solved, yielding the finite reciprocal range \(\xi_{\rm FB}\).

For compactness, denote by
\begin{equation}
\mathcal R_{\rm FB}[C]
=
\left(
\mathcal L_F^{(0,x)}
+
\mathcal L_B^{(0,y)}
\right)C
+
\mathcal J_{\rm FB}[C]
\label{eq:pair-full-generator}
\end{equation}
the complete right-hand side of Eq.~\eqref{eq:pair-equation-compact}.
Setting \(r=0\) gives the diagonal evolution
\begin{equation}
\partial_t\rho_{\rm BSM}(X_{\rm cm},t)
=
-\mathcal R_{\rm FB}[C](X_{\rm cm},0,t).
\label{eq:exact-diagonal-equation}
\end{equation}
Differentiating the factorization,
\(\partial_tC=-F\,\partial_t\rho_{\rm BSM}
-\rho_{\rm BSM}\,\partial_tF\),
and eliminating \(\partial_t\rho_{\rm BSM}\) with
Eq.~\eqref{eq:exact-diagonal-equation} yields
\begin{align}
\rho_{\rm BSM}\,\partial_tF
&=
-\mathcal R_{\rm FB}[C](X_{\rm cm},r,t)
\nonumber\\
&\quad+
F(X_{\rm cm},r,t)\,
\mathcal R_{\rm FB}[C](X_{\rm cm},0,t).
\label{eq:exact-relative-subtraction}
\end{align}
This equation is exact.  Its right-hand side vanishes identically at
\(r=0\), so the normalization \(F(X_{\rm cm},0,t)=1\) is preserved by
construction: the diagonal weight evolves only through
Eq.~\eqref{eq:exact-diagonal-equation}, while
Eq.~\eqref{eq:exact-relative-subtraction} reorganizes the pair kernel
at fixed diagonal.  Its interaction part,
\begin{align}
\mathcal I_{\rm FB}[F](X_{\rm cm},r,t)
&=
-{1\over\rho_{\rm BSM}(X_{\rm cm},t)}
\Big[
\mathcal J_{\rm FB}(X_{\rm cm},r,t)
\nonumber\\
&\qquad-
F(X_{\rm cm},r,t)\,
\mathcal J_{\rm FB}(X_{\rm cm},0,t)
\Big],
\label{eq:relative-overlap-feedback}
\end{align}
satisfies \(\mathcal I_{\rm FB}[F](X_{\rm cm},0,t)=0\) and is
therefore the diagonal-preserving component of the reciprocal
interaction kernel.  It collects the cross-covariance seed generated
by \(\mathcal N_{\rm FB}\), which nucleates the short-range overlap
at the onset of reciprocal organization, and the density-dependent
Bohm-Fisher feedback of \(\mathcal B_{\rm FB}\), which makes
\(\mathcal I_{\rm FB}\) a state-dependent functional of the
accumulated pair correlations.

The infrared closure is organized by the resolved pair-creation scale
\(\ell\).  For \(|r|\lesssim\ell\), the microscopic overlap core, the
complete nonlinear kernel is retained.  For \(|r|\gtrsim\ell\) the
relative profile varies slowly and the action of the feedback admits
the local expansion
\begin{equation}
\mathcal I_{\rm FB}[F]
=
-\mu_{\rm FB}F
+
\Delta D\,\nabla_r^2F
+
O\left(F^2,\nabla_r^4F\right),
\label{eq:interaction-relative-expansion}
\end{equation}
which defines \(\mu_{\rm FB}\) and \(\Delta D\) as its two leading
coefficients; both depend parametrically on the selected reference
family.  Using
\(\nabla_x^2+\nabla_y^2=\tfrac12\nabla_{X}^2+2\nabla_r^2\), each
reciprocal member diffusing with \(\Dq\), the projected fluctuation
hierarchy supplies the bare relative diffusivity
\(D_{\rm rel}^{(0)}=2\Dq=\hbar/m\), and the effective coefficient
\begin{equation}
D_{\rm eff}
=
D_{\rm rel}^{(0)}
+
\Delta D
\label{eq:effective-relative-diffusion}
\end{equation}
governs the propagation of relative deformations.  The infrared
relative dynamics reads
\begin{equation}
\partial_tF
=
D_{\rm eff}\nabla_r^2F
-
\mu_{\rm FB}F
+
\cdots
\qquad
|r|\gtrsim\ell,
\label{eq:relative-overlap-dynamics}
\end{equation}
with small-\(k\) dispersion
\begin{equation}
\lambda_{\rm rel}(k;k_n)
=
-\mu_{\rm FB}
-
D_{\rm eff}k^2
+
O(k^4),
\label{eq:relative-response}
\end{equation}
where \(k\) is conjugate to \(r\) and \(k_n\) is the spectral scale
of the selected reference state.  The intrinsic reciprocal response
is separated from this finite-size scale by the ordered limits
\begin{equation}
\mu_{\rm FB}
=
-\lim_{k_n\to0}
\lambda_{\rm rel}(0;k_n),
\label{eq:intrinsic-overlap-mass}
\end{equation}
the relative \(k=0\) limit being taken before the large-system limit
\(k_n\to0\) of the reference family.

The sign of \(\mu_{\rm FB}\) selects the infrared regime:
\(\mu_{\rm FB}>0\) yields a stable finite-range reciprocal sector,
\(\mu_{\rm FB}=0\) a gapless relative sector, and \(\mu_{\rm FB}<0\)
a long-wavelength relative instability.  The finite-range organized
branch therefore corresponds to \(\mu_{\rm FB}>0\).  The expansion
identifies the channel through which this coefficient enters the
relative dynamics; its sign, magnitude, and time dependence are
dynamical outputs of the full joint law.

In the stationary finite-range regime and outside the core, the
profile obeys
\begin{equation}
\left(
-D_{\rm eff}\nabla_r^2
+
\mu_{\rm FB}
\right)
F_\star(r)
=
0,
\qquad
|r|\gtrsim\ell,
\label{eq:stationary-relative-equation}
\end{equation}
where \(F_\star(r)\) denotes the stationary relative profile in the
locally homogeneous relative sector.  The microscopic core fixes the
matching amplitude of the infrared tail, consistently with the exact
normalization \(F(X_{\rm cm},0,t)=1\).  The large-distance solution
is
\begin{equation}
F_\star(r)
\propto
r^{-(d-1)/2}
\exp\left(-{r\over\xi_{\rm FB}}\right),
\qquad
\xi_{\rm FB}
=
\sqrt{{D_{\rm eff}\over\mu_{\rm FB}}},
\label{eq:xiFB}
\end{equation}
reducing in one spatial dimension to
\(F_\star(r)\propto\ee^{-|r|/\xi_{\rm FB}}\) outside the core, with
unit prefactor in the idealization of a pointlike overlap core.

For a stationary zero-current well satisfying the diagonal matching
condition \(\rho_{\rm BSM}(X_{\rm cm})=R_n^2(X_{\rm cm})\), the
near-diagonal pair kernel becomes
\begin{equation}
C_{\rm FB}^\star(X_{\rm cm},r)
\simeq
-R_n^2(X_{\rm cm})\,F_\star(r).
\label{eq:stationary-well-pair-kernel}
\end{equation}
Its diagonal carries the Born density, and its relative tail decays
over the finite range \(\xi_{\rm FB}\).  Under diagonal matching and
for \(\mu_{\rm FB}>0\), the Born profile and the finite reciprocal
range are thus the diagonal and off-diagonal projections of the same
pair kernel: reciprocal spectral cancellation supplies the matched
collective mode, and the dressed joint dynamics selects its diagonal
weight and off-diagonal response.
\subsection{Ensemble averaging and effective loss of ergodicity}
\label{sec:ergodicity-overlap}

The separation between the collective diagonal and the relative overlap
also clarifies the statistical meaning of the localized sector.  The
 organized connected weight
\begin{equation}
\rho_{\rm BSM}(X_{\rm cm},t)
=
-C_{\rm FB}(X_{\rm cm},0,t)
\end{equation}
is obtained after averaging over the stochastic realizations.  In the
matched stationary sector it is required to approach
\begin{equation}
\rho_{\rm BSM}(X_{\rm cm})
=
R_n^2(X_{\rm cm}),
\end{equation}
and therefore retains the extended Schr\"odinger-Nagasawa envelope.

\begin{figure}[!t]
\includegraphics[width=\columnwidth]
{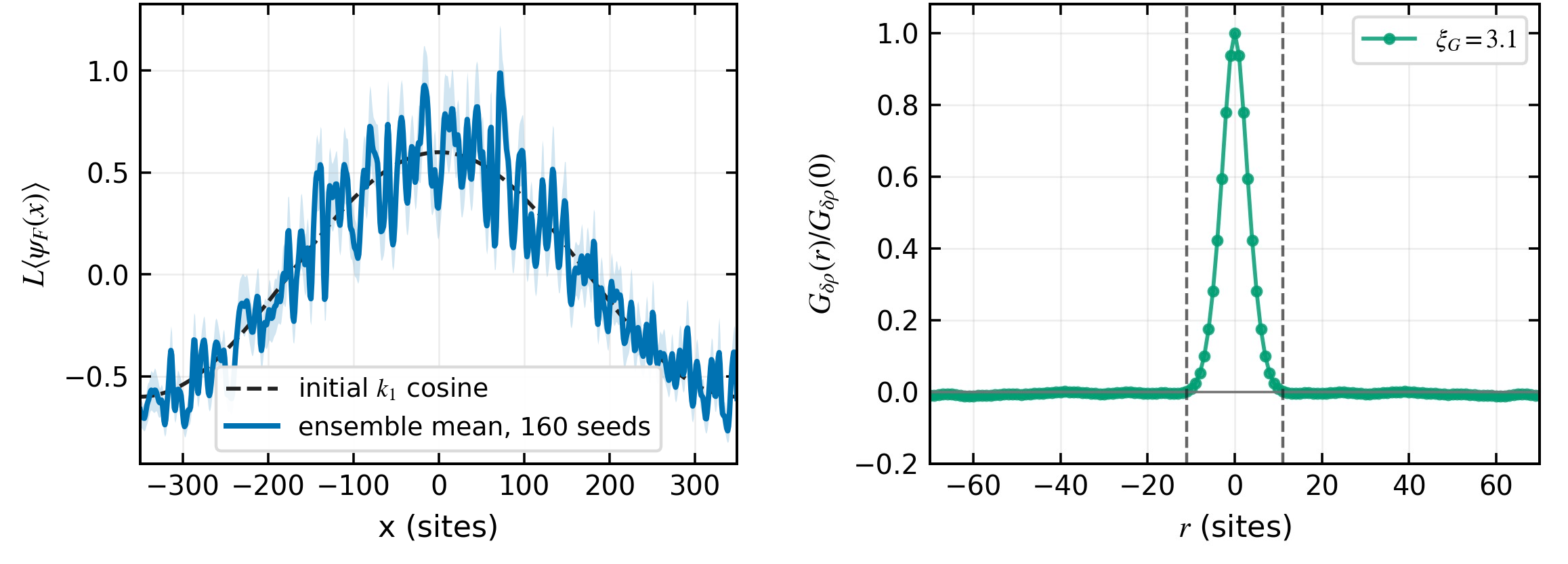}
\caption{
BSM-MC illustration of the separation between collective and relative
observables in the organized reciprocal sector.  Left: normalized
pair-amplitude estimator averaged over the stochastic ensemble,
compared with the prescribed Schr\"odinger-Nagasawa profile.  The
collective observable retains the extended spectral envelope.  Right:
normalized pair-density diagnostic as a function of relative
separation, showing the finite-range structure generated by reciprocal
locking.  The same stochastic dynamics therefore reproduces the Born
envelope at the collective level while retaining a localized structure
in the relative sector. This separation illustrates the effective loss
of ergodicity associated with the selection of a correlated
forward-backward configuration in individual realizations.
}
\label{fig:mean-vs-correlations}
\end{figure}

A single organized realization contains additional information.  Once
the reciprocal overlap has nucleated, the forward and backward fields
are locked on a correlated structure with a finite relative extent.
This information is carried by the dependence of the pair kernel on
\begin{equation}
r=x-y
\end{equation}
and is described in the infrared by
\begin{equation}
{C_{\rm FB}(X_{\rm cm},r)\over
 C_{\rm FB}(X_{\rm cm},0)}
=
F_\star(r)
\sim
\exp
\left(
-{|r|\over\xi_{\rm FB}}
\right).
\label{eq:conditional-relative-profile}
\end{equation}
This separation between the collective and relative observables is
illustrated in Fig.~\ref{fig:mean-vs-correlations}.  The ensemble
pair-amplitude is compared with the extended Schr\"odinger-Nagasawa profile,
while the relative pair-density diagnostic resolves the finite-range
structure generated by reciprocal locking.

The two diagnostics therefore probe different statistical levels of the
same stochastic state. Under diagonal matching, the signed ensemble
covariance gives the collective Born density through
$\rho_{\rm BSM}=-C_{\rm FB}(X_{\rm cm},0)$, whereas a
realization-resolved correlation diagnostic retains information about the
finite relative structure. A detector-level measurement map requires an
additional dynamical construction.

This distinction motivates an effective ergodicity-breaking interpretation
of the organized reciprocal sector.  Before reciprocal locking, the
ensemble samples the extended stochastic sector without a persistent
relative structure.  After nucleation, each organized realization
selects one correlated configuration, and the information contained in
this selected configuration cannot be reconstructed from the one-point
ensemble density alone.  The term ``effective loss of ergodicity'' refers
here to this persistence of realization-level information.  The matched
stationary construction imposes
\begin{equation}
-C_{\rm FB}(X_{\rm cm},0)
=
R_n^2(X_{\rm cm})
\end{equation}
while the realization-sensitive information survives in the
off-diagonal dependence of \(C_{\rm FB}\).

Equivalently, reducing the stochastic state to its diagonal,
\begin{equation}
C_{\rm FB}(X_{\rm cm},r)
\longrightarrow
C_{\rm FB}(X_{\rm cm},0),
\end{equation}
removes the relative coordinate and therefore the observable associated
with the localized overlap.  
A Born diagonal and finite relative structure are consequently
compatible at the level of kinematics.  Their simultaneous selection requires a dynamical closure beyond the present kinematic analysis.

\subsection{Mass dependence of the reciprocal length}
\label{sec:relative-mass-scaling}

The mass dependence of the relative size separates into the explicit
kinematic scaling of the stochastic diffusion and the dynamical scaling
of the overlap feedback.

At leading projected order,
\begin{equation}
D_{\rm rel}^{(0)}
=
{\hbar\over m},
\end{equation}
whereas the complete coefficient
\begin{equation}
D_{\rm eff}
=
D_{\rm rel}^{(0)}
+
\Delta D
\end{equation}
contains the correlation dressing.

When this dressing preserves the leading mass scaling of the relative
diffusion,
\begin{equation}
\xi_{\rm FB}(m)
\sim
\left[
{\hbar\over
m\,\mu_{\rm FB}(m)}
\right]^{1/2}.
\label{eq:relative-length-mass-scaling}
\end{equation}

More generally, if the dressed reciprocal coefficient follows
\begin{equation}
\mu_{\rm FB}(m)
\propto
m^\beta,
\end{equation}
then
\begin{equation}
\xi_{\rm FB}(m)
\propto
m^{-(1+\beta)/2}.
\label{eq:relative-size-general-beta}
\end{equation}

The exponent \(\beta\) characterizes the mass dependence generated by
the correlation feedback and remains undetermined at the present
level.  The explicit \(1/m\) dependence of the bare relative diffusion
and the dynamical mass dependence of
\(\mu_{\rm FB}\) are therefore cleanly separated.

\subsection{Standing modes and branch-resolved transport}
\label{sec:branch-transport}

The relative localization and the transport of the reciprocal pair
sector concern distinct coordinates.  The scale
\(\xi_{\rm FB}\) characterizes the internal forward-backward
separation, while the phase structure of the selected
Schr\"odinger-Nagasawa state controls the transport of its collective
spectral branch.

For a real stationary eigenmode of the infinite well,
\begin{equation}
\nabla S_n
=
0
\end{equation}
and the mean probability current vanishes.  The same spectral state has
energy
\begin{equation}
E_n
=
{\hbar^2k_n^2\over2m}
\end{equation}
and therefore carries the momentum scale
\begin{equation}
\sqrt{
\left\langle p^2\right\rangle
}
=
\hbar k_n
\qquad
\left\langle p\right\rangle
=
0.
\label{eq:standing-spectral-momentum}
\end{equation}

The standing eigenmode admits the two branch-resolved spectral momenta
\begin{equation}
p_+
=
+\hbar k_n
\qquad
p_-
=
-\hbar k_n,
\label{eq:spectral-momentum}
\end{equation}
associated with opposite phase gradients
\begin{equation}
\nabla S_\pm
=
p_\pm.
\end{equation}

\begin{figure*}[!t]
\includegraphics[width=\textwidth]
{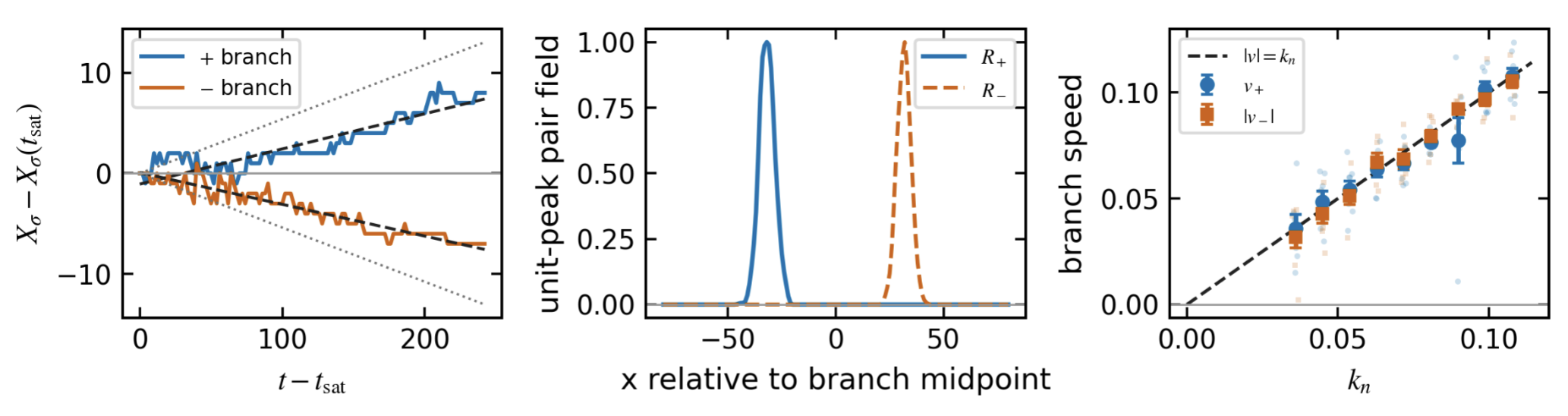}
\caption{
BSM-MC diagnostic of branch-resolved transport. Starting from a
mean-field profile with spectral scale $k_n$, the two propagating
branches cannot be separated when the real standing mode is used
directly; branch selection is forced here by imposing opposite phase
gradients $S_\pm=\pm\hbar k_nx$. Left: center-of-mass evolution of
the two branch populations after the overlap saturation time
\(t_{\rm sat}\), defined as the time at which the numerical
projection gate reaches its plateau; this label describes the
simulation protocol and is not an analytic saturation claim. Dashed lines show the predicted classical drifts
$\pm\hbar k_n t/m$ and dotted lines the residual diffusive envelope.
Middle: unit-peak spatial profiles of the two counter-propagating
branches. Right: measured cluster velocity versus $k_n$ (faint
symbols: individual realizations; solid symbols: ensemble means, with
statistical error bars); the dashed line is the de~Broglie prediction
$|v_\pm|=\hbar k_n/m$, in the reduced units $\hbar=m=1$ of the
simulation.
}
\label{fig:branch-velocity}
\end{figure*}

The corresponding branch velocities are
\begin{equation}
v_\pm
=
{p_\pm\over m}
=
\pm
 {\hbar k_n\over m}.
\label{eq:branch-velocity-prediction}
\end{equation}
The branch-resolved prediction
\eqref{eq:branch-velocity-prediction} is tested numerically in
Fig.~\ref{fig:branch-velocity}, where the two current-carrying sectors
are isolated by imposing opposite phase gradients.

For a branch carrying a local paired density \(\rho_\pm\), the
corresponding drift contribution to the collective current is
\begin{equation}
J_{\rm cm}^{(\pm)}
=
\rho_\pm v_\pm.
\end{equation}

The stationary well state contains the two branches symmetrically.
Their opposite drift currents therefore cancel,
\begin{equation}
J_{\rm cm}^{(+)}
+
J_{\rm cm}^{(-)}
=
0,
\end{equation}
while the internal spectral momentum magnitude remains finite,
\begin{equation}
|p_\pm|
=
\hbar k_n.
\end{equation}

The zero current of the standing eigenmode thus results from the
symmetric superposition of the two counter-propagating spectral
branches.  A nonzero collective drift appears when one branch is
selected, as in free propagation, a local WKB regime, or a Galilean
boost of the reciprocal state.

In the numerical diagnostic, this selection is implemented through
\begin{equation}
S_\pm
=
\pm\hbar k_nx
\end{equation}
and the measured velocities are compared with
\begin{equation}
v_\pm
=
\pm
{\hbar k_n\over m}.
\end{equation}

The relation
\begin{equation}
p_\pm
=
\pm\hbar k_n
\end{equation}
is the spectral momentum decomposition of the stationary
Schr\"odinger-Nagasawa family.  The dynamical relation between this
branch momentum, the candidate reciprocal coefficient
\(\mu_{\rm FB}\), and autonomous propagation of the correlated sector
remains to be determined by the complete coupled dynamics.

The reciprocal pair construction therefore separates three structures
within a single stochastic kernel.  Reciprocal spectral cancellation
selects the collective stationary mode,
\begin{equation}
-C_{\rm FB}(X_{\rm cm},0)
=
R_n^2(X_{\rm cm}),
\end{equation}
which gives, when the full stationary
kernel satisfies diagonal matching, the Born density on the diagonal.  The correlation feedback
generated by the cross covariance and the Bohm/Fisher response acts in
the diagonal-preserving relative sector and introduces the infrared
channel
\begin{equation}
\lambda_{\rm rel}(k)
=
-\mu_{\rm FB}
-
D_{\rm eff}k^2
+
\cdots,
\end{equation}
whose conditional positive finite-range branch carries
\begin{equation}
\xi_{\rm FB}
=
\sqrt{
{D_{\rm eff}\over\mu_{\rm FB}}
}.
\end{equation}
Finally, the phase structure of the selected spectral family resolves
the standing state into the two transport branches
\begin{equation}
v_\pm
=
\pm
{\hbar k_n\over m}.
\end{equation}

The matched Born density, the finite reciprocal range, and branch-resolved transport thus belong respectively to the diagonal, relative, and phase sectors of the same forward-backward stochastic construction.

\section{Discussion}
\label{sec:discussion}

\subsection{Interpretational status}

Throughout this paper the branching ensemble is given a minimal
reading: it is a statistical representation of the possible
continuations allowed by the same local variational structure that
yields the Schr\"odinger equation in the mean.  Its elements need not
be interpreted as a population of ordinary particles evolving in our
single observed world.  The elementary outgoing continuations are
treated as distinct alternatives, but only their first moment is
encoded by the Schr\"odinger-Nagasawa fields: the ensemble itself
carries additional structure -- a genealogy and connected
correlations -- that has no representative at the level of the wave
function.

Read in this light, the construction admits a compact reversed
summary.  In the scaling limit of infinite branching rate, the
forward superprocess samples the set of continuations compatible with
the local dynamics; the backward superprocess, which carries the
terminal normalization, projects this set onto the measurement; and
the Bohm-Fisher rate, built from the local curvature of the ensemble
amplitude, couples the two sectors through an effective interaction
on this weighted space of possibilities.   The first moment reproduces the Schr\"odinger evolution at prescribed
rate, and is held at that reference in the centered reciprocal closure.
The connected second moment describes the organized sector whose
conditional infrared structure is derived here.

A stronger ontological reading is nevertheless possible. One may regard
the elementary continuations as physically realized branches, or as
configurations instantiated in distinct worlds, in the spirit of the
relative-state interpretation
\cite{everett_many-worlds_1957,wallace_emergent_2012}. The analogy should
not be pushed too far: Everettian branches arise from the unitary,
deterministic evolution of the wave function under decoherence, whereas
the branches considered here are realizations of a stochastic branching
hierarchy, carrying a genealogy and connected correlations that have no
counterpart in the relative-state formulation. In that stronger reading,
the connected second moment would characterize statistical correlations
across an ensemble of physically realized branches, rather than merely
correlations between possible histories. We do not rely on this stronger
reading here. The results of the present paper require only the
measure-valued statistical structure of the branching process; whether
the branches are regarded as bookkeeping devices, possible continuations,
or ontologically real continuations is left open.

At prescribed rate, independent forward and backward marginals give
$C_{\rm FB}=0$, while their smooth first moments reconstruct
$\phid\phi=\rho$.  Once the Bohm/Fisher rate responds to
$\rho_\omega=\Phi_B\Phi_F$, the joint dynamics generates the signed
connected kernel
$C_{\rm FB}=\mathbb E_\omega[\psi_F\psi_B]$.  The associated organized
weight is $\rho_{\rm BSM}=-C_{\rm FB}(x,x)$.  Stationary recovery of the
prescribed Born profile is the diagonal condition
$\rho_{\rm BSM}^{\star}=\rho$.  This hierarchy has the structure of a
nonequilibrium statistical mechanics for a branching ensemble.

A final qualification concerns nodes and phase quantization.
As in other real-variable and stochastic representations of quantum
mechanics, the present construction is formulated on nodal domains where
$\rho>0$ and assumes the circulation or single-valuedness conditions
required to reconstruct a global wave function. We therefore do not claim
to solve the Wallstrom objection here~\cite{Wallstrom1994}: the branching
hierarchy is introduced only after the Schr\"odinger-Nagasawa equivalence,
including its boundary and phase conditions, has been imposed. Passing
from a single stochastic trajectory to a branching measure does not by
itself supply the missing topological condition. Accordingly, the results
of the present paper remain conditional on the usual quantum admissibility
conditions. Determining whether branching dynamics can provide an
independent mechanism for phase quantization would require a separate
analysis on multiply connected domains and lies beyond the scope of this
work. Existing proposals within stochastic mechanics are discussed in
Refs.~\cite{derakhshani_2018_thesis,derakhshani_2019_zbwI,
derakhshani_2019_zbwII,kuipers_2023}; for a recent mathematical analysis of the
Madelung-Wallstrom problem, see Ref.~\cite{reddiger_towards_2023}.

\subsection{Scope of the moment-hierarchy result}

Three limitations should be stated explicitly. First, branching Brownian
clustering itself is known. The novelty claimed here is the
 quantum-statistical dictionary relating the prescribed
Schr\"odinger-Nagasawa first moments to branching genealogy, together
with the collective/relative decomposition and the conditional
screening relation of the reciprocal connected sector. Second, the numerical results are diagnostics of
specified effective Monte Carlo closures, not independent simulations
of the full joint stochastic law and not experimental tests of quantum
mechanics.
Third, the relative coefficient \(\mu_{\rm FB}\) is introduced at the
infrared level.  The stationary locking condition fixes
\(\epsilon_\star=1\), while the dynamical approach to this value and the
 sign, magnitude, and evolution of \(\mu_{\rm FB}\) remain open
beyond the present closure. Their autonomous selection is not assumed
to have been established by the conditional screening result.

For reference, Table~\ref{tab:status} summarizes the logical status of
the successive steps and separates the exact results, representation
choices, effective closures, conditional conclusions, and open
dynamical-selection problems.

\begin{table*}[t]
\caption{\label{tab:status}Logical status and principal outputs of the successive steps in the construction.}
\footnotesize
\begin{ruledtabular}
\begin{tabular}{p{0.27\textwidth}p{0.34\textwidth}p{0.29\textwidth}}
Step & Status & Output\\
\hline
Schr\"odinger-Nagasawa transform & exact, under standard regularity and
phase conditions & real forward/backward diffusion pair\\
Branching interpretation & representation choice, operational &
ensemble of possible continuations\\
Forward/backward marginal superprocesses &
exact at frozen prescribed rate, on each nodal domain &
positive genealogical fields with
\(\sqrt{\Phi_a}\) branching noise\\
Superprocess scaling & exact measure-valued limit &
$N,\lambda_N\to\infty$, $\lambda_N/N\to \lambda_0$\\
Pair-source normalization & offspring-law dependent source intensity &
$\ktwo$ sets the amplitude and formation scale of connected pairs\\
Branching Brownian-motion
(BBM) moment hierarchy & exact for local branching; effective once the
Bohm/Fisher rate is frozen on the mean field & connected second moment,
Eq.~\eqref{eq:G-bbm}\\
Confined and free-space clustering & analytic within the frozen-rate
closure & control parameter $a$, three regimes, critical dimension
$d_c=2$\\
One-body uncertainty & standard Fisher-Cram\'er-Rao consequence of the
Born density & ordinary Heisenberg bound\\
Pair dispersion & second-moment observable, not a Hilbert-space variance &
relative spread of connected continuations
under the pair-weighted branching ensemble\\
Joint forward-backward law & specified at the level required by the
connected second-moment hierarchy & admissible block covariance,
\(\mathbb N\succeq0\)\\
Effective reciprocal cross covariance & admissible parametrization of a
dressed joint kernel; not a bare input &
\(\mathcal N_{\rm FB}=-\epsilon q\), \(0\le\epsilon\le1\),
\(q=2\nu_2R\) on the smooth reference, and
\(\epsilon_\star=1\) under stationary $UU$-source suppression\\
Bohm zero-mode compensation & Ward identity of the matched Bohm response &
\(Q[c\rho]=Q[\rho]\); the density-rescaling direction carries no intrinsic gap\\
Relative infrared coefficient &
conditional infrared
parameterization; sign and magnitude are dynamical outputs &
\(\mu_{\rm FB}>0\) is the additional condition for screening\\
Relative overlap sector & exact diagonal subtraction followed by a conditional infrared
Green problem &
\(\xi_{\rm FB}^{2}=D_{\rm eff}/\mu_{\rm FB}\) if
\(\mu_{\rm FB}>0\); bare reference half-kernel
\(\widehat\Gamma_S^{(0)}=\nu_2R\mathbb I\)\\
Branch transport & kinematic, branch-resolved statement &
current-carrying branches drift with $v_B=\nabla S/m$\\
Infinite-well benchmark & parameter-free spectral diagnostic &
genuine eigenmodes extended; spectral branch scale
$p_{\rm spec}=\hbar k_n$\\
 Stationary connected-weight matching &
condition on the signed connected reciprocal kernel &
\(-C_{\rm FB}^\star(x,x)=\rho_{\rm BSM}^\star(x)=\rho(x)\)
and a stationary
diagonal pair equation\\
Dynamical approach to \(\epsilon_\star=1\) and self-consistent evolution
of \(\mu_{\rm FB}\) & open problem beyond the present closure & requires a
full response-field formulation\\
\end{tabular}
\end{ruledtabular}
\end{table*}

The frozen hierarchy is useful because it isolates universal features of
branching correlations. It probes a prescribed one-sector branching dynamics,
in which localization of genealogical correlations is the standard
subcritical branching mechanism. The reciprocal sector introduced in
Sec.~\ref{sec:reciprocal-overlap} is different. It uses a joint forward-backward
second-order law, an anti-correlated reciprocal covariance, and a diagonal-preserving
relative response. The frozen control parameter and the reciprocal gap are
therefore analogous diagnostics, but they should not be identified. In
particular, the present paper does not assert
an identification of
\(\mu_{\rm FB}\) with the one-sector branching rate. The frozen calculation is a benchmark for
branching clustering; \(\mu_{\rm FB}\) instead denotes the local infrared
coefficient of the organized diagonal-preserving overlap response.

The centered pair equation and its diagonal subtraction identify the
relative response sector; its closure is dynamical.  If the full feedback generates
\(\mu_{\rm FB}>0\), its infrared Green problem is screened with the
conditional length given by Eq.~\eqref{eq:xiFB}.  Diffusion then tends to
separate the two continuations, whereas the positive response coefficient
suppresses relative deformations.

When the screened branch is realized, the resulting object is a connected
reciprocal-overlap sector with a finite internal relative length.  This length
characterizes the internal organization of the pair kernel rather than the
localization of a wave function or a de~Broglie wave packet.  The conditional
internal length and the center-of-mass transport have
distinct origins: the latter, when present, is carried by the current-carrying
phase branches of the first moment, as established in
Sec.~\ref{sec:branch-transport}. The self-consistent matching between the
overlap gap, the branch momentum, and the autonomous propagation of the
correlated sector requires the full joint dynamics.

The scope of the present paper is deliberately structural: the  organized connected weight is defined as the negative diagonal of the centered reciprocal
kernel, its pair equation is separated into collective and relative parts,
its stationary matching to the Born density is stated explicitly, and the
screened infrared form of the relative branch is obtained for
\(\mu_{\rm FB}>0\). Turning this
structural result into a falsifiable collapse or measurement model
requires an additional dynamical selection principle and an explicit map
from the pair correlator to detector outcomes.

The selection established in the present paper is spectral and statistical at fixed dynamics, together with the sign selection of the anticorrelated covariance through suppression of the linear realization-level density channel; the autonomous selection of the gap amplitude remains an open problem for the full joint dynamics. The forward-backward construction propagates
the continuations admitted by the prescribed drift, diffusion, boundary
data, and action functional, while recurrently compatible continuations
receive the largest paired weight. At the one-point level, the
Schr\"odinger eigenvalue fixes the reciprocal spectral weighting of the
forward and backward fields. At the connected two-point level, a finite forward-backward overlap range corresponds in the minimal infrared theory to \(\mu_{\rm FB}>0\); the self-consistent magnitude and time dependence of this coefficient are determined by the full joint dynamics.

The role of $\ktwo$ should be read narrowly. In the one-sector
superprocess it fixes the strength of the local genealogical pair source,
hence the amplitude and formation scale of connected correlations; it
does not impose the structural form of the localized reciprocal sector.
The latter uses the reciprocal source normalization
\(\nu_2=\ktwo/2\), with local
background weight \(R\). If the dressed relative
coefficient is positive, its competition with relative diffusion produces
the screened overlap length. Appendix
\ref{app:pair-source-normalization} explains how $\ktwo$ is related to
offspring-law dependent pair production in standard branching
conventions.

The stochastic covariance must likewise be distinguished at three
levels.  The exact bare marginal law is the field-dependent multiplicative
matrix
\begin{equation}
\Gamma_S^{(0)}[\Phi_S]
=
\nu_2
\begin{pmatrix}
\Phi_{S,F}&0\\
0&\Phi_{S,B}
\end{pmatrix}
\end{equation}
of Eq.~\eqref{eq:bare-genealogical-covariance}.  Its evaluation on the
smooth reference is the bare quadratic half-kernel
\begin{equation}
\widehat\Gamma_S^{(0)}
=
\nu_2R\,\mathbb I,
\end{equation}
while the remaining field dependence constitutes the multiplicative
branching vertex.  Neither level contains a bare forward-backward cross
covariance.

The anti-correlated entry parameterized by \(\epsilon\) in
Sec.~\ref{sec:overlap-nucleation} instead represents an admissible local
 form of an effective reciprocal covariance. The departure of its
half-kernel from the reference value is parameterized as
\begin{equation}
\widehat\Gamma_S^{\rm eff}
=
\widehat\Gamma_S^{(0)}
+
\Sigma_C,
\end{equation}
so that a reciprocal cross entry, if dynamically generated, belongs to
the correlation self-energy \(\Sigma_C\) rather than to the bare source.
The transformation of \(\nu_F\) and \(\nu_B\) shows that this distinction
is independent of the reciprocal energy gauge.  The
positive-semidefinite bound, the negative sign favored by suppression of
the linear realization-level channel, the exact overlap budget, and the
conditional screened relation
\(\xi_{\rm FB}^2=D_{\rm eff}/\mu_{\rm FB}\) are consequently statements
about admissibility and response structure; finite-amplitude saturation and the value of \(\mu_{\rm FB}\) belong to the response-field dynamics.

\subsection{Condition for stationary  matching of the
organized connected weight}
\label{sec:discussion-renewal}

 The closure studied here fixes the marginal first moments to the
selected Schr\"odinger-Nagasawa fields and places the reciprocal
 organization in the centered second moment, without excluding changes
to the first moments in a more general interacting dynamics. The relevant object is
\begin{equation}
C_{\rm FB}(x,y,t)
=
\mathbb E_\omega[\psi_F(x,t)\psi_B(y,t)],
\end{equation}
with  organized connected weight
\begin{equation}
\rho_{\rm BSM}(x,t)=-C_{\rm FB}(x,x,t).
\end{equation}
For a target Schr\"odinger state with Born density \(\rho\), stationary
recovery is therefore the pair-level
matching condition
\begin{equation}
-C_{\rm FB}^{\star}(x,x)
=
\rho(x),
\qquad
\mathcal R_{\rm FB}[C_{\rm FB}^{\star}](x,0)
=
0.
\label{eq:stationary-diagonal-matching}
\end{equation}
The first equality fixes the target diagonal; the second is the stationary
diagonal equation obtained from Eq.~\eqref{eq:exact-diagonal-equation}.
Equivalently, the first condition reads
\(\rho_{\rm BSM}^\star=\rho\).

The stationary normalized relative profile \(F_\star\) and the diagonal
condition express complementary requirements.  The complete joint
dynamics must keep both of them fixed.  The nonlinear expectation
\(\mathbb E_\omega[Q[\rho_\omega]\Phi_a]
\neq Q[\mathbb E_\omega\rho_\omega]\mathbb E_\omega[\Phi_a]\)
feeds the connected equations while the centering condition preserves the
selected first moment.  Establishing the stationary matching condition
requires the full response-field dynamics.  In the convention used here,
the exact budget
$\mathbb E_\omega[\rho_\omega]=\rho-\rho_{\rm BSM}$  relates the mean reciprocal co-occupation to the organized connected
weight at fixed first moments.

\subsection{Relation to collapse models, decoherence, and open problems}

The nonlinear term \(\mathcal B_{\rm FB}\) in the centered
pair equation contains the feedback of the fluctuating reciprocal product
on the pair kernel.  It is therefore the natural place for a
density-coupled correction once the joint sector is populated.  The
quantity that must be tested against the Born profile is its contribution
to the diagonal evolution, not a product reconstructed from two separately
averaged marginals.

This structure is reminiscent of density-coupled collapse equations such
as the Ghirardi-Rimini-Weber
(GRW) and continuous spontaneous localization (CSL) models
\cite{Ghirardi1986,Bassi2013} and of
gravity-related proposals \cite{Diosi1989,Penrose1996}. The analogy is
structural only, and the mechanism is conceptually distinct from a
postulated stochastic collapse. No universal collapse noise, collapse
rate, state-reduction law, or detector map is inserted as an external law.
The stochasticity comes from the branching hierarchy, and the candidate
nonlinear correction is generated by the Bohm/Fisher response to its
mixed pair kernel.

Turning this structural mechanism into a quantitative collapse model
would require an autonomous selection law, a calibrated noise covariance,
and an explicit map from the pair correlator to detector outcomes. Such
predictions would then have to be confronted with existing experimental
bounds \cite{Donadi2021}.

The proposed restriction of ergodic exploration should also be distinguished from
environment-induced and intrinsic decoherence
\cite{stamp_environmental_2012,schlosshauer_quantum_2019}, and from
decoherent-histories classicalization in isolated systems
\cite{strasberg_first_2024}. None of these mechanisms is included in the
local pair-source model considered here.

Several problems are left open, in a definite order of importance. The
first is the self-consistent approach of the reciprocal covariance strength
to \(\epsilon_\star=1\) and the dynamical determination of the
diagonal-preserving infrared coefficient \(\mu_{\rm FB}\). The present paper
characterizes the frozen-rate hierarchy and 
shows how a positive reciprocal gap produces a screened relative sector; its sign, magnitude, and dynamics require a self-consistent field-theoretic treatment of the branching hierarchy.

Further open issues include the nonperturbative
diagonal/off-diagonal Bohm/Fisher feedback; the covariance of the
branching noise under a mass-conservation constraint
\cite{BurdzyHolystMarch2000}; the stationary pair-kernel matching condition
\eqref{eq:stationary-diagonal-matching}; and the construction of a
quantitative map from the pair correlator to detector outcomes.
Extracting from this program a quantitative, falsifiable prediction is
the decisive open problem.

\section{Conclusion}

We have studied a branching stochastic representation in which the
Schr\"odinger-Nagasawa diffusion pair forms the first-moment sector of a
branching ensemble of possible continuations.  In the common reciprocal
basis, the centered fields define the signed connected kernel
\(C_{\rm FB}=\mathbb E_\omega[\psi_F\psi_B]\), and the  organized connected weight is
\(\rho_{\rm BSM}=-C_{\rm FB}(x,x)\).  Independent bare genealogies give
\(C_{\rm FB}=0\); the joint Bohm/Fisher dynamics subsequently builds the
anticorrelated sector.  Stationary reciprocal matching requires
\(\rho_{\rm BSM}^{\star}=\rho=\phid\phi\).  Each marginal remains a
positive superprocess with noise proportional to the square root of its
realized branching mass.  Expressing these masses in the common reciprocal
basis makes the construction energy-gauge independent and yields the exact
covariance \(\nu_2R\mathbb I\) plus the multiplicative branching vertex.

Within the frozen-rate hierarchy, the connected pair sector reproduces the
standard clustering structure of branching Brownian motion. In confinement
one obtains the supercritical, critical, and subcritical regimes, while in
free space the marginal behavior inherits the critical dimension
\(d_c=2\). Genuine stationary Schr\"odinger eigenmodes remain on the
extended side of the frozen hierarchy, so the construction does not turn
ordinary quantum stationary states into localized clusters.

We then considered the effect of fluctuations of the Bohm/Fisher term. The
forward-backward construction was specified at the level required by the
connected second-moment hierarchy through an admissible block covariance. Its
effective reciprocal cross-covariance channel carries an anti-correlated
sign, because this is the sign that suppresses the linear
fluctuation of the
realization-level reciprocal product. In
the matched sector, the homogeneity of the Bohm
functional, \(Q[c\rho]=Q[\rho]\), implies a zero-mode compensation: the
one-body Bohm spectral contribution cannot itself be identified with an
intrinsic restoring rate of the matched density-rescaling mode. The
intrinsic local scale instead belongs to the full dressed diagonal-preserving
forward-backward response, in which branching fluctuations and
Bohm/Fisher feedback are coupled.

For nonzero marginal noise power, the stationary suppression condition
$N_{UU}^{\rm eff}\to0$ selects $\epsilon_\star=1$.  This value describes
perfect anticorrelation of the instantaneous sources.  The corresponding
locking of the accumulated fields is characterized independently by
$\alpha\to-1$ and $X_{\rm FB}\to1$.

The centered pair equation separates the BSM diagonal from
the relative profile.  If the dressed relative response has
\(\mu_{\rm FB}>0\), its infrared sector is screened and
Eq.~\eqref{eq:xiFB} gives
\[
    \xi_{\rm FB}^{2}
    =
    \frac{D_{\rm eff}}{\mu_{\rm FB}} .
\]
 Within the infrared closure, Eq.~\eqref{eq:xiFB} follows from the
small-$k$ response for \(\mu_{\rm FB}>0\). The sign and magnitude of
this effective coefficient are not selected by that relation.
This result concerns the screened reciprocal correlation sector. The
momentum scale \(\hbar k_n\) enters through the branch decomposition of the
first-moment spectral family: a real stationary well eigenmode has zero
net current because two counter-propagating branches are symmetrically
occupied.

The approach to \(\epsilon_\star=1\), the self-consistent value and scaling
of \(\mu_{\rm FB}\), and the dynamical matching between relative
overlap, branch transport, and measurement outcomes remain open. So does the verification of
the stationary pair-kernel condition
\eqref{eq:stationary-diagonal-matching}, under which the  organized connected weight $\rho_{\rm BSM}$ recovers the prescribed Born profile.  These questions call for a self-consistent response-field treatment
with full propagators. The present paper establishes the moment-hierarchy
dictionary, the admissible reciprocal second-moment channel, and the
conditional infrared form of a screened relative sector. These results
provide a basis for investigating whether the resulting correlation
mechanism, supplemented by self-consistent gap dynamics and an explicit
detector coupling, can support a quantitative account of quantum
measurement and collapse.

\section{Acknowledgments}

E.D. conceived the project, designed the theoretical framework, and supervised all stages of the work. B.B. made central contributions to the closure of the hierarchy, the physical interpretation of the framework and the numerical aspects. A.L. and L.T. participated in extensive discussions that shaped the direction of this work throughout. C.D. contributed to the BSM-MC numerical implementations. The authors are grateful to Thibaut Pellerin for his contributions to the mathematics of superprocesses during an earlier stage of this project and also warmly thank Alain Mazzolo and Cheikh Diop for their valuable advice and insightful discussions respectively on stochastic processes and reactor physics in connection with this work.

\appendix
\section{Schr\"odinger-Nagasawa equivalence}
\label{app:nagasawa}

Starting from $\psi=R\exp(\ii S/\hbar)$, the Schr\"odinger equation gives
\begin{align}
\partial_t\rho
+\nabla\!\cdot\!\Big(\rho\,\frac{\nabla S}{m}\Big)&=0,
\label{eq:continuity-app}\\
\partial_tS+\frac{(\nabla S)^2}{2m}+V+Q&=0 .
\label{eq:HJ-app}
\end{align}
For $\phi=R\,\ee^{-S/\hbar}$ one computes
\begin{align}
\partial_t\phi&=\ee^{-S/\hbar}
\Big(\partial_tR-\frac{R}{\hbar}\partial_tS\Big),\\
\nabla^2\phi&=\ee^{-S/\hbar}
\Big[\nabla^2R-\frac{2}{\hbar}\nabla R\cdot\nabla S
-\frac{R}{\hbar}\nabla^2S
+\frac{R}{\hbar^2}(\nabla S)^2\Big] .
\end{align}
Using Eqs.~\eqref{eq:continuity-app} and \eqref{eq:HJ-app}, the terms
reorganize into Eq.~\eqref{eq:forward}; the calculation for
$\phid=R\,\ee^{S/\hbar}$ gives Eq.~\eqref{eq:backward}. Conversely,
multiplying the two real equations by the inverse transformations recovers
the continuity and Hamilton-Jacobi equations, and therefore the
Schr\"odinger equation up to the usual single-valuedness conditions on
$S$ \cite{Wallstrom1994}.

\section{Pair-source normalization}
\label{app:pair-source-normalization}

In a standard branching convention, if $K$ descendants are produced at a
local event, the first offspring moment is
\begin{equation}
\nu_1=\mathbb E[K],
\end{equation}
whereas the source of distinct ordered correlated pairs is controlled by
the second factorial moment
\begin{equation}
\nu_{2,f}=\mathbb E[K(K-1)] .
\end{equation}
This moment is offspring-law dependent and is not fixed by $\nu_1$ in
general. For example, a binary death/splitting process may give
$\nu_{2,f}=\nu_1$ only as a special case of its event convention; a
deterministic offspring number gives
$\nu_{2,f}=\nu_1^2-\nu_1$; and a Poisson offspring number gives
$\nu_{2,f}=\nu_1^2$.

The paper therefore uses the dimensionful coefficient $\ktwo$ in the
source term $\ktwo n$ of Eq.~\eqref{eq:G-bbm}. In the measure-valued
scaling of Sec.~\ref{sec:bbm}, it is the finite limiting combination
\begin{equation}
\frac{\lambda_N}{N}\nu_{2,f,N}
\longrightarrow
\ktwo ,
\end{equation}
to leading order in the near-critical limit. Thus $\ktwo$ retains the
microscopic offspring-law dependence of pair production after the
elementary mass $1/N$ and the diverging event rate have been combined
in the superprocess limit. Its precise normalization depends on the
replacement convention and on the treatment of self-correlations. A
minimal Poisson source-block convention can also be used to fix this
normalization; this is a convention for the effective source block, not
a generic identity between offspring moments.

\section{Connected pair equation in a reflecting interval}
\label{app:bbm}

In this appendix \(\beta\) denotes a generic constant
one-sector rate; the quantum specialization is obtained by setting it
to the fixed-gauge constant projection of \(\beta_{\rm q}\).

For test functions \(f\) and \(g\), define
\[
C_t(f,g)
=
\operatorname{Cov}_{\omega}
\bigl(
\langle\varrho_t,f\rangle,
\langle\varrho_t,g\rangle
\bigr).
\]
Applying It\^o's product rule to the martingale problem
\eqref{eq:superprocess-mart}, and using the polarized covariation
associated with Eq.~\eqref{eq:superprocess-qv},
and writing
\(n_t(\dd x)\equiv n(x,t)\dd x\) for the mean measure, one obtains
\begin{equation}
\frac{\dd}{\dd t}C_t(f,g)
=
C_t(\Lop f,g)
+
C_t(f,\Lop g)
+
\ktwo\langle n_t,fg\rangle .
\label{eq:app-weak-pair}
\end{equation}
When the covariance admits a density and
\(\Lop=\Dq\nabla^2+\beta\) is self-adjoint under the prescribed boundary
conditions, Eq.~\eqref{eq:app-weak-pair} is precisely
Eq.~\eqref{eq:G-bbm}.

Let \(P_{t,s}(x|z)\) be the one-body propagator generated by \(\Lop\).
For \(G_{\rm br}(x,y,0)=0\), the Green-function solution is
\begin{equation}
G_{\rm br}(x,y,t)
=
\ktwo
\int_0^t\dd s\int\dd z\,
P_{t,s}(x|z)
P_{t,s}(y|z)
n(z,s).
\label{eq:Gbr-conv}
\end{equation}
Thus, a pair observed at \((x,y)\) at time \(t\) descends from a
branching event occurring at position \(z\) and time \(s\). Notice that
the general source contains the evolved mean density \(n(z,s)\), rather
than the initial density \(n_0(z)\).

For a constant rate \(\beta\) in the reflecting interval \([-L,L]\),
the one-body propagator is
\[
P_{t,s}(x|z)
=
\ee^{\beta(t-s)}
G_D(x,t-s|z).
\]
If the initial mean density is uniform, \(n_0(x)=c_0\), then
\(n(x,s)=c_0\ee^{\beta s}\). Setting \(\tau=t-s\), using the semigroup
property and the mode expansion of Eq.~\eqref{eq:green-reflect}, define
\begin{equation}
I_n(t)
=
\int_0^t\dd\tau\,
\ee^{(\beta-2n^2/\tau_D)\tau},
\qquad n\geq0.
\label{eq:app-In}
\end{equation}
The uniform-density solution can then be written as
\begin{align}
G_{\rm br}(x,y,t)
&=
\ktwo c_0\ee^{\beta t}
\Bigg[
\frac{I_0(t)}{2L}
\nonumber\\
&\quad+
\frac{1}{L}
\sum_{n\geq1}
c_n(x)c_n(y)I_n(t)
\Bigg].
\label{eq:Gbr-modes}
\end{align}
The factor \(2\) in the nonuniform-mode decay contained in \(I_n\)
comes from the propagation of the two descendants away from their
common branching event. For a nonnegative local source and vanishing
initial covariance, Eq.~\eqref{eq:Gbr-conv} also shows that
\(G_{\rm br}(x,y,t)\geq0\), so that it can be used directly as the pair
weight in Eq.~\eqref{eq:r2def}.

Substituting Eq.~\eqref{eq:Gbr-modes} into Eq.~\eqref{eq:r2def} and
performing the spatial integrals gives
\begin{equation}
\mean{r^2}_{x,y}(t)
=
\frac{2L^2}{3}
-
\frac{64L^2}{\pi^4}
\frac{
\displaystyle
\sum_{\substack{n\geq1\\n\ {\rm odd}}}
I_n(t)/n^4
}{
I_0(t)
}.
\label{eq:app-r2-finite}
\end{equation}
For \(\beta<0\), setting \(a=-\beta\tau_D>0\) and taking
\(t\to\infty\), the standard odd-mode sums yield
Eq.~\eqref{eq:r2-bbm}. For \(\beta=0\) and \(\beta>0\), the nonuniform
modes become negligible relative to \(I_0(t)\), yielding respectively
Eqs.~\eqref{eq:r2crit} and \eqref{eq:r2sup}.

\section{Velocity moments and the Bohm decomposition}
\label{app:velocity}

For $\psi=R\,\ee^{\ii S/\hbar}$ with reflecting (or vanishing-current)
boundary conditions, the first velocity moment is the mean Bohm velocity,
$\mean{v}=\int\rho\,(\partial_xS/m)\,\dd x=\mean{v_B}$, the boundary term
$\propto\int\partial_xR^2\,\dd x$ vanishing. The second moment splits as
\begin{equation}
\mean{v^2}=\frac{\hbar^2}{m^2}\!\int(\partial_xR)^2\dd x
+\int R^2\Big(\frac{\partial_xS}{m}\Big)^2\dd x
=\mean{u^2}+\mean{v_B^2},
\label{eq:v2split}
\end{equation}
with $u=(\hbar/m)\,\partial_xR/R$ the osmotic velocity; an integration by
parts identifies $\mean{u^2}=(2/m)\mean{Q}$, whence
Eq.~\eqref{eq:sigmav}.

\section{Free-space Gaussian quadratures}
\label{app:free}

For the localized initial condition the kernel
$\int\dd z\,G_D(x,t'|z)G_D(y,t'|z)G_D(z,t-t'|0)$ is a product of Gaussians
whose quadrature gives, with $u=t'/t$,
\begin{align}
G_{\rm br}(x,y,t)
&\propto
t^{1-d}
\int_0^1\dd u\,
\frac{\ee^{-tu/\bar\tau_c}}{[u(2-u)]^{d/2}}
\nonumber\\
&\quad\times
\exp\!\Bigg[
-\frac{(x-y)^2}{4\Dq t\,u(2-u)}
-\frac{2\,x\!\cdot\!y}{4\Dq t\,(2-u)}
\Bigg].
\label{eq:Gbr-free-localized}
\end{align}
The moments follow from the two Gaussian quadratures in
relative/center-of-mass variables,
$I_1(u)=4d\Dq tu\,(8\pi\Dq tu)^{d/2}(2\pi\Dq t(2-u))^{d/2}$ and
$I_2(u)=(8\pi\Dq tu)^{d/2}(2\pi\Dq t(2-u))^{d/2}$, yielding
Eq.~\eqref{eq:r2free} for both uniform and localized initial data. The
asymptotic
one-sector branching profile, with decay scale
\(\ell_{\rm br}=\sqrt{2\Dq\bar\tau_c}\), is the saddle-point (large-$r$)
evaluation of Eq.~\eqref{eq:gfree}, equivalent to the modified-Bessel
representation
\(g_{\rm br}(r)\propto
r^{1-d/2}K_{d/2-1}(r/\ell_{\rm br})\), where \(K_\nu\) is the
modified Bessel function of the second kind.

\section{Common-filtration convention for the reciprocal martingales}
\label{app:common-filtration}

The backward Schr\"odinger-Nagasawa field is specified by terminal
conditioning.  To use It\^o product rules for equal-time forward-backward
correlators, we work on a finite interval \([0,T]\) and introduce the reversed
representative
\begin{equation}
\check\Phi_B(s)=\Phi_B(T-s),
\qquad
s=T-t .
\end{equation}
The pair \((\Phi_F(t),\check\Phi_B(s))\) is then represented on a common
increasing filtration for the purpose of defining quadratic covariations.
For test functions \(f,g\), let \(M_F^f\) and
\(\check M_B^g\) be the martingales obtained by testing the forward and
reversed-backward measures against \(f\) and \(g\), respectively.
The second-order joint law is specified by
\begin{equation}
d\langle M_F^f,\check M_B^g\rangle
=
\langle f,\mathcal N_{\rm FB}g\rangle\,dt .
\end{equation}
The last bracket denotes the kernel pairing
\(\int\dd x\,\dd y\,f(x)\mathcal N_{\rm FB}(x,y,t)g(y)\).
After this covariation is formed, the result is relabelled at the common
physical time \(t\).  This convention is what is used in
Eq.~\eqref{eq:pair-equation-compact}.  It fixes the equal-time
second-moment budget, but it does not claim to determine all higher
finite-dimensional distributions of a non-Gaussian joint process.

\section{Linear admissibility of the anti-correlated projected branch}
\label{app:anti-correlated}

The reciprocal structure and the
linearized projected dynamics provide two admissibility arguments for a
negative projected branch; they do not determine the nonlinear covariance
strength. Introducing
\begin{equation}
\psi_{\rm s}
=
\frac{\psi_F+\psi_B}{2},
\qquad
\psi_{\rm a}
=
\frac{\psi_F-\psi_B}{2},
\end{equation}
the realization-level pair-density fluctuation reads
\begin{equation}
\delta\rho_\omega
=
2R\psi_{\rm s}
+
\psi_{\rm s}^{2}
-
\psi_{\rm a}^{2}.
\label{eq:density-symmetric-antisymmetric}
\end{equation}
The symmetric branch $\psi_B=\psi_F$ modifies the realization-level
reciprocal product at linear order.  The anticorrelated branch
$\psi_B=-\psi_F$ eliminates this linear density channel and leaves the
bilinear contribution.  Its ensemble average is the negative connected
kernel whose signed diagonal defines \(\rho_{\rm BSM}\).

The latter is also the infinitesimal direction of the reciprocal
transformation
\begin{equation}
\phi\longrightarrow \ee^\chi\phi,
\qquad
\phi^\dagger\longrightarrow \ee^{-\chi}\phi^\dagger,
\end{equation}
which leaves \(\rho=\phi^\dagger\phi\) invariant. Linearization gives
\begin{equation}
\delta\phi=\chi\phi,
\qquad
\delta\phi^\dagger=-\chi\phi^\dagger,
\end{equation}
and hence, in the regular fluctuation variables,
\begin{equation}
\psi_B=-\psi_F.
\end{equation}
Stochastic increments restricted to this reciprocal channel satisfy
\(\dd\psi_B=-\dd\psi_F\) and therefore have a negative cross
covariance.

The same sign is also the linearly well-posed choice
within the projected equation used for the diagnostic. For the stationary-well reference, $Q_0=E_n$ is the constant Bohm potential on each nodal domain. Freezing $\alpha$ on the fluctuation time scale gives
\begin{equation}
\partial_t\psi
=
-\alpha\Dq\nabla^2\psi
-
\alpha\frac{Q_0}{\hbar}\psi.
\label{eq:linear-projected-stability}
\end{equation}
For a Fourier mode
\begin{equation}
\psi(x,t)
=
\psi_k
\exp\left[
\lambda_\alpha(k)t+\ii kx
\right],
\end{equation}
where \(k\) is the wave number and \(\psi_k\) its amplitude,
the growth rate is
\begin{equation}
\lambda_\alpha(k)
=
\alpha
\left(
\Dq k^2-\frac{Q_0}{\hbar}
\right).
\label{eq:projected-dispersion}
\end{equation}

For the symmetric branch, $\alpha\rightarrow+1$, one has
\begin{equation}
\lambda_{\rm s}(k)
\rightarrow
\Dq k^2-\frac{Q_0}{\hbar},
\end{equation}
which is unbounded from above as $k\to\infty$. This branch is therefore
anti-diffusive and ultraviolet unstable. This instability concerns the
effective forward equation for $\psi$ itself, not the intrinsic
backward-time propagation of $\phi^\dagger$, which plays no role once
the \((\psi_F,\psi_B)\) decomposition of
Eq.~\eqref{eq:regular-fields} is performed. By contrast, for the
anti-correlated branch, $\alpha\rightarrow-1$,
\begin{equation}
\lambda_{\rm a}(k)
=
\frac{Q_0}{\hbar}
-
\Dq k^2,
\end{equation}
so that short wavelengths are damped and any instability is restricted
to the finite band $k^2<k_n^2$. In the stationary-well benchmark, this is
the long-wavelength band below the spectral scale of the reference mode.

The well-posedness of the projected continuum equation thus requires
$\alpha<0$. 
This linear argument excludes the positive projected branch within this effective equation; the degree of projection, \(|\alpha|\), \(\epsilon\), and the nonlinear saturation state are dynamical quantities beyond its scope.

\section{Monte Carlo protocol}
\label{app:mc-protocol}
\begin{table*}[t]
\caption{\label{tab:numerical-parameters}Numerical parameters reported for
each BSM-MC figure. The table records the role of each class of parameter;
numerical values are run dependent and accompany the released figure scripts.}
\footnotesize
\begin{ruledtabular}
\begin{tabular}{p{0.22\textwidth}p{0.33\textwidth}p{0.35\textwidth}}
Class & Examples & Role\\
\hline
Resolution and sampling & lattice size, grid
spacing, time step, sector populations, number of realizations &
Controls finite-population and discretization
effects\\
Projected diffusion gate & move probability, relaxation time, hopping
bias &
Numerical relative-sector transport, not bare
\(\Dq\)\\
Pair source and locking & %
pair-locking probability, attempted replacements, noise
amplitude, shared-birth rate & Implements the effective projected
pair-source closure\\
Bohm regularization & support smoothing, Fourier cutoff, active mask,
denominator floor, clipping & Controls ultraviolet noise in the lattice Bohm
signal\\
\end{tabular}
\end{ruledtabular}
\end{table*}

All projected-closure BSM-MC figures are generated from the
same forward-backward Monte Carlo code on the periodic lattice
specified below. The frozen-rate reflecting-interval benchmark of
Fig.~\ref{fig:a-transition} is separate: it uses the one-sector
reflecting walk with the reciprocal locking and Bohm-feedback gates
disabled. The projected simulation is a constant-population particle representation of the effective projected closure, used for diagnostics; it is neither a direct finite-difference solver for the continuum stochastic equation nor a direct implementation of the admissible reciprocal covariance of Sec.~\ref{sec:overlap-nucleation}.  At
each time step the code stores the two
integer occupation fields on a periodic lattice of
\(N_x\) sites,
\begin{equation}
F_i(t),\qquad B_i(t),\qquad
i=0,\ldots,N_x-1,
\end{equation}
with fixed populations
\begin{equation}
\sum_i F_i=N_1,\qquad \sum_i B_i=N_2 .
\end{equation}
The diagnostic forward and backward fluctuation fields are
\begin{align}
\psi_{F,i}(t)
&=
\frac{F_i(t)}{N_1}-\frac{1}{N_x},\\
\psi_{B,i}(t)
&=
\frac{B_i(t)}{N_2}-\frac{1}{N_x}.
\end{align}
The backward field is represented numerically by an independent positive
forward-walker population. This positive representation is only a Monte
Carlo device. The reciprocal sign of the backward sector is inserted in
the feedback variable, so that the physical attractive feedback
corresponds to the sign used in the continuum closure.

The numerical status is therefore narrower than the analytic construction.
The continuous second-moment law admits an anti-correlated overlap channel
through the effective cross covariance of
Eq.~\eqref{eq:local-reciprocal-covariance}.  The present BSM-MC code instead uses positive populations and shared
positive births, and then inserts the reciprocal sign in the projected
diagnostic variable.  The figures should consequently be read as tests and
visualizations of the specified late projected closure, not as an independent
demonstration that the continuous joint law dynamically selects \(\epsilon\) or
\(\mu_{\rm FB}\).

The diffusion scales are kept conceptually distinct.  The bare
Schr\"odinger-Nagasawa first-moment diffusion is
\(\Dq=\hbar/(2m)\).
The projected relative response has leading coefficient
\(D_{\rm rel}^{(0)}=2\Dq=\hbar/m\) on the anti-correlated branch, whereas
the screened shape is governed by the dressed coefficient \(D_{\rm eff}\)
defined by the \(k^2\) term of Eq.~\eqref{eq:relative-response}.
The correlation-dependent move probability below is a numerical gate for
the projected relative sector; it should not be read as turning on the
bare Nagasawa diffusion.

\emph{Order parameter and pair field.}
The scalar correlation variable is computed at every time step from the
overlap of the two background-subtracted sectors:
\begin{equation}
\alpha_{\rm raw}(t)
=
\frac{\sum_i\psi_{F,i}(t)\psi_{B,i}(t)}
     {\sum_i\psi_{F,i}^{2}(t)} .
\end{equation}
The positive feedback gate is the clipped and baseline-subtracted quantity
\begin{equation}
g_\alpha(t)
=
\left[
\frac{\alpha_{\rm raw}(t)-\alpha_{\rm raw}(0)}
{1-\alpha_{\rm raw}(0)}
\right]_{0}^{1},
\end{equation}
where $[\cdot]_0^1$ denotes clipping to the interval $[0,1]$. The signed
coefficient corresponding to the continuum convention is
\begin{equation}
\alpha_{\rm phys}(t)=-g_\alpha(t).
\end{equation}
Thus the magnitude of the feedback is measured from the instantaneous
overlap, while the anti-correlated backward sign is applied explicitly.

The realization-wise closure error is computed from the same numerical
fields as
\begin{equation}
\varepsilon_{{\rm cl},\omega}(t)
=
1-
\frac{
\left[
\sum_i\psi_{F,i}(t)\psi_{B,i}(t)
\right]^2
}{
\left[
\sum_i\psi_{F,i}^{2}(t)
\right]
\left[
\sum_i\psi_{B,i}^{2}(t)
\right]
}.
\end{equation}
It is the discrete realization-wise counterpart of
the residual projection diagnostic
in Eq.~\eqref{eq:residual-overlap-fraction}.

The compatible pair density used for the Bohm feedback is
\begin{equation}
\rho_{{\rm pair},i}(t)=F_i(t)B_i(t),\qquad
R_i(t)=\sqrt{\rho_{{\rm pair},i}(t)} .
\end{equation}
This is a realization-level lattice product.  Its pointwise
ensemble average estimates the residual reciprocal product
\(\mathbb E_\omega[\rho_\omega]\).  The numerical counterpart of the BSM
paired density is instead the signed centered overlap
\(-\mathbb E_\omega[\psi_{F,i}\psi_{B,i}]\), consistently with
\(\rho_{\rm BSM}=-C_{\rm FB}(x,x)\).  In the projected BSM-MC closure, the positive product $F_iB_i$ is the auxiliary local support used by the Bohm feedback, whereas the  organized connected weight is the signed centered quantity. At exact diagonal matching, Eq.~\eqref{eq:averaged-product-budget} gives $\mathbb E_\omega[\rho_\omega]=0$, so the terminal matching state is carried by the connected anticorrelated sector rather than by positive reciprocal co-occupation.  The prescribed Born profile is
denoted by \(\rho\).
This choice makes the Bohm feedback sensitive to locations where both
sectors occupy the same region, not to a single-sector density alone.  A principal-cluster-mass diagnostic, defined as the mass of the
largest connected component of this pair density after mild
correlation smoothing, is also recorded by the code; it is not shown
in the figures of the present paper.  The signed overlap plots use
$\sum_i\psi_{F,i}\psi_{B,i}$, whereas the cluster mass uses the positive
pair support. The shared-fluctuation intensity $I_s(t)$ reported in the overlap
diagnostics is the integrated magnitude of the signed shared component
of the two background-subtracted sectors (the shaded area in the
corresponding figure), normalized by its saturation value $I_s^{f}$,
so that $I_s/I_s^{f}\to1$ in the locked regime.
For the left panel of
Fig.~\ref{fig:mean-vs-correlations}, the displayed positive
pair-amplitude estimator is
\begin{equation}
\widehat R_{\rm pair}(x_i,t)
=
\frac{\mathbb E_\omega[\sqrt{F_i(t)B_i(t)}]}
{\left\|\mathbb E_\omega[\sqrt{F(t)B(t)}]\right\|_2},
\label{eq:numerical-pair-amplitude-estimator}
\end{equation}
where \(\|\cdot\|_2\) is the lattice \(L^2\) norm. It is a normalized
one-point estimator of the positive realization-level pair amplitude.
It monitors the collective reference profile, while the signed centered
overlap monitors the organized BSM sector.

When displayed as a function of time, the ensemble-averaged diagnostic
\(\mathbb E_\omega[|\alpha_{\rm phys}(t)|]\) is compared with an Avrami
(JMAK-type) sigmoidal fit,
\begin{equation}
|\alpha_{\rm phys}|_{\rm fit}(t)
=
|\alpha_{\rm phys}|_\infty
\left\{
1-\exp[-(t/t_0)^n]
\right\},
\end{equation}
where \(|\alpha_{\rm phys}|_\infty\) is the fitted
plateau, \(t_0\) the characteristic formation time, and \(n\) the
Avrami exponent,
used as a phenomenological guide to the transient shape of the projected numerical diagnostic.

\emph{Bohm feedback.}
At each step the code evaluates a regularized lattice Bohm signal from
the pair amplitude,
\begin{equation}
Q_i
=
-\,\frac{\Delta R_{q,i}}{R_{q,i}+Q_{\rm floor}},
\end{equation}
where $\Delta$ is the periodic lattice Laplacian and $R_q$ is a weakly
regularized version of $R$, and \(Q_{\rm floor}>0\) is a
small denominator floor.  The regularization has three roles only:
avoiding division by empty sites, removing unsupported isolated spikes,
and suppressing ultraviolet lattice noise produced by differentiating a
particle histogram.  Concretely, the code uses support smoothing, a
spectral low-pass filter, an active-support mask, and a symmetric clip of
$Q$.  No deterministic target profile is inserted.

The low Fourier modes of $Q R_q$ are also projected onto the pair
amplitude.  This gives a diagnostic expansion
\begin{equation}
Q R_q \simeq a_0 R_q + a_2 \Delta R_q + a_4 \Delta^2 R_q ,
\end{equation}
where \(a_0,a_2,a_4\) are the fitted projection
coefficients, from which the code records a massive part $a_0$, an effective diffusive
part, and a coherence measure.  In the production run used for the
article, this projection is used primarily as a diagnostic of the Bohm
closure.  The physical diffusion probability is controlled separately by
the measured correlation gate described below.

\emph{Diffusion controlled by the measured correlation.}
The walkers diffuse on the periodic lattice by a nearest-neighbour move.
The total move probability is a dynamical function of the measured
correlation and of the pair activity:
\begin{equation}
p_{\rm move}(t)
=
p_{\rm move}\!\left[g_\alpha(t),{\cal A}(t),Q(t)\right],
\end{equation}
Here \({\cal A}(t)\) is the scalar pair-activity
diagnostic and \(Q(t)\) is the scalar summary of the regularized
lattice Bohm signal used by the move gate.
The gate is updated with an exponential
moving relaxation in time.  Let \(p_R(t)\) and \(p_L(t)\)
denote the right- and left-move probabilities. For the unbiased runs,
\begin{equation}
p_R(t)=p_L(t)=\frac{p_{\rm move}(t)}{2}.
\end{equation}
Thus when the projected sectors are uncorrelated the numerical relative-motion
gate is almost closed, while in the late correlated part of the run the move probabilities
approach $p_R\simeq p_L\simeq 1/2$.  This implements the intended ordering
of the effective projected closure: the relative shape dynamics is activated
with forward-backward correlation.  It does not state that the bare
Schr\"odinger-Nagasawa diffusion
\(\Dq\) is absent at the first-moment level.

For drift diagnostics the same rule is kept, but the left/right split is
biased by a parameter \(p\in[0,1]\):
\begin{equation}
p_R(t)=p\,p_{\rm move}(t),\qquad
p_L(t)=(1-p)\,p_{\rm move}(t).
\end{equation}
The laboratory profile is measured in the fixed lattice frame.  The
comoving profile is obtained by centering each realization on the circular
center of the instantaneous pair cluster before averaging.  This
separates broadening caused by motion of the cluster center from the
intrinsic width of the correlated pair cloud.

\emph{Branching, population control, and noise.}
After diffusion, each sector is resampled at fixed population. This
Monte Carlo population-control step is a numerical device used to limit
population fluctuations and should not be confused either with physical
reactor-power control or with the discretization parameter $N$ of the
analytic superprocess limit.
The
selection probability is a normalized local fitness built from the pair
density, the Bohm signal, the current branching strength, and the
correlation gate.  The population control subtracts only the spatially
uniform part needed to keep $N_1$ and $N_2$ fixed; it does not impose the
position of the cluster.

The correlated pair source is implemented by drawing shared births from a
Poisson process.  If $P_i(t)$ is the normalized pair-selection
probability, the number of shared correlated births at site $i$ is drawn
as
\begin{equation}
n^{\rm corr}_i(t)
\sim
{\rm Poisson}\!\left[
\sigma_\eta(t)^2\,N_{\rm rep}(t)\,p_{\rm pair}(t)\,P_i(t)
\right],
\end{equation}
where $N_{\rm rep}$ is the attempted replacement number and $p_{\rm pair}$
is the pair-locking gate.  The noise amplitude follows the factor used in
the continuum equation,
\begin{equation}
\sigma_\eta(t)=\min\{\exp(E t/\hbar),\sigma_{\rm max}\}.
\end{equation}
Here \(E\) is the imposed spectral-energy scale and
\(\sigma_{\rm max}\) is the numerical cap on the noise amplitude.
This is the discrete origin of
the exponentially rescaled stochastic source in the common
reciprocal basis.  The law of correlated draws can accordingly be verified as a direct
histogram of the integer Poisson births before the numerical plateau; this
diagnostic is recorded by the code but not shown here.

\emph{Separation of the linear and quadratic channels.}
For comparison with the continuum equation, the code also performs a
local regression of the measured time derivative of $\psi$ onto the four
operators
\begin{equation}
\alpha_{\rm phys}\,\Delta\psi,\qquad
\alpha_{\rm phys}\,\Delta\!\left(\frac{\psi^2}{R}\right),\qquad
\alpha_{\rm phys}\,\psi,\qquad
\alpha_{\rm phys}\,\frac{\psi^2}{R}.
\end{equation}
This diagnostic is important because the first and third terms
renormalize the linear diffusion and mass, whereas the second and fourth
terms 
monitor nonlinear derivative and local contributions separately.
They are not combined into a single effective coefficient in the diagnostic plots; their scaling analysis lies beyond the present diagnostic closure.

\emph{Two-point correlation.}
The correlation figure is computed from the pair density itself.
Let \(\rho_{\rm pair}(x,t)\) denote the spatial
interpolation of the lattice pair density
\(\rho_{{\rm pair},i}(t)=F_i(t)B_i(t)\);

This field represents one realization of the reciprocal
product.  Its ensemble average estimates
\(\mathbb E_\omega[\rho_\omega]\), while the signed centered covariance
estimates \(C_{\rm FB}\).  Stationary diagonal matching is tested through
\(-C_{\rm FB}(x,x)=\rho\).
In
center and relative coordinates,
\begin{equation}
X_{\rm cm}
=\frac{x+y}{2},\qquad r=x-y,
\end{equation}
the plotted one-dimensional cut fixes the barycenter at the cluster
center and evaluates
\begin{equation}
\begin{aligned}
G_{\rho\rho}(r)
&\propto
\,
\mathbb E_\omega\!\Bigl[\\[-2pt]
&\quad
\rho_{\rm pair}\!\left(
X_{\rm cm}+r/2,t
\right)
\rho_{\rm pair}\!\left(
X_{\rm cm}-r/2,t
\right)
\Bigr] .
\end{aligned}
\end{equation}
The right panel of Fig.~\ref{fig:mean-vs-correlations} shows the
uncentered estimator \(G_{\rho\rho}\), which is distinct from
\(G_\psi\) and \(\Gamma_{\rm FB}\). Consequently, the correlation length extracted from $G_{\rho\rho}$ can
differ from the apparent width of the raw one-sector field $\psi(x)$: the
former measures correlated pair support, whereas the latter also contains
single-sector wandering and residual background fluctuations.

\emph{Velocity and branch diagnostics.}
The velocity tests use two distinct protocols.  In the biased-diffusion
diagnostic a left/right hopping bias is imposed and the lab-frame and
comoving cluster widths are compared.  In the phase-branch diagnostic no
constant drift velocity is inserted directly.  Instead the initial
density is a cosine mode and the two branches are guided by opposite
phases
\begin{equation}
S_\pm(x)/\hbar=\pm k_n x .
\end{equation}
The code then measures the cluster-center velocity and the velocity
deduced from the Fourier phase drift. Writing
\(\widehat{\delta\rho}_{FB}(k,t)\) for the spatial Fourier coefficient
at wave number \(k\) of the forward-backward pair-density deviation,
\begin{equation}
v_{\rm phase}
=
-\frac{1}{k_n}\frac{d}{dt}
\arg \widehat{\delta\rho}_{FB}(k_n,t).
\end{equation}
The plotted test is whether the surviving branch speed follows the
de~Broglie scaling $|v|\simeq k_n$ in the reduced units of the simulation.
This diagnostic is used only to check the propagation of selected
branches; it is not used to infer the autonomous propagation of the
correlated sector, which remains an open problem
(Sec.~\ref{sec:branch-transport}).

\emph{Numerical status and controls.}
The numerical material should be separated into two categories. First, a
direct test of the early reciprocal covariance can be performed by
implementing stochastic increments that realize the local covariance of
Eq.~\eqref{eq:local-reciprocal-covariance} and verifying the short-time
budget \(dP/dt\simeq N_{\rm FB}\) as \(\epsilon\) is varied. This test does not require
the late projected Bohm gate. Second, the figures in the present manuscript
belong to the effective projected-closure category: they explore the behavior
of the late reciprocal sector once the overlap gate, pair source, population
control, smoothing, and regularized Bohm signal have been specified.

For this second category, the quantitative validation program comprises: variation of population size, grid spacing,
time step, Bohm smoothing length, spectral cutoff, active-support mask,
regularization floor, clipping threshold, pair-locking probability, and shared
birth rate; null runs without common births, without pair locking, without
clipping, with constant diffusion, and with the correlation-controlled
diffusion gate; and a response experiment in which a stationary relative profile \(F\) is perturbed by a diagonal-preserving mode in order to extract the leading
relaxation rate identified with \(\mu_{\rm FB}\). Within the present paper the BSM-MC figures serve as diagnostics of the specified closure; extracting the reciprocal gap requires the response experiment above together with a full response-field calculation.

\section{Code and data availability}
\label{app:code-data}

The BSM-MC scripts, figure-generation notebooks, run parameters, and raw
diagnostic data will be made available in an open repository or archival
supplement. The repository will identify which quantities are imposed by the
projected closure, which are measured diagnostics, and which are fitted
response coefficients.


\bibliographystyle{apsrev4-2}
\bibliography{references}

\end{document}